\documentclass[12pt,a4paper]{article}

\usepackage{algorithm}
\usepackage{algpseudocode}
\usepackage{geometry}
\RequirePackage{amsmath,amssymb}
\RequirePackage{mathrsfs}
\RequirePackage{amsfonts,bm}
\RequirePackage{dsfont}
\RequirePackage{braket}
\RequirePackage{tabularx}
\usepackage{multirow}
\RequirePackage{nicefrac}
\RequirePackage{graphicx}
\RequirePackage{booktabs}
\RequirePackage{colortbl}
\RequirePackage{xcolor}
\RequirePackage[symbol]{footmisc}
\usepackage{mathtools}
\usepackage{comment}
\usepackage{blkarray}
\usepackage{stackengine}
\usepackage{float}
\usepackage{rotating}
\usepackage{hhline}
\usepackage{tikz}
\usepackage{multirow}
\usepackage[section]{placeins}

\def\AEF{A.E. Faraggi}

\def\IJMP#1#2#3{{\it Int.\ J.\ Mod.\ Phys.}\/ {\bf A#1} (#2) #3}

\def\EJP#1#2#3{{\it Eur.\ Phys.\ Jour.}\/ {\bf C#1} (#2) #3}

\def\JHEP#1#2#3{{\it JHEP}\/ {\bf #1} (#2) #3}
\def\NPB#1#2#3{{\it Nucl.\ Phys.}\/ {\bf B#1} (#2) #3}
\def\PLB#1#2#3{{\it Phys.\ Lett.}\/ {\bf B#1} (#2) #3}
\def\PRD#1#2#3{{\it Phys.\ Rev.}\/ {\bf D#1} (#2) #3}
\def\PRL#1#2#3{{\it Phys.\ Rev.\ Lett.}\/ {\bf #1} (#2) #3}
\def\PRT#1#2#3{{\it Phys.\ Rep.}\/ {\bf#1} (#2) #3}
\def\etal{{\it et al\/}}

\def\beq{\begin{equation}}
\def\eeq{\end{equation}}
\def\beqn{\begin{eqnarray}}
\def\eeqn{\end{eqnarray}}

\usepackage{empheq}
\usepackage[most]{tcolorbox}
\usepackage{blkarray}
\newtcbox{\mymath}[1][]{%
    nobeforeafter, math upper, tcbox raise base,
    enhanced, colframe=blue!30!black,
    colback=blue!30, boxrule=1pt,
    #1}

\newcommand{\CC}[2]{C{#1\atopwithdelims[]#2}}
\newcommand{\Z}[2]{Z{#1\atopwithdelims[]#2}}

\newcommand{\ba}{\begin{eqnarray}}
\newcommand{\ea}{\end{eqnarray}}
\DeclareRobustCommand{\sqbinom}{\genfrac[]{0pt}{}}

\newcommand{\vth}{\vartheta} 
\newcommand{\vthb}{\bar{\vartheta}} 
\newcommand{\smb}[2]{\hspace{-0.075cm}\left[\begin{smallmatrix}#1\\#2\end{smallmatrix} \right]}

\newcommand{\unrevised}[1]{{\leavevmode\color{red}[#1]}}

\numberwithin{equation}{section}

\begin{document}
\begin{titlepage}
\samepage{
\setcounter{page}{1}
\rightline{LTH-1292}
\rightline{February 2022}

\vfill
\begin{center}
  {\Large \bf{
     Towards Classification \\ \medskip 
     of $\mathcal{N}=1$ and $\mathcal{N}=0$ Flipped $SU(5)$ 
     \\ \bigskip
    Asymmetric $\mathbb{Z}_2 \times \mathbb{Z}_2$ Heterotic String Orbifolds }}

\vspace{1cm}
\vfill

{\large Alon E. Faraggi$^{1,2}$\footnote{E-mail address: alon.faraggi@liverpool.ac.uk}, Viktor G. Matyas$^{1}$\footnote{E-mail address: viktor.matyas@liverpool.ac.uk} and 
 Benjamin Percival$^{1}$\footnote{E-mail address: benjamin.percival@liverpool.ac.uk}}

\vspace{1cm}

{\it $^{1}$ Dept.\ of Mathematical Sciences, University of Liverpool, Liverpool
L69 7ZL, UK\\}
\vspace{.08in}

{\it $^{2}$ Dept. of Particle Physics and Astrophysics, Weizmann Institute,
Rehovot 76100, Israel\\}

\vspace{.025in}
\end{center}

\vfill
\begin{abstract}
\noindent
The free fermionic classification method provides a powerful tool to investigate string vacua, which led to the discovery of spinor--vector duality and exophobic string models. 
We extend the classification methodology to both $\mathcal{N}=1$ and $\mathcal{N}=0$ Flipped $SU(5)$ $\mathbb{Z}_2 \times \mathbb{Z}_2$ heterotic string orbifolds with asymmetric shifts. The impact of the asymmetric assignments on the phenomenological characteristics of these models is investigated. Of particular interest is the analysis of untwisted moduli fixing for various choices of asymmetric boundary conditions. Two classes of vacua with different characteristics are systematically investigated with help from SAT/SMT algorithms, which are shown to increase search efficiency by up to two orders of magnitude, as well as providing useful tools to find contradictions between various phenomenological criteria. The general form of the partition function for the space of models is explained and given for two specific example models for different choices of asymmetric boundary conditions. Additionally, the distribution of one-loop cosmological constant contributions for samples in the two different classes of models are depicted and discussed.

\end{abstract}

\smallskip}

\end{titlepage}

\section{Introduction}\label{intro}

The heterotic string models in the free fermionic formulation are
among the string models studied in most detail to date. These
models correspond to $\mathbb{Z}_2 \times \mathbb{Z}_2$ toroidal orbifolds \cite{z2z2}
and detailed dictionaries can be used to translate between the
fermionic and bosonic representations \cite{panosdic}. Models in the
free fermionic formulation are defined in terms of a set of boundary condition
basis vectors and the associated Generalised GSO (GGSO) phases in the one-loop partition function \cite{fff}, which is constrained
by modular invariance. 
The early phenomenological constructions consisted of isolated examples,
produced by a trial and error method, with different unbroken $SO(10)$ subgroups, 
and the canonical GUT embedding of the electroweak hypercharge
\cite{fsu5, fny, slm, alr, lrs}.

Over the past two decades, 
systematic computerised classification methods of the fermionic
$\mathbb{Z}_2 \times \mathbb{Z}_2$ orbifold have been
developed. The classification method works with a fixed set of basis
vectors and the enumeration of the string vacua is obtained by
random generation of sets of GGSO projection coefficients. 
This program is progressive and each step introduces new details to the
analysis of the string vacua. 
The first step entailed classification
of $\mathcal{N}=1$ supersymmetric (SUSY) vacua with unbroken $SO(10)$ gauge symmetry 
\cite{fknr,fkr1}, which led to the discovery of Spinor-Vector Duality (SVD) \cite{fkr2, svd}. 
The classification methodology was subsequently extended to various subgroups of $SO(10)$ in \cite{acfkr,frs,slmclass,lrsclass}. It led to the discovery of exophobic string vacua \cite{acfkr}, and provides a fishing tool to extract models with particular phenomenological characteristics \cite{su62,frzprime}. 

The classification program was extended further to the case of non-supersymmetric (non-SUSY) heterotic string models in refs. \cite{so10tclass, PStclass, type0, type0bar}. This new direction fits in with the recent renewed interest in non-SUSY string phenomenology and new approaches to SUSY breaking in string theory. These include mechanisms to break supersymmetry via brane constructions \cite{NSUSYBranes1,NSUSYBranes2,NSUSYBranes3,NSUSYBranes4} or via a stringy Scherk–Schwarz mechanism \cite{SS1,SS2,SS3,CDC2,CDC4}. The non-SUSY landscape has also been further explored via coordinate-dependent compactifications \cite{CDC2,CDC4,CDC1,CDC3} which allows for interpolations between various SUSY and non-SUSY theories \cite{Interpol1,Interpol2,Interpol3,Interpol4}. This type of approach opens up the possibility to analyse the possible suppression of the one-loop cosmological constant via some interpolating parameters \cite{CoCSuppression1,CoCSuppression2,CoCSuppression3,CoCSuppression4}. From the point of view of the heterotic free fermionic classification program, two classes of non-SUSY models were developed, which are labelled as $S$--models and ${\tilde S}$--models, where $S$-models can be viewed as compactifications of the tachyon-free $SO(16)\times SO(16)$ vacuum \cite{SO16}, whereas the ${\tilde S}$--models correspond to compactifications of a tachyonic ten dimensional vacuum \cite{tach10d,spwsp,stable}. From the classification perspective, the novel aspect in these models is the proliferation of sectors that can \textit{a priori} produce tachyonic states \cite{aafs}. Phenomenologically interesting models are those that are tachyon-free, {\it i.e.} in which for a given choice of GGSO projection coefficients all the tachyonic states are projected out. This step substantially increases the sparsity of viable models, necessitating adaptation of the classification methodology towards more sophisticated computational tools. Already in the supersymmetric classifications of $SU(3)\times SU(2)\times U(1)^2$ and $SU(3)\times U(1)\times SU(2)^2$ string vacua, phenomenologically viable models were found with probability of the order $10^{-11}$, which occurs due to the proliferation of chiral exotic sectors. Allowing for non-SUSY configurations that are tachyon-free would decrease this probability further by several orders of magnitude. 

The application of fertility conditions to the $SO(10)$ level was shown to be an effective tool for increasing search efficiency in refs. \cite{slmclass, ferlrs} for these supersymmetric cases.
Furthermore, genetic algorithms have been applied in the space of Pati-Salam free fermionic models in \cite{genalgo} to improve the efficiency of fishing out viable models. Other machine learning approaches to the string landscape have become common in recent years and have been employed in various other string constructions \cite{MLReview}.
More recently, it was shown that an algorithm ideally suited for the free fermionic classification
program is provided by Satisfiability Modulo Theories (SMTs), which can impose constraints on spaces of free fermionic models and resolve them using highly optimised Boolean operations which, in the set-up of ref. \cite{fpsw}, was shown to shorten the
computer running time by up to three orders of magnitude. 

The classification methodology so far discussed has solely been developed for models with
symmetric boundary conditions. The heterotic string in general,
and the free fermionic models in particular, allow for more general
assignments of boundary conditions, which are asymmetric
between the left and the right-moving worldsheet fermions. 
These can be complicated assignments that realise the non-Abelian
gauge symmetries at higher level Kac-Moody algebra \cite{Reps}, 
or more mundane assignments that leave the gauge symmetries at level
$k=1$. Although symmetric in the $\mathbb{Z}_2\times \mathbb{Z}_2$ twists, these asymmetric assignments produce asymmetric shift orbifold models, which amount to non-geometric compactifications, a review of which is given in ref. \cite{nongeomreview}.  Completing a first step towards the extension of the classification methodology to such asymmetric orbifolds is the objective of this paper. We choose to study models with Flipped $SU(5)$ (FSU5) gauge symmetry for both the $\mathcal{N}=0$ and $\mathcal{N}=1$ cases.

There are several profound phenomenological implications of choosing such asymmetric boundary condition assignments rather than symmetric ones.
Of crucial importance to us is how they help to realise moduli fixing \cite{moduli}, top-quark Yukawa couplings from the untwisted sector \cite{yukawa} and doublet-triplet splitting \cite{dtsm}. Furthermore, we note that the
early free fermionic constructions \cite{fny,slm} do utilise asymmetric
boundary conditions, which gave rise to a stringy explanation of the
hierarchical top-bottom quark mass splitting \cite{tqmp}.

The fixing of some of the three complex and
K\"ahler structures that comprise the moduli space of the
$\mathbb{Z}_2 \times \mathbb{Z}_2$ orbifold is of particular significance in the context of investigating the one-loop potential generating the (leading order) vacuum energy of a string model. This is of key interest in this work since we classify non-supersymmetric configurations. Various works on non-supersymmetric string vacua have attempted to use Scherk-Schwarz supersymmetry breaking \cite{SS1,SS2,SS3,CDC2,CDC4} and a so-called `super no-scale' condition \cite{KP,ADM} to argue for a suppression of the one-loop cosmological constant. Florakis and Rizos demonstrated the existence of free fermion models with positive vacuum energy at the minimum of the potential for one of the radii \cite{fr1,fr2}.
However, in order to argue for stability of the vacua one needs to incorporate all moduli into the analysis, which is far too cumbersome in the symmetric orbifold case to be performed. This is where asymmetric orbifolds come into their own, as they give some control over the fixing of certain moduli. 


Our paper is organised as follows, in Section 2 we overview the key aspects of free fermionic model building. In Section 3 we explain the translation of free fermionic constraints into the language of Boolean algebra. Then we turn to explaining the construction of Flipped $SU(5)$ asymmetric orbifold models for classification in Section 4. In Section 5 we classify the asymmetric pairings of the internal fermions according to key characteristics such as the number of untwisted moduli they preserve. Following this, Section 6 details generic features of the FSU5 models we classify including the structure of their partition functions, whilst Section 7 and Section 8 deal with classifying specific example Classes of models and their classification results. Finally, in Section 9 we give conclusions.

\section{Model Building in the Free Fermionic Formulation}\label{FFModelBuilding}
The basic idea of the free fermionic formulation is to build consistent models of the heterotic string directly in four dimensions such that the additional degrees of freedom required to cancel the conformal anomaly are free fermions propagating on the string worldsheet. In particular, the spacetime lightcone coordinates $X^\mu(z,\bar{z})$ and $\psi^\mu(z)$, $\mu=1,2$, are accompanied by 18 additional internal holomorphic Majorana-Weyl fermions: $\chi^I,y^I,w^I(z)$, $I=1,...,6$, and 48 antiholomorphic Majorana-Weyl fermions $\bar{\Phi}(\bar{z})^a$. Of these, 32 are the gauge degrees of freedom of the 16-dimensional gauge lattice of the heterotic string, which we complexify into 16 fermionic fields $\overline{\psi}^{1,...,5},\ \overline{\eta}^{1,2,3}$ and $\overline{\phi}^{1,...,8}$.  The other 12 antiholomorphic real fermions we denote by $\bar{y}^I,\bar{w}^I$, $I=1,..,6$.

The real free fermions $\{y^I,w^I \ | \ \bar{y}^I,\bar{w}^I \}$ can be thought of as the (fermionic) coordinates of the internal six-dimensional manifold. A pair $\{y^I,w^I\}$ of the holomorphic internal fermions can be replaced by a chiral boson $X^I_L$ through the bosonisation equation $i\partial X^I_L=y^Iw^I$ and similarly for the antiholomorphic side. 

An important starting point for consistent model building is ensuring that there is an $N=1$ superconformal algebra on the string worldsheet from the holomorphic (supersymmetric) degrees of freedom. This can be done by realising worldsheet supersymmetry non-linearly through the worldsheet supercurrent 
\beq \label{scurrent}
T_F(z)=i\psi^\mu \partial X^\mu (z)+i\sum^6_{I=1}\chi^I y^I w^I
\eeq 
with conformal weight $(\frac{3}{2},0)$. This results in a local enhanced symmetry group $SU(2)^6$, the adjoint representation of which is given by the six $SU(2)$-triplets $\{\chi^I,y^I,w^I\}$. With this definition it is ensured that the $N=1$ superconformal algebra holds for the holomorphic degrees of freedom. 

Models in the free fermionic formalism can now be constructed by considering the toroidal worldsheet and defining a set of $N$ basis vectors, $\bm{v_i}\in \mathcal{B}$, specifying boundary conditions, $v(f)\in (-1,1]$, according to 
\beq 
\bm{v_i}=\{v(\psi^\mu),...,v(\bar{\phi}^8)\}
\eeq 
for each free fermion, $f$, as it is parallel transported around the two non-contractible loops of the torus. Ramond (R) boundary conditions corresponds to $v(f)=1$, while Neveu-Schwarz (NS) is $v(f)=0$. For the models under consideration in this paper we also have some cases of complex boundary conditions such that $v(f)=\pm\frac{1}{2}$. 

The partition function of the free fermions can be written as
\beq 
Z_f=\sum_{\bm{\alpha},\bm{\beta}} \CC{\bm{\alpha}}{\bm{\beta}} \Z{\bm{\alpha}}{\bm{\beta}},
\eeq 
where $\bm{\alpha},\bm{\beta}$ are linear combinations of the basis vectors, $\CC{\bm{\alpha}}{\bm{\beta}}$ are Generalised GSO (GGSO) phases and $Z  {\bm{\alpha} \atopwithdelims [] \bm{\beta}}$ will be products of Jacobi theta functions. More details on the form of this partition function for the models we consider are given in Section \ref{PF}.
The specification of the GGSO phases in agreement with modular invariance is thus a key component of defining a consistent free fermionic model. A general GGSO phase between sectors $\bm{\alpha}=m_i\bm{v_i}$ and $\bm{\beta}=n_i\bm{v_i}$ can be decomposed into the GGSO phases between basis vectors, $\bm{v_i}$, through the equation 
\beq \label{GGSOdecomp}
\CC{\bm{\alpha}}{\bm{\beta}}= \Gamma(\bm{\alpha},\bm{\beta}) \prod_{i,j}\CC{\bm{v_i}}{\bm{v_j}}^{m_in_i},
\eeq 
where the details of the prefactor $\Gamma(\bm{\alpha},\bm{\beta})$ can be found, for example, in ref. \cite{fff}. 

Once the basis vectors and GGSO phases are specified, the modular invariant Hilbert space $\mathcal{H}$ of states $\ket{S_{\bm{\alpha}}}$ is found through implementing the one-loop GGSO projection according to:
\begin{equation}
    \mathcal{H}=\bigoplus_{\bm{\alpha}\in \Xi}\prod^{N}_{i=1}
    \left\{ e^{i\pi \bm{v_i}\cdot F_{\bm{\alpha}}}\ket{S_{\bm{\alpha}}}=\delta_{\bm{\alpha}}
    \CC{\bm{\alpha}}{\bm{v_i}}^*
    \ket{S_{\bm{\alpha}}}\right\}\mathcal{H}_{\bm{\alpha}},
\end{equation}
where $F_{\bm{\alpha}}$ is the fermion number operator, $\Xi$ is the additive group given by the span of the basis vectors and $\delta_{\bm{\alpha}}=1,-1$ is the spin-statistics index.

The sectors, $\bm{\alpha}$, in the model can be characterised according to their left and
right moving vacuum separately
\begin{align}\label{massform}
\begin{split}
M_L^2&=-\frac{1}{2}+\frac{\bm{\alpha}_L \cdot\bm{\alpha}_L}{8}+N_L\\
M_R^2 &=-1+\frac{\bm{\alpha}_R \cdot \bm{\alpha}_R}{8}+N_R 
\end{split}
\end{align}
where $N_L$ and $N_R$ are sums over left and right moving oscillator frequencies,
respectively
\begin{align}
    N_L&=\sum_{\lambda}\nu_\lambda+\sum_{\lambda^*}\nu_{\lambda^*} \\
    N_R&=\sum_{\bar{\lambda}}\nu_{\bar{\lambda}}+\sum_{\bar{\lambda}^*}\nu_{\bar{\lambda}^* } 
\end{align}
where $\lambda$ is a holomophic oscillator and $\bar{\lambda}$ is an antiholomorphic oscillator and the frequency is defined through the boundary condition in the sector $\bm{\alpha}$
\beq \label{freq}
\nu_\lambda = \frac{1+\alpha(\lambda)}{2}, \ \ \ \nu_{\bar{\lambda}} = \frac{1-\alpha(\lambda)}{2}.
\eeq 
Physical states must satisfy the
Virasoro matching condition, $M_L^2=M_R^2$, such that massless states are those with $M_L^2=M_R^2=0$ and on-shell tachyons arise for sectors with $M_L^2=M_R^2<0$. 

The fermionisation of the worldsheet degrees of freedom employed in the free fermionic construction demands that the heterotic string is constructed at the self-dual point in the moduli space where the radii are fixed to $R=\sqrt{\alpha'/2}$. At this point the theory is consistent and, as was shown in \cite{fff}, the modular invariance constraints for one-loop and higher-loop amplitudes can be completely solved. A key advantage of the free fermionic formulation is that non-geometric constructions \cite{nongeomreview}
such as the asymmetric orbifolds studied in this paper may be realised naturally. However, being fixed at the self-dual point in the moduli space becomes restrictive when we wish to study stability issues that arise for non-supersymmetric string models. For example, it becomes essential to introduce moduli dependence when investigating the non-trivial one-loop potential arising in the absence of supersymmetry or to understand the type of supersymmetry breaking within the string model. To tackle these issues it therefore requires a translation of the free fermionic theory into a bosonic orbifold construction using the tools reviewed in \cite{florakis} and employed in \cite{fr1,fr2}. 

For the purposes of initiating a systematic classification of asymmetric orbifolds, the free fermionic construction is the perfect starting point since it provides great computational power through the simplicity of representation for worldsheet boundary conditions, allowing for easy algebraic expressions for key features of the spectrum and modular properties of the partition function. It then allows for large spaces of string vacua to be analysed with computer programs dealing with these simple algebraic expressions. Notably, the binary nature of the worldsheet boundary conditions in the free fermionic construction allows for an immediate translation of phenomenological constraints into a language interpretable by powerful SAT/SMT computer algorithms trained on Boolean expressions. This method will be explored in the next section.

\section{SMTs and Free Fermionic Classification}\label{SMTClass}
The advanced computer algorithms Satisfiability Modulo Theories (SMTs) were introduced as a tool for exploring the string landscape in \cite{fpsw} where they increased the efficiency of solving constraints from free fermionic models by three orders of magnitude. Compared with the more primitive SAT solvers that can only interpret Boolean formulae, SMTs allow for operations over non-Boolean types such as integers, reals,
bitvectors, and arrays. There is a well-known trade off between efficiency and expressibility however, since a Boolean encoding of a problem will drastically increase efficiency compared with even a simple encoding in terms of integers. This too was demonstrated for free fermionic constraints in \cite{fpsw}. 

It is a great advantage of the free fermionic construction that phenomenological constraints reduce to elementary algebraic expressions in terms of the GGSO phases $\CC{\bm{v_i}}{\bm{v_j}}$, which are binary inputs. Therefore the reduction to Boolean expression is almost immediate but it is worthwhile giving some details on how to perform this reduction before we go on to the specifics of the asymmetric orbifold models we seek to classify.  Before doing so, it is worth highlighting that a quite different application of the SMT solver is employed in Section \ref{Pairings} to classify the asymmetric pairings of the internal fermions $\{y^I,w^I \ | \ \bar{y}^I,\bar{w}^I\}$ according to various characteristics they impose on the Class of asymmetric orbifold models. This pairing classification could be performed with more familiar programming tools in reasonable computing time since there are only 24 inputs. However, the Z3 SMT tool \cite{Z3} we use makes it very easy to impose the constraints and completely classifies the pairings in $\sim 25$ seconds, which would be tough to beat with other approaches. This application demonstrates how versatile the SMT tool can be in helping with a variety of problems that require large sets of constraints to be satisfied.

\subsection{Boolean Reduction}\label{BoolReduc}

In the classification program of free fermionic models, phenomenological criteria are typically constructed through selection rules on certain sectors. Typically these are massless sectors or, for non-supersymmetric models, the on-shell tachyonic sectors when ensuring that the models are tachyon-free. Depending on the mass formulae (\ref{massform}), the level-matching condition may necessitate that there are left-moving oscillators, $\lambda$, or right-moving oscillators, $\bar{\lambda}$, acting on the ket vector of the sector $\bm{\alpha}$, which is given by the combined degenerate Ramond vacua of all fermions with $\alpha(f)=1$ in $\bm{\alpha}$. 

Given a basis, to determine whether a particular sector $\bm{\alpha}$ survives for a model depends solely on the GGSO phase configuration. In particular, taking a sector with no oscillators $\ket{\bm{\alpha}}$ and employing the notation of \cite{fr2}, the survival/projection condition is encapsulated in the generalised projector
\beq \label{GProj}
\mathbb{P}_{\bm{\alpha}}=\prod_{\bm{\xi}\in \Upsilon(\bm{\alpha})}\frac{1}{2}\left( 1+\delta_{\bm{\alpha}} \CC{\bm{\alpha}}{\bm{\xi}}\right)
\eeq 
where 
\beq 
\delta_{\bm{\alpha}}=\begin{cases} +1 \ \ \text{if }  \alpha(\psi^\mu)=0 \ \ \iff \ \ \text{sector is bosonic}\\
-1 \ \ \text{if }  \alpha(\psi^\mu)=1 \ \ \iff \ \ \text{sector is fermionic}.
\end{cases}
\eeq 
The $\Upsilon(\bm{\alpha})$ is defined as a minimal linearly independent set of vectors $\bm{\xi}$ such that $\bm{\xi} \cap \bm{\alpha}=\emptyset$. To check whether the sector $\bm{\alpha}$ is projected simply amounts to checking $\mathbb{P}_{\bm{\alpha}}=0$.

In the presence of a right-moving oscillator, $\bar{\lambda}$, for example, this generalised projector is modified to
\beq \label{GProjOScill}
\mathbb{P}_{\bm{\alpha}}=\prod_{\bm{\xi}\in \Upsilon(\bm{\alpha})}\frac{1}{2}\left( 1+\delta_{\bm{\alpha}} \delta^{\bar{\lambda}}_{\bm{\xi}} \CC{\bm{\alpha}}{\bm{\xi}}\right)
\eeq 
such that 
\beq 
\delta^{\bar{\lambda}}_{\bm{\xi}}=\begin{cases}
+1 \ \ \text{if } \ \bar{\lambda} \in \bm{\xi} \\
-1 \ \ \text{if } \ \bar{\lambda}\notin \bm{\xi}.
\end{cases}
\eeq 
and there would be an analogous insertion of $\delta^\lambda_{\bm{\xi}}$ for a left-moving oscillator $\lambda$. 

The translation of these generalised projector equations that build up to form our phenomenological criteria into a Boolean language ripe for a SAT or SMT solver is immediate. For example, the constraint $\mathbb{P}_{\bm{\alpha}}=0$ for the case (\ref{GProj}) can be translated into a Boolean constraint by taking the set 
\beq 
C=\left\{\delta_{\bm{\alpha}} \CC{\bm{\alpha}}{\bm{\xi}_1},...,\delta_{\bm{\alpha}} \CC{\bm{\alpha}}{\bm{\xi}_n}\right\},
\eeq 
with $n=|\Upsilon(\bm{\alpha})|$. Using eq. (\ref{GGSOdecomp}) we can rewrite the phases $\CC{\bm{\alpha}}{\bm{\xi}_i}$ in terms of a product of basis GGSO phases $\CC{\bm{v_i}}{\bm{v_j}}$. We then associate to each of these a Boolean. In the case of a real GGSO phase this can be taken to be
\beq \label{Bool}
B_{ij}=\begin{cases}
\texttt{True} ~~\text{ if }  \CC{\bm{v_i}}{\bm{v_j}}=-1 \\
\texttt{False } \text{if }  \CC{\bm{v_i}}{\bm{v_j}}=+1 \\
\end{cases}
\eeq
where $i,j=1,...,N$ such that $N=|\mathcal{B}|$ is the number of basis vectors. Then the product of such basis GGSO phases can be rewritten in Boolean language as an exclusive or, $\bar{\lor}$, that returns $\texttt{True}$ if there is an odd number of $\texttt{True}$ $\CC{\bm{v_i}}{\bm{v_j}}$'s in the product, and $\texttt{False}$ if even. In this way, each entry $\delta_{\bm{\alpha}} \CC{\bm{\alpha}}{\bm{\xi}_i}$ of $C$ can be recast as a Boolean clause constructed through the $\bar{\lor}$ of basis GGSO phases. There is then the complication of dealing with any imaginary basis GGSO phases. This is easy to resolve by consistently taking $\pm i$ as the binary to assign Boolean values to. 
Then the set $C$ can be redefined as a set of Boolean clauses.

Once this reduction to Booleans is complete, the constraint $\mathbb{P}_{\bm{\alpha}}=0$ from (\ref{GProj}) is equivalent to imposing
\beq 
\neg(C_1 \land ... \land C_n).
\eeq 
where $C_i\in C$, $i=1,...,n$. It is precisely this kind of constraint that can be added to a constraint system for an SAT/SMT solver such as Microsoft's open source solver Z3 that we employ in this work. 

The absence of tachyonic sectors for non-supersymmetric models is a repeated application of this constraint for all on-shell tachyonic sectors. A phenomenological constraint such as checking for three generations requires a couple of additional steps that need encoding into SMT language. Once again, sectors giving rise to the fermion generations need checking for survival via $\mathbb{P}_{\bm{\alpha}}\neq 0$, then additional GGSO projection determining the chirality of the sectors under the relevant observable gauge factors need encoding. This can be done as a natural extension of the projections done in $\mathbb{P}_{\bm{\alpha}}$. Then, checking for 3 generations can be handled easily through the use of a Boolean adder. 

\section{Asymmetric Orbifold Classification Set-up}\label{setup}
In this work we begin the task of extending the classification methodology to the space of asymmetric orbifolds. There are several appealing features of such asymmetric orbifolds, including the presence of an untwisted Doublet-Triplet splitting mechanism, realistic Yukawa couplings and the projection of untwisted moduli. 

We can take the NAHE-set \cite{NAHE} as the starting point for classifying large spaces of asymmetric orbifolds, which is the set: 
\begin{align}
    \begin{split}
    \label{NAHE}
    \bm{\mathds{1}}&=\{\psi^\mu,\
    \chi^{1,\dots,6},y^{1,\dots,6}, w^{1,\dots,6}\ | \ \overline{y}^{1,\dots,6},\overline{w}^{1,\dots,6},
    \overline{\psi}^{1,\dots,5},\overline{\eta}^{1,2,3},\overline{\phi}^{1,\dots,8}\},\\
    \bm{S}&=\{{\psi^\mu},\chi^{1,\dots,6}  \ \}\\
    \bm{b_1}&=\{\psi^\mu,\chi^{12},y^{34},y^{56}\; | \; \overline{y}^{34},\overline{y}^{56},\overline{\psi}^{1,\dots,5},\overline{\eta}^1\},\\
    \bm{b_2}&=\{\psi^\mu,\chi^{34},y^{12},w^{56}\; | \; \overline{y}^{12},\overline{w}^{56},\overline{\psi}^{1,\dots,5},\overline{\eta}^2\},\\
    \bm{b_3}&=\{\psi^\mu,\chi^{56},w^{12},w^{34}\; | \; \overline{w}^{12},\overline{w}^{34},\overline{\psi}^{1,\dots,5},\overline{\eta}^3\}
    \end{split}
\end{align}
which gives rise to an $SO(10)$ symmetric GUT and due to the $\bm{S}$ vector can realise $\mathcal{N}=1$ supersymmetry for appropriate choices of GGSO phases. We will then choose to add the additional basis vectors:
\begin{align}
    \begin{split}
    \bm{x}&=\{\bar{\psi}^{1,...,5},\bar{\eta}^{1,2,3}\}\\
    \bm{z_1}&=\{ \bar{\phi}^{1,...,4} \}
    \end{split}
\end{align}
such that $\bm{z_1}$ reduces the dimension of the Hidden gauge group and the $\bm{x}$ vector induces the enhancement $SO(10)\times U(1)\rightarrow E_6$ for certain choices of GGSO phases which can be seen as taking us from the space of vacua with $(2,0)$ worldsheet supersymmetry to those with $(2,2)$. 

The untwisted gauge group is 
\beq 
SO(10)\times SO(4)^3\times U(1)^3\times SO(8)\times SO(8)
\eeq  
at this level, with the three $SO(4)$ factors arising from the three groups of internal fermions from the $\bm{b_k}$, $k=1,2,3$, such that the NS sector gauge bosons can be written 
\begin{align}
&\psi^\mu \{ \bar{y}^{3,4,5,6}\}\{ \bar{y}^{3,4,5,6}\} \ket{0}_{NS},\nonumber \\ 
&\psi^\mu \{ \bar{y}^{1,2},\bar{w}^{5,6}\}\{ \bar{y}^{1,2},\bar{w}^{5,6}\} \ket{0}_{NS} \\  
&\psi^\mu \{ \bar{w}^{1,2,3,4}\}\{ \bar{w}^{1,2,3,4}\} \ket{0}_{NS}. \nonumber
\end{align}
The NAHE-set naturally implements a $\mathbb{Z}_2\times \mathbb{Z}_2$ orbifolding through the twist vectors $\bm{b_k}$ that leave an untwisted moduli space of
\beq 
\left(\frac{SO(2,2)}{SO(2)\times SO(2)}\right)^3
\eeq 
where each of the three factors is parameterised by the moduli scalar fields from the NS sector
\beq \label{hij}
h_{ij}=\ket{\chi^i}_L\otimes \ket{\bar{y}^j \bar{w}^j}_R=
\begin{cases} 
(i,j=1,2)\\
(i,j=3,4)\\
(i,j=5,6)
\end{cases}.
\eeq 
In free fermionic models these untwisted moduli are in one to one correspondence with marginal operators that generate Abelian Thirring Interactions. For the NAHE-set the only such marginal operators left invariant are 
\beq \label{JiJj}
J^i_L(z)\bar{J}_R^j(\bar{z})=:y^iw^i::\bar{y}^j \bar{w}^j: =
\begin{cases} 
(i,j=1,2)\\
(i,j=3,4)\\
(i,j=5,6).
\end{cases}
\eeq
From this it is straightforward to observe that the projection or retention of moduli is governed by the boundary conditions of the set of 12 internal real fermions $\{y^I,w^I \ | \ \bar{y}^I,\bar{w}^I\}$. In particular, we note that if the basis remains left-right symmetric in these internal fermions then all the untwisted moduli of the NAHE-set are  retained. This is a central reason for attempting to classify asymmetric orbifolds models where the internal real fermions $\{y^I,w^I \ | \ \bar{y}^I,\bar{w}^I\}$ are not left-right symmetric. 

In order to make the connection between the fields $h_{ij}$ and the familiar three K\"ahler and three complex structure moduli of the $\mathbb{Z}_2\times \mathbb{Z}_2$ orbifold we can construct six complex moduli from the six real ones of eq. (\ref{hij}). For the first complex plane we can write
\begin{align}
    \begin{split}
        H_1^{(1)}&=\frac{1}{\sqrt{2}}(h_{11}+ih_{21})=\frac{1}{\sqrt{2}}\ket{\chi^1+i\chi^2}_L\otimes \ket{\bar{y}^1 \bar{w}^1}_R\\
        H_2^{(1)}&=\frac{1}{\sqrt{2}}(h_{12}+ih_{22})=\frac{1}{\sqrt{2}}\ket{\chi^1+i\chi^2}_L\otimes \ket{\bar{y}^2 \bar{w}^2}_R
    \end{split}
\end{align}
which can then be combined to define the K\"ahler and complex structure moduli for the first complex plane
\begin{align}
    \begin{split}
        T_1&=\frac{1}{\sqrt{2}}(H_1^{(1)}-iH_2^{(1)})=\frac{1}{\sqrt{2}}\ket{\chi^1+i\chi^2}_L\otimes \ket{\bar{y}^1 \bar{w}^1-i\bar{y}^2 \bar{w}^2}_R\\
        U_1&=\frac{1}{\sqrt{2}}(H_1^{(1)}+iH_2^{(1)})=\frac{1}{\sqrt{2}}\ket{\chi^1+i\chi^2}_L\otimes \ket{\bar{y}^1 \bar{w}^1+i\bar{y}^2 \bar{w}^2}_R
    \end{split}
\end{align}
and similarly for $T_{2,3}$ and $U_{2,3}$. 

We choose to classify Flipped SU(5) models such that a single basis vector both breaks the $SO(10)$ GUT and assigns asymmetric pairings to the internal fermions. This vector can then be taken to be of the general form
\beq \label{GeneralGamma}
\bm{\gamma} = \bm{A}+\{ \bar{\psi}^{1,...,5}=\bar{\eta}^{1,2,3}=\bar{\phi}^{1,2,6,7}=\frac{1}{2}\}+\bm{B}.
\eeq
where $\bm{A}$ ensures that the internal fermions are not symmetrically paired and $\bm{B}$ assigns appropriate boundary conditions to the hidden complex fermions
\beq 
\bm{B}=\{B(\bar{\phi}^3),B(\bar{\phi}^4),B(\bar{\phi}^5),B(\bar{\phi}^8)\}
\eeq 
where we choose real boundary conditions $B(\bar{\phi}^{3,4,5,8})=0,1$
so as to be consistent with the modular invariance rules
\begin{align}
&N_{\gamma}\bm{\gamma}\cdot \bm{\gamma}=0 \ \text{mod} \ 8.\\
&N_{z_1\gamma}\bm{z_1}\cdot \bm{\gamma}=0 \ \text{mod} \ 4
\end{align}
where $N_{\gamma}$ is the smallest positive integer for which $N_{\gamma}\bm{\gamma} = 0$ and $N_{z_1\gamma}$ is
the least common multiple of $N_{z_1}$ and $N_{\gamma}$. 

The supercurrent constraint (\ref{scurrent}) imposes a different constraint on these boundary conditions depending on whether $\bm{\gamma}$ is fermionic or bosonic. 
In the bosonic case we can write $\bm{A}$ as
\beq \label{Abos}
\bm{A}=\{A(y^1),...,A(y^6),A(w^1),...,A(w^6)\ | \ A(\bar{y}^1),...,A(\bar{y}^6),A(\bar{w}^1),...,A(\bar{w}^6)\}
\eeq 
and (\ref{scurrent}) thus imposes that the boundary condition of the holomorphic internal fermions are
\beq \label{BosBCs}
(y^I,w^I)=(1,1) \ \text{or} \ (0,0), \ \ \ I=1,...,6
\eeq
to ensure a consistent supercurrent. On the other hand, if $\bm{\gamma}$ is fermionic then we choose $\bm{A}$ to be of the form
\beq \label{Aferm}
\bm{A}=\{\psi^\mu,\chi^{12},A(y^1),...,A(y^6),A(w^1),...,A(w^6)\ | \ A(\bar{y}^1),...,A(\bar{y}^6),A(\bar{w}^1),...,A(\bar{w}^6)\}
\eeq 
and the supercurrent consistency imposes that
\beq \label{FermBCs}
(y^I,w^I)=
\begin{cases} 
(0,0) \ \text{or} \ (1,1), \ \ \ I=1,2 \\
(1,0) \ \text{or} \ (0,1), \ \ \ I=3,...,6
\end{cases}
\eeq
and similar for the cases where $A(\chi^{34})=1$ or $A(\chi^{56})=1$ and $A(\chi^{12})=0$.

The next step towards classifying Flipped SU(5) asymmetric orbifolds is the addition of the symmetric shift vectors:
\beq 
\bm{e_i}=\{y^i,w^i\ | \ \bar{y}^i,\bar{w}^i \}, \ \ \ i=1,...,6
\eeq 
so long as they are consistent with the choice of $\bm{\gamma}$, in the sense that they satisfy the modular invariance rule
\beq 
N_{\gamma e_i} \bm{\gamma} \cdot \bm{e_i} = 0 \ \ \text{mod } 4. 
\eeq 
 In the previous classifications of symmetric orbifolds all six $\bm{e_i}$'s are present in the basis to impose the 12 symmetric pairings between $\{y^I,w^I\}$ and $\{\bar{y}^I,\bar{w}^I\}$ to form 12 Ising model operators. One corollary of this symmetric pairing is that the rank of the untwisted gauge group from the holomorphic sector takes its minimal value of $16$. However, asymmetric pairings will generate up to six additional $U(1)$'s from the pairing of two antiholomophic internal fermions $\{\bar{y}^I,\bar{w}^I\}$. 

Putting this all together, we can write the basis we take as a starting point for exploring the space of asymmetric orbifolds as 
\begin{align}
    \begin{split}
    \label{basis}
    \bm{\mathds{1}}&=\{\psi^\mu,\
    \chi^{1,\dots,6},y^{1,\dots,6}, w^{1,\dots,6}\ | \ \overline{y}^{1,\dots,6},\overline{w}^{1,\dots,6},
    \overline{\psi}^{1,\dots,5},\overline{\eta}^{1,2,3},\overline{\phi}^{1,\dots,8}\},\\
    \bm{S}&=\{{\psi^\mu},\chi^{1,\dots,6}  \ \}\\
    \bm{e_i} &= \{y^i,w^i\ | \ \bar{y}^i,\bar{w}^i \}, \ \ \ i \subset \{1,2,3,4,5,6\}\\
    \bm{b_1}&=\{\psi^\mu,\chi^{12},y^{34},y^{56}\; | \; \overline{y}^{34},\overline{y}^{56},\overline{\psi}^{1,\dots,5},\overline{\eta}^1\},\\
    \bm{b_2}&=\{\psi^\mu,\chi^{34},y^{12},w^{56}\; | \; \overline{y}^{12},\overline{w}^{56},\overline{\psi}^{1,\dots,5},\overline{\eta}^2\},\\
    \bm{b_3}&=\{\psi^\mu,\chi^{56},w^{12},w^{34}\; | \; \overline{w}^{12},\overline{w}^{34},\overline{\psi}^{1,\dots,5},\overline{\eta}^3\}\\
    \bm{z_1}&=\{\bar{\phi}^{1,2,3,4}\}\\
    \bm{x}&=\{\bar{\psi}^{1,...,5},\bar{\eta}^{1,2,3}\}\\
    \bm{\gamma} &=\bm{A}+ \{
\bar{\psi}^{1,...,5}=\bar{\eta}^{1,2,3}=\bar{\phi}^{1,2,5,6}=\frac{1}{2}\}+\bm{B} 
    \end{split}
\end{align}

We furthermore note the existence of the following important linear combination of hidden fermions
\beq \label{z2}
\bm{z_2}=\bm{\mathds{1}} +\sum_{k=1}^3 \bm{b_k}+\bm{z_1}=\{\bar{\phi}^{5,6,7,8}\}.
\eeq 
and the combination generating the internal fermions
\beq \label{G}
\bm{G}=\bm{S}+\sum_{k=1}^3 \bm{b_k}+\bm{x}=\{y^I,w^I \ | \ \bar{y}^I,\bar{w}^I\}, \ \ \ I=1,2,3,4,5,6.
\eeq 

Our approach towards this classification will be two-fold. The first step is to classify the asymmetric pairings within $\bm{\gamma}$ given through the $\bm{A}$ vector in both the bosonic case (\ref{Abos}) and fermionic case (\ref{Aferm}) with respect to their impact on important characteristics of the resultant models such as the number of retained moduli. The details are presented in the next section. This step is new to the classification program due to the asymmetric orbifolding. The second step is to pick a particular pairing and perform a classification of the resultant space of vacua according to their phenomenological features, such as the number of particle generations at the Flipped $SU(5)$ level. 

\section{Classification of Asymmetric Pairings} \label{Pairings}

Due to the centrality of the pairings of the internal fermions $\{y^I,w^I \ | \ \bar{y}^I,\bar{w}^I\}$ in determining important features of the class of asymmetric orbifold models, a useful first step towards classifying the asymmetric orbifolds is to classify their possible pairings defined through the vector $\bm{A}$. The key criteria we can classify these pairings according to will be the untwisted moduli they retain 
and their number of possible chiral generations. 

A convenient tool for classifying these pairings is to use an SAT/SMT solver such as Z3, as discussed in Section \ref{SMTClass}, 
where the input is a list of 24 Boolean variables determining the boundary conditions $\{A(y^{1,...,6},w^{1,...,6}) | A(\bar{y}^{1,...,6},\bar{w}^{1,...,6})\}$ within $\bm{A}$. This is sufficient for both the bosonic case (\ref{Abos}) and the fermionic case (\ref{Aferm}) with the respective boundary conditions (\ref{BosBCs}) and (\ref{FermBCs}) from the supercurrent condition. Imposing the relevant supercurrent constraint, as well as ensuring the pairing is asymmetric and consistent with the NAHE set allows us to generate all possible pairings as output from the SAT/SMT solver. 

\subsection{Asymmetric Pairings and 
Three Generations} \label{Asymm}
One key phenomenological feature impacted by the choice of pairings in $\bm{A}$ is on the number of observable spinorial sectors that are required to give rise to the particle generations. 
In order to explore this further it will be helpful to define two quantities which result from a choice of pairings $\bm{A}$. Firstly we have
\begin{align}\label{ED}
&\bm{E}=(E_1,E_2,E_3,E_4,E_5,E_6) \ \ \text{s.t. } \ \begin{cases}
E_i=1 \ \text{ if } \ A(y^i)=A(w^i)=A(\bar{y}^i)=A(\bar{w}^i)=0\\ 
E_i=0 \ \text{else}
\end{cases}
\end{align}
for $i=1,...,6$. This simply quantifies which of the $\bm{e_i}$ symmetric shift vectors remain in the basis. We can note that any asymmetric pairing automatically makes two $\bm{e_i}$ incompatible with modular invariance constraints and therefore
\beq 
\max \left(\sum_i E_i\right) =4.
\eeq
The second quantity we can define is
\begin{align}
&\bm{\Delta}=(\Delta_1,\Delta_2,\Delta_3) \ \ \text{s.t. } \ \begin{cases}
    \Delta_1=0 \ \ \text{if} \ A(y^{3456})=A(\bar{y}^{3456})\\
    \Delta_1=1 \ \ \text{else}
    \end{cases}\\
\end{align}
and similarly for $\Delta_2$ and $\Delta_3$. This notation has been employed, for example, in \cite{moduli} and \cite{towards}. With this notation defined we can now consider the fermion generations. 

At the level of the NAHE-set $\{\mathds{1},\bm{S},\bm{b_1},\bm{b_2},\bm{b_3}\}$, the sectors $\bm{b_1},\ \bm{b_2}$ and $\bm{b_3}$, if present in the massless spectrum, give rise to sixteen copies of the $\mathbf{16}$ or $\overline{\mathbf{16}}$ of $SO(10)$ due to the degeneracy of the sets of internal fermions 
$\{y^{3,4,5,6} \ | \ \bar{y}^{3,4,5,6},\bar{\eta}^1\}, \ \{y^{1,2},w^{5,6}\ |\ \bar{y}^{1,2},\bar{w}^{5,6},\bar{\eta}^2\}$ and $\{w^{1,2,3,4}\ | \ \bar{w}^{1,2,3,4},\bar{\eta}^3\}$, respectively. The addition of $\bm{x}$ reduces the degeneracy to eight copies of $\mathbf{16}$ or $\overline{\mathbf{16}}$ by separating out the $\bar{\eta}^k$ for each plane. 

In the classification program for symmetric orbifolds, the basis contains all six symmetric shift $\bm{e_i}$ vectors. These symmetric shifts completely remove the degeneracy on the three orbifold planes and the sectors giving rise to observable spinorial states from the $\mathbf{16}/\overline{\mathbf{16}}$ of $SO(10)$ are
\begin{align}
    \begin{split}
        \bm{F}^1_{pqrs} &= \bm{b_1}+p\bm{e_3}+q\bm{e_4}+r\bm{e_5}+s\bm{e_6}\\ 
        \bm{F}^2_{pqrs} &= \bm{b_2}+p\bm{e_1}+q\bm{e_2}+r\bm{e_5}+s\bm{e_6}\\ 
        \bm{F}^3_{pqrs} &= \bm{b_3}+p\bm{e_1}+q\bm{e_2}+r\bm{e_3}+s\bm{e_4}.
    \end{split}
\end{align}
such that any sector $\bm{F}^k_{pqrs}$, $k=1,2,3$, in the massless spectrum produces exactly one $\mathbf{16}$ or $\overline{\mathbf{16}}$. 

This picture requires adjustment for the case of the Flipped $SU(5)$ asymmetric orbifolds generated by the basis of eq. (\ref{basis}). In particular, the number and degeneracy of each group of sectors $\bm{F}^k_{pqrs}$ will vary according to the pairing choice $\bm{A}$. More specifically, we will see that the degeneracies of each plane can be written as a function of $\bm{E}$ and $\bm{\Delta}$.

The impact of the inclusion of an $\bm{e_i}$ vector in the basis (\ref{basis}) on the degeneracy of each orbifold plane can be seen to reduce the degeneracy of the orbifold plane $k=1,2,3$ by a factor two if $\bm{e_i} \cap \bm{b_k} \neq \emptyset$. Similarly, an asymmetric pairing in one of the three planes, {\it i.e.} $\Delta_k=1$, will also reduce the degeneracy by a factor 2. 

We can now write the degeneracies as a vector
\beq\label{D}
\bm{D}=(D_1,D_2,D_3)
\eeq  
for each orbifold plane such that
\begin{align}\label{Ds}
    D_1&=\frac{8}{2^{\Delta_1+E_3+E_4+E_5+E_6}}\\
    D_2&=\frac{8}{2^{\Delta_2+E_1+E_2+E_5+E_6}}\\
    D_3&=\frac{8}{2^{\Delta_3+E_1+E_2+E_3+E_4}},
\end{align}
and we note that 
\beq 
\min \left(D_k\right)=\frac{1}{2}
\eeq 
which when true tells us that the sectors $\bm{F}^k_{pqrs}$ will give rise to one component of the FSU5 representations of the $\mathbf{16}$ or $\overline{\mathbf{16}}$ and not the whole $SO(10)$ representation. In particular, since the decomposition under $SU(5)\times U(1)$ is
\begin{align}
\label{decomp}
\mathbf{16}=&\left(\mathbf{10}, +\frac{1}{2}\right)+\left(\mathbf{\bar{5}}, -\frac{3}{2} \right) + \left(\mathbf{1},~~\frac{5}{2}\right) \\ 
\mathbf{\overline{16}}=&\left(\mathbf{\overline{10}}, -\frac{1}{2}\right)+\left(\mathbf{5}, +\frac{3}{2}\right) +\left(\mathbf{1},-\frac{5}{2}\right) 
\end{align}
sectors $\bm{F}^k_{pqrs}$ with $D_k=\frac{1}{2}$ will generate either the states with representation $\left(\mathbf{10}, +\frac{1}{2}\right)$ or those transforming under $\left(\mathbf{\bar{5}}, -\frac{3}{2} \right) + \left(\mathbf{1},\frac{5}{2}\right)$, in the case of the sector being from $\mathbf{16}$. 

Once we calculate the degeneracies $(D_1,D_2,D_3)$ from $\bm{A}$ we can immediately check a necessary, but certainly not sufficient, condition for the presence of odd and, in particular, three generations, which is simply 
\beq \label{OddGen}
\exists \ k\in\{1,2,3\}: \  D_k\leq 1.
\eeq 
A sufficient condition for the presence of three generations is presented in Section \ref{CIA} but the condition (\ref{OddGen}) can be checked immediately from the pairing choice $\bm{A}$ so will be tested for in the classification of pairings performed in this section.




\subsection{Asymmetric Pairings and Retained Moduli}
As mentioned in Section \ref{setup}, the moduli scalar fields (\ref{hij}) are in one to one correspondence with the marginal operators (\ref{JiJj}). From the form of these operators we can immediately derive conditions on their retention/projection depending on the boundary condition assignments from $\bm{A}$. The result is
\beq 
J^i_L(z)\bar{J}^j_R(\bar{z})\begin{cases}
\ \text{retained if  } \ \  \left[A(y^i)+A(w^i)+A(\bar{y}^j)+A(\bar{w}^j)\right] \ \text{mod } 2 =0\\
\ \text{projected if  } \ \left[A(y^i)+A(w^i)+A(\bar{y}^j)+A(\bar{w}^j)\right] \ \text{mod } 2 =1.
\end{cases}
\eeq 
It will be useful when constructing the pairing classification Tables \ref{PairingsClassB} and \ref{PairingsClassF} to write the number of retained moduli in each orbifold plane as a triple 
\beq \label{M}
\bm{M}=(M_1,M_2,M_3).
\eeq 

\subsection{Results for Classification of Pairings}\label{PairingTables}
The result of the classification of asymmetric pairings with a bosonic $\bm{A}$ are summarised in Table \ref{PairingsClassB} and with fermionic $\bm{A}$ for Table \ref{PairingsClassF}. The data most important to consider is the number of untwisted moduli retained in each plane (\ref{M}) and whether odd number generations are possible through checking (\ref{OddGen}).  
The Z3 SMT classifies all the asymmetric pairings in each case, bosonic and fermionic, in approximately 20 seconds. 

\begin{table}[!htb]
\footnotesize
\centering
\begin{tabular}{|c|c|c|}
\hline
Untwisted Moduli in each Torus & Odd Number Generations Possible & Frequency\\
\hline
(2, 2, 0) & No & 992 \\ \hline
(2, 0, 2) & No & 992 \\ \hline
(0, 2, 2) & No & 992 \\ \hline
(4, 2, 2) & No & 824 \\ \hline
(2, 4, 2) & No & 824 \\ \hline
(2, 2, 4) & No & 824 \\ \hline
(0, 0, 0) & No & 256 \\ \hline
(4, 0, 0) & No & 244 \\ \hline
(0, 4, 0) & No & 244 \\ \hline
(0, 0, 4) & No & 244 \\ \hline
(4, 4, 0) & No & 200 \\ \hline
(4, 2, 2) & Yes & 200 \\ \hline
(4, 0, 4) & No & 200 \\ \hline
(2, 4, 2) & Yes & 200 \\ \hline
(2, 2, 4) & Yes & 200 \\ \hline
(0, 4, 4) & No & 200 \\ \hline
(4, 4, 4) & No & 146 \\ \hline
(4, 4, 4) & Yes & 94 \\ \hline
(4, 4, 0) & Yes & 56 \\ \hline
(4, 0, 4) & Yes & 56 \\ \hline
(0, 4, 4) & Yes & 56 \\ \hline
(2, 2, 0) & Yes & 32 \\ \hline
(2, 0, 2) & Yes & 32 \\ \hline
(0, 2, 2) & Yes & 32 \\ \hline
(4, 0, 0) & Yes & 12 \\ \hline
(0, 4, 0) & Yes & 12 \\ \hline
(0, 0, 4) & Yes & 12 \\ \hline
\end{tabular}
\caption{\label{PairingsClassB} \emph{Possible moduli and whether odd number generations are possible for all bosonic type asymmetric pairings of internal fermions.}} 
\end{table}

\begin{table}[!htb]
\footnotesize
\centering
\begin{tabular}{|c|c|c|}
\hline
Untwisted Moduli in each Torus & Odd Number Generations Possible & Frequency\\
\hline
(2, 4, 2) & No & 1024 \\ \hline
(2, 2, 4) & No & 1024 \\ \hline
(2, 2, 0) & No & 1024 \\ \hline
(2, 0, 2) & No & 1024 \\ \hline
(0, 2, 2) & No & 1024 \\ \hline
(4, 2, 2) & No & 976 \\ \hline
(0, 4, 4) & No & 256 \\ \hline
(0, 4, 0) & No & 256 \\ \hline
(0, 0, 4) & No & 256 \\ \hline
(0, 0, 0) & No & 256 \\ \hline
(4, 4, 0) & No & 244 \\ \hline
(4, 0, 4) & No & 244 \\ \hline
(4, 0, 0) & No & 244 \\ \hline
(4, 4, 4) & No & 228 \\ \hline
(4, 2, 2) & Yes & 48 \\ \hline
(4, 4, 4) & Yes & 12 \\ \hline
(4, 4, 0) & Yes & 12 \\ \hline
(4, 0, 4) & Yes & 12 \\ \hline
(4, 0, 0) & Yes & 12 \\ \hline
\end{tabular}
\caption{\label{PairingsClassF} \emph{Possible moduli and whether odd number generations are possible for all fermionic type asymmetric pairings of internal fermions.}}
\end{table}
\pagebreak

Having classified the possible FSU5 pairings we can now move to the second step of the asymmetric orbifold classification where we fix the pairing and, therefore, the basis vectors and then classify the space of asymmetric orbifold models in reference to phenomenological characteristics. 

\section{Class-Independent Analysis}\label{CIA}
A class of Flipped $SU(5)$ models is defined through the basis (\ref{basis}) with a specific choice of $\bm{A}$. This choice of $\bm{A}$ tells us a concomitant consistent $\bm{B}$ and number of $\bm{e_i}$ vectors quantified by $\bm{E}$. 
Two such classes will be investigated in Section \ref{ClassA} and Section \ref{ClassB}. Before inspecting a specific class, it is worth seeing what we can say about all classes of models derived from the generic basis (\ref{basis}) since several features will be the same for all models. 
\subsection{Supersymmetry Constraints and Class Parameter Space}\label{SUSYSpace}
We seek to classify both $\mathcal{N}=0$ and $\mathcal{N}=1$ models and so it is important to define a necessary and sufficient condition for the presence of $\mathcal{N}=1$ supersymmetry. To do this we first note that the gravitini and gaugini arise from 
\begin{align}
    \partial \overline{X}^\mu \ket{\bm{S}}&\\
    \{\bar{\lambda}^a\}\{\bar{\lambda}^b\}\ket{\bm{S}}&
\end{align}
respectively. Therefore the following GGSO phases are fixed as follows
\beq \label{SUSYconstraint}
\CC{\bm{S}}{\bm{e_i}}=\CC{\bm{S}}{\bm{z_1}}=\CC{\bm{S}}{\bm{x}}=\CC{\bm{S}}{\bm{\gamma}}=-1
\eeq
in order to preserve one gravitino. Furthermore we note that the phases $\CC{\mathds{1}}{\bm{S}}$ and $\CC{\bm{S}}{\bm{b_k}}$, $k=1,2,3$, determine the chirality of the degenerate Ramond vacuum $\ket{\bm{S}}$ and the gravitino is retained so long as
\beq \label{SUSYChiral}
\CC{\mathds{1}}{\bm{S}}=\CC{\bm{S}}{\bm{b_1}}\CC{\bm{S}}{\bm{b_2}}\CC{\bm{S}}{\bm{b_3}}
\eeq 
which can, without loss of generality, be fixed to
\beq \label{SUSYChiral2}
\CC{\mathds{1}}{\bm{S}}=\CC{\bm{S}}{\bm{b_1}}=\CC{\bm{S}}{\bm{b_2}}=\CC{\bm{S}}{\bm{b_3}}=-1
\eeq 
for a scan of $\mathcal{N}=1$ vacua.

The number of independent GGSO phases for a class of models will be determined from the number of basis vectors, $N$, which can be written as 
\beq 
N=8+\sum_i E_i.
\eeq 
Taking into account the constraints (\ref{SUSYconstraint}) and (\ref{SUSYChiral}) for $\mathcal{N}=1$ models there are 
\beq 
\frac{N(N-1)}{2}-7-\sum_i E_i
\eeq 
independent GGSO phases\footnote{We can fix $\CC{\mathds{1}}{\mathds{1}}=+1$ without loss of generality and all other phases are determined from modular invariance rules}.
The space of $\mathcal{N}=0$ vacua can be defined as the space of models violating either condition (\ref{SUSYconstraint}) or (\ref{SUSYChiral}). In ref. \cite{aafs} breaking supersymmetry with different phases is discussed and it is noted how different breakings affect the spectra. If desired, we can restrict the breaking to just shifts beyond the $\mathbb{Z}_2\times \mathbb{Z}_2$ orbifold sectors by preserving condition (\ref{SUSYChiral}), such that $\bm{b_1}$, $\bm{b_2}$ and $\bm{b_3}$ still preserve supersymmetry, then breaking would originate from the vectors beyond the NAHE-set through violating condition (\ref{SUSYconstraint}).

\subsection{Phenomenological Features}\label{MIPheno}
\subsubsection*{Observable Spinorial Representations}
As discussed in Section \ref{Asymm} the twisted sectors such as those giving rise to the spinorial $\mathbf{16}/\overline{\mathbf{16}}$ representations of $SO(10)$ are impacted by the choice of $\bm{A}$. To write these $\bm{F}^k_{pqrs}$ for a particular $\bm{A}$ we must first note the presence of the following possible linear combinations of the vector (\ref{G}), arising for certain $\bm{E}$
\beq 
\begin{cases}
\bm{e_{3456}}=\bm{G}+\bm{e_1}+\bm{e_2}=\{y^{3456},w^{3456} \ | \ \bar{y}^{3456},\bar{w}^{3456}\} \ \ \text{for } \ \bm{E}=(1,1,0,0,0,0)\\
\bm{e_{1256}}=\bm{G}+\bm{e_3}+\bm{e_4}=\{y^{1256},w^{1256} \ | \ \bar{y}^{1256},\bar{w}^{1256}\} \ \ \text{for } \ \bm{E}=(0,0,1,1,0,0)\\
\bm{e_{1234}}=\bm{G}+\bm{e_5}+\bm{e_6}=\{y^{1234},w^{1234} \ | \ \bar{y}^{1234},\bar{w}^{1234}\} \ \ \text{for } \ \bm{E}=(0,0,0,0,1,1).
\end{cases}
\eeq 
Then we can write the sectors giving rise to the fermion generations as
\begin{align}\label{spin16sAll}
    \begin{split}
        \bm{F}^1_{pqrst} =& \bm{b_1}+pE_3\bm{e_3}+qE_4\bm{e_4}+rE_5\bm{e_5}+sE_6\bm{e_6} \\
        &+tE_1E_2(1-E_3)(1-E_4)(1-E_5)(1-E_6)\bm{e_{3456}}\\ 
        \bm{F}^2_{pqrst} =& \bm{b_2}+pE_1\bm{e_1}+qE_2\bm{e_2}+rE_5\bm{e_5}+sE_6\bm{e_6} \\
        &+tE_3E_4(1-E_1)(1-E_2)(1-E_5)(1-E_6)\bm{e_{1256}}\\ 
        \bm{F}^3_{pqrst} =& \bm{b_3}+pE_1\bm{e_1}+qE_2\bm{e_2}+rE_3\bm{e_3}+sE_4\bm{e_4} \\
        &+tE_5E_6(1-E_1)(1-E_2)(1-E_3)(1-E_4)\bm{e_{1234}}.
    \end{split}
\end{align}
where $p,q,r,s,t\in \{0,1\}$. 

In order to write down the number of $\mathbf{16}$ and $\overline{\mathbf{16}}$,
$N_{16}$ and $N_{\overline{16}}$, as a function of the GGSO coefficients we can construct the generalised projectors for these sectors $\mathbb{P}_{\bm{F}^k_{pqrst}}$, $k=1,2,3$, such that 
\begin{align}\label{FUpsilons}
    \begin{split}
        \Upsilon(\bm{F}^1_{pqrst})&=\{\bm{x}+2\bm{\gamma},\bm{z_1},\bm{z_2},E_1\bm{e_1},E_2\bm{e_2}\}\\
        \Upsilon(\bm{F}^2_{pqrst})&=\{\bm{x}+2\bm{\gamma},\bm{z_1},\bm{z_2},E_3\bm{e_3},E_4\bm{e_4}\}\\
        \Upsilon(\bm{F}^3_{pqrst})&=\{\bm{x}+2\bm{\gamma},\bm{z_1},\bm{z_2},E_5\bm{e_5},E_6\bm{e_6}\}
    \end{split}
\end{align}
where we recall that the vector $\bm{z_2}=\{\overline{\phi}^{5,6,7,8}\}$ is
the combination defined in eq. (\ref{z2}). 

In order to determine whether a sector will give rise to a $\mathbf{16}$ or a
$\overline{\mathbf{16}}$ we can first define the chirality phases
\begin{align}\label{ChProjs}
\begin{split}
\bm{X}^1_{pqrs0} &= -\text{ch}(\psi^\mu)\CC{\bm{F}^1_{pqrs0}}{\bm{b_2}+rE_5\bm{e_5}+sE_6\bm{e_6}}^*\\
\bm{X}^2_{pqrs0} &= -\text{ch}(\psi^\mu)\CC{\bm{F}^2_{pqrs0}}{\bm{b_1}+rE_5\bm{e_5}+sE_6\bm{e_6}}^*\\
\bm{X}^3_{pqrs0} &= -\text{ch}(\psi^\mu)\CC{\bm{F}^3_{pqrs0}}{\bm{b_1}+pE_3\bm{e_3}+qE_4\bm{e_4}}^*
\end{split}
\end{align}
where $\text{ch}(\psi^\mu)$ is the spacetime fermion chirality 
and we note that the sectors $\bm{F}^k_{00001}$ do not have a chirality and, instead, give rise to $D_k/2$ copies of both the $\mathbf{16}$ and the $\overline{\mathbf{16}}$.

With these definitions we can write compact
expressions for $N_{16}$ and $N_{\overline{16}}$
\begin{align}\label{N16s}
\begin{split}
N_{16} &= \frac{1}{2}\sum_{\substack{k=1,2,3 \\ p,q,r,s=0,1}} 
D_k\mathbb{P}_{\bm{F}^k_{pqrs0}}\left(1 + \bm{X}^k_{pqrs0}\right) +\frac{D_k}{2}\mathbb{P}_{F^k_{00001}}\\
N_{\overline{16}} &= \frac{1}{2}\sum_{\substack{k=1,2,3 \\ p,q,r,s=0,1}}
D_k\mathbb{P}_{\bm{F}^k_{pqrs0}}\left(1 - \bm{X}^k_{pqrs0}\right)+\frac{D_k}{2}\mathbb{P}_{F^k_{00001}}. 
\end{split}
\end{align}
Since the $SO(10)$ breaking projection $\bm{\gamma}$ decomposes the $\mathbf{16}$/$\overline{\mathbf{16}}$ representations into those of $SU(5)\times U(1)$ according to eq. (\ref{decomp}), we can write a compact expression for each of the FSU5 quantum numbers. These of course depend on the degeneracies (\ref{Ds}) and can be written
\begin{align} \label{FSU5Ns}
\begin{split}
    n_{10}&=\sum_{\substack{k=1,2,3 \\ p,q,r,s=0,1}} \frac{1}{2^{2-\Delta_k}}D_k
\mathbb{P}_{\bm{F}^k_{pqrs0}}\left(1 + \bm{X}^k_{pqrs0}\right)\left(1 +(1-\Delta_k) \CC{\bm{F}^k_{pqrs0}}{\bm{\gamma}}\right) +\frac{D_k}{2}\mathbb{P}_{F^k_{00001}} \\
    n_{\bar{5}}&=\sum_{\substack{k=1,2,3 \\ p,q,r,s=0,1}} \frac{1}{2^{2-\Delta_k}}D_k
\mathbb{P}_{\bm{F}^k_{pqrs0}}\left(1 + \bm{X}^k_{pqrs0}\right)\left(1 -(1-\Delta_k) \CC{\bm{F}^k_{pqrs0}}{\bm{\gamma}}\right)+\frac{D_k}{2}\mathbb{P}_{F^k_{00001}}\\
    n_{\overline{10}}&=\sum_{\substack{k=1,2,3 \\ p,q,r,s=0,1}} \frac{1}{2^{2-\Delta_k}}
\mathbb{P}_{\bm{F}^k_{pqrs0}}\left(1 - \bm{X}^k_{pqrs0}\right)\left(1 +(1-\Delta_k) \CC{\bm{F}^k_{pqrs0}}{\bm{\gamma}}\right)+\frac{D_k}{2}\mathbb{P}_{F^k_{00001}}\\
    n_{5}&=\sum_{\substack{k=1,2,3 \\ p,q,r,s=0,1}} \frac{1}{2^{2-\Delta_k}}D_k
\mathbb{P}_{\bm{F}^k_{pqrs0}}\left(1 - \bm{X}^k_{pqrs0}\right)\left(1 -(1-\Delta_k) \CC{\bm{F}^k_{pqrs0}}{\bm{\gamma}}\right)+\frac{D_k}{2}\mathbb{P}_{F^k_{00001}}.
\end{split}
\end{align}
The number of generations for a model is then
\beq \label{gens}
n_g=n_{10}-n_{\overline{10}}=n_{\bar{5}}-n_5.
\eeq 
From this we can construct a necessary condition for three generation models to exist once $\bm{A}$ is specified 
\begin{align}\label{3gen}
     \exists \ \CC{\bm{v_i}}{\bm{v_j}}: \ \ &\sum_{\substack{k=1,2,3 \\ p,q,r,s=0,1}} \frac{1}{2^{1-\Delta_k}}D_k
\mathbb{P}_{\bm{F}^k_{pqrs0}}\bm{X}^k_{pqrs0}\left(1 +(1-\Delta_k) \CC{\bm{F}^k_{pqrs0}}{\bm{\gamma}}\right)=3\\
\text{and } &\sum_{\substack{k=1,2,3 \\ p,q,r,s=0,1}} 2^{\Delta_k}D_k
\mathbb{P}_{\bm{F}^k_{pqrs0}}\bm{X}^k_{pqrs0}(1-\Delta_k) \CC{\bm{F}^k_{pqrs0}}{\bm{\gamma}}=0.
\end{align}
Checking that there exists a solution to this equation for a class of models and enumerating such solutions can be done easily by inputting this constraint into an SMT solver such as Z3. 
\subsubsection*{Heavy Higgs}
Another key representation for phenomenology is the presence of a Higgs breaking the $SU(5)\times U(1)$ that we call the Heavy Higgs. This arises from the representation $\left(\mathbf{10}, +\frac{1}{2}\right)+\left(\mathbf{\overline{10}}, -\frac{1}{2}\right)$. 
The relevant sectors are 
\begin{align}\label{HeavyBs}
    \bm{B}^{k}_{pqrst}&=\bm{S}+\bm{F}^k_{pqrst}, \ \ \ \ k=1,2,3
\end{align}
which in the case of $\mathcal{N}=1$ supersymmetric models are the (bosonic) superpartners of the spinorials $\mathbf{16}/\overline{\mathbf{16}}$ sectors (\ref{spin16sAll}). We note that the generalised projectors for these sectors $\mathbb{P}_{\bm{B}^k_{pqrst}}$, $k=1,2,3$, can be constructed such that $\Upsilon(\bm{B}^k_{pqrst})$ equals $\Upsilon(\bm{F}^k_{pqrst})$ from eq. (\ref{FUpsilons}).

We note that with a Heavy Higgs the FSU5 GUT can be broken and the particles of Standard Model arise from the decomposition of the FSU5 representations (\ref{decomp}) under $SU(3)\times SU(2)\times U(1)$
\begin{align}
    \begin{split}\label{SMdecomp}
\left(\mathbf{\bar 5},-\frac{3}{2}\right)&=\left(\mathbf{\bar 3},\mathbf{1},-\frac{2}{3}\right)_{u^c}+\left(\mathbf{1},\mathbf{2},-\frac{1}{2}\right)_L,\\
\left(\mathbf{10},+\frac{1}{2}\right)&=\left(\mathbf{3},\mathbf{2},+\frac{1}{6}\right)_Q+\left(\mathbf{\bar 3},\mathbf{1},+\frac{1}{3}\right)_{d^c}+ \left(\mathbf{1},\mathbf{1},0\right)_{\nu^c},\\
\left(\mathbf{1},+\frac{5}{2}\right)&= \left(\mathbf{1},\mathbf{1},+1 \right)_{e^c},     
    \end{split}
\end{align}
where $L$ is the lepton–doublet; $Q$ is the quark–doublet; $d^c$, $u^c$, $e^c$ and $\nu^c$ are the quark and lepton singlets.
\subsubsection*{Light Higgs}
The light Higgs representations are electroweak Higgs doublets. In $\mathcal{N}=1$ supersymmetric
models, a pair is required to give masses to up-- and down--quark, respectively. In models
in which spacetime supersymmetry is broken entirely at the string level, this may be relaxed. However, as the models descend from $\mathcal{N}=1$ supersymmetric models, they retain some
of this underlying structure and mass terms at leading order are generated to the 
respective Higgs doublets pairs. We therefore require the existence of a pair of light
Higgs multiplets also in $\mathcal{N}=0$ models. We further note the existence of a doublet--triplet
splitting mechanism in the untwisted sector of the asymmetric models \cite{dtsm}. This 
mechanism is operational in asymmetric models with the breaking pattern 
$SO(10)\rightarrow SO(6)\times SO(4)$ and is therefore not relevant in the flipped 
$SU(5)$ models that are of interest here. We note, however, that in flipped $SU(5)$ models
the untwisted sector produces three pairs in $\mathbf{5}+\mathbf{\bar{5}}$ representation of $SU(5)$, 
which contain electroweak Higgs doublets that may serve as light Higgs multiplets. 
However, we note that the generation of hierarchical fermion masses typically necessitates utilisation of Higgs doubelts that arise from twisted sectors 
\cite{fsu5masses, ckm}. We therefore examine here the conditions for obtaining 
vectorial representations in the twisted sectors.

Sectors giving rise to vectorial $\mathbf{10}$ representations, that include the twisted Light Higgs, can be written 
\begin{align}\label{Vs}
    \begin{split}
        \bm{V}^k_{pqrst} &= \bm{S}+\bm{F}^k_{pqrst}+\bm{x}
    \end{split}
\end{align}
where the states are of the form $\{\bar{\lambda}\}_{\frac{1}{2}} \ket{\bm{V}^k_{pqrst}}$, $k=1,2,3$, meaning that they have a single antiholomorphic oscillator of frequency $\frac{1}{2}$, as defined in eq. (\ref{freq}), accompanying the degenerate Ramond vacuum. The SM Higgs will arise when this sector with $\bar{\lambda}=\bar{\psi}^a$, $a\in\{1,2,3,4,5\}$, is retained in the massless spectrum of a model. 
For these sectors the generalised projector $\mathbb{P}_{\{\bar{\psi}^a\}\bm{V}^k_{pqrst}}$ takes the general form of eq. (\ref{GProjOScill}) and $\Upsilon(\bm{V}^k_{pqrst})$ will be the same as $\Upsilon(\bm{F}^k_{pqrst})$ from eq. (\ref{FUpsilons}).

We note that any surviving sector gives rise to a vectorial $\mathbf{10}$ decomposing under $SU(5)\times U(1)$ according to
\beq 
\mathbf{10}=(\mathbf{5},-1)+(\mathbf{\bar{5}},+1).
\eeq 
These two representations taken together can be identified as the SM Higgs breaking the electroweak gauge group. 
Therefore the number of Light Higgses from the twisted sectors is given by
\beq 
n_{5h}=\# \left[ (\mathbf{5},-1)+(\mathbf{\bar{5}},+1)\right]. 
\eeq 
\subsubsection*{Tachyonic Sectors}
Since we include non-supersymmetric models in our classification it is vital we check for the absence of on-shell tachyons in order to ensure the stability of our models for a 4D Minkowski background. In order to do this we encode the GGSO projections for all on-shell tachyonic sectors. Many tachyonic sectors can arise due to $\bm{e_i}$ vectors, certain $\bm{\gamma}$ combinations and other class-dependent combinations and therefore are dependent on the choice of $\bm{A}$ and require class-by-class analysis.  
However, we will always have the untwisted tachyon
\beq 
\{\bar{\lambda}\}\ket{0}_{NS}
\eeq 
that is projected for all models through the $\bm{S}$ projection. In addition, the following on-shell tachyonic sectors arise for all classes of models 
\begin{equation}\label{MItachs}
\footnotesize
T=\begin{Bmatrix}
\ket{\bm{z_1}} & \ket{\bm{z_2}} & \ket{\bm{x}+2\bm{\gamma}} \\
\ket{\bm{z_1}+\bm{x}+2\bm{\gamma}} & \ket{\bm{z_2}+\bm{x}+2\bm{\gamma}} & \ket{\bm{z_1}+\bm{z_2}+\bm{x}+2\bm{\gamma}} 
\end{Bmatrix}
\end{equation}

All of these sectors, $\bm{t}\in T$, must be projected from the spectrum through appropriate definitions of their generalised projectors $\mathbb{P}_{\bm{t}}=0$. Once we specify the vector $\bm{A}$ we can then determine the further class-dependent tachyonic sectors and ensure their projection. 

\subsubsection*{Enhancements}

Additional space-time vector bosons may arise in all models derived from the basis (\ref{basis}). The following enhancements arise independent of the class
\begin{equation}\label{Enhs}
\begin{Bmatrix}
\psi^\mu \{\bar{\lambda}\}_{\frac{1}{2}}: &\ket{\bm{z_1}} & \ket{\bm{z_2}} & \ket{\bm{x}+2\bm{\gamma}} & \ket{\bm{z_1}+\bm{x}+2\bm{\gamma}} & \ket{\bm{z_2}+\bm{x}+2\bm{\gamma}}\\
\psi^\mu: & \ket{\bm{x}} & \ket{\bm{z_1}+\bm{z_2}} &&
\end{Bmatrix}
\end{equation}
with the following subset being enhancements to the observable gauge factors
\begin{equation}\label{ObsEnhs}
H=\begin{Bmatrix}
\psi^\mu \{\bar{\psi}^a\}: &\ket{\bm{z_1}} & \ket{\bm{z_2}} & \ket{\bm{x}+2\bm{\gamma}} \\
\psi^\mu \{\bar{\psi}^a\}:& \ket{\bm{z_1}+\bm{x}+2\bm{\gamma}} & \ket{\bm{z_2}+\bm{x}+2\bm{\gamma}}&\ket{\bm{x}+2\bm{\gamma}+\bm{z_1}+\bm{z_2}}\\
\psi^\mu: & \ket{\bm{x}} &&&
\end{Bmatrix}.
\end{equation}
with $a=1,2,3,4,5$.
Therefore, from these sectors we can restrict our analysis to models with observable gauge group $SU(5) \times U(1) \times U(1)_{i=1,2,3}$ 
by imposing 
\beq \label{NoObsEnhMI}
\forall \ \bm{h}\in H: \ \mathbb{P}_{\bm{h}}=0.
\eeq 
In this case, for these generalised projectors we have
\begin{align}
\begin{split}
\Upsilon(\bm{z_1})&=\{\bm{S},E_1\bm{e_1},E_2\bm{e_2},E_3\bm{e_3},E_4\bm{e_4},E_5\bm{e_5},E_6\bm{e_6},\bm{x},\bm{b_1},\bm{b_2},\bm{z_2}\}\\
\Upsilon(\bm{z_2})&=\{\bm{S},E_1\bm{e_1},E_2\bm{e_2},E_3\bm{e_3},E_4\bm{e_4},E_5\bm{e_5},E_6\bm{e_6},\bm{x},\bm{b_1},\bm{b_2},\bm{z_1}\}\\
\Upsilon(\bm{x}+2\bm{\gamma})&=\{\bm{S},E_1\bm{e_1},E_2\bm{e_2},E_3\bm{e_3},E_4\bm{e_4},E_5\bm{e_5},E_6\bm{e_6},\bm{x},\bm{x}+2\bm{\gamma}+\bm{z_1}+\bm{z_2}\}\\
\Upsilon(\bm{x}+2\bm{\gamma}+\bm{z_1})&=\{\bm{S},E_1\bm{e_1},E_2\bm{e_2},E_3\bm{e_3},E_4\bm{e_4},E_5\bm{e_5},E_6\bm{e_6},\bm{x},\bm{x}+2\bm{\gamma}+\bm{z_2}\}\\
\Upsilon(\bm{x}+2\bm{\gamma}+\bm{z_2})&=\{\bm{S},E_1\bm{e_1},E_2\bm{e_2},E_3\bm{e_3},E_4\bm{e_4},E_5\bm{e_5},E_6\bm{e_6},\bm{x},\bm{x}+2\bm{\gamma}+\bm{z_1}\}\\
\Upsilon(\bm{x}+2\bm{\gamma}+\bm{z_1}+\bm{z_2})&=\{\bm{S},E_1\bm{e_1},E_2\bm{e_2},E_3\bm{e_3},E_4\bm{e_4},E_5\bm{e_5},E_6\bm{e_6},\bm{x},\bm{x}+2\bm{\gamma}\}\\
\Upsilon(\bm{x})&=\{\bm{S},E_1\bm{e_1},E_2\bm{e_2},E_3\bm{e_3},E_4\bm{e_4},E_5\bm{e_5},E_6\bm{e_6},\bm{z_1},\bm{z_2}\}
\end{split}
\end{align}
Additional enhancements may arise depending on the specific form of $\bm{\gamma}$ which can be analysed class-by-class. 

\subsubsection*{Exotics}
Another important consideration for ensuring reasonable phenomenology is the absence of chiral exotics. The exotics sectors in general depend on the class, in particular on the exact form of $\bm{\gamma}$ since combinations of $\bm{\gamma}$ will be those that can generate exotics.

However, we can note here the following exotic sectors with $(\bm{\alpha}_L\cdot \bm{\alpha}_L,\bm{\alpha}_R\cdot \bm{\alpha}_R)=(4,4)$
\begin{equation}\label{Exots}
\begin{Bmatrix}
 \{\bar{\psi}^{a}\}_{\frac{1}{2}}: &\ket{\bm{S}+\bm{z_1}} & \ket{\bm{S}+\bm{z_2}} & \ket{\bm{S}+\bm{x}+2\bm{\gamma}} & \ket{\bm{S}+\bm{z_1}+\bm{x}+2\bm{\gamma}} & \ket{\bm{S}+\bm{z_2}+\bm{x}+2\bm{\gamma}}
\end{Bmatrix}
\end{equation}
where $a\in [1,...,5]$. We note that these are the would-be gaugini of the enhancements (\ref{Enhs}).  These sectors will not contribute to a chiral anomaly as they are automatically vector-like. 
It will then be necessary to analyse the other exotics at the level of a particular class of vacua.


\subsection{Asymmetric Pairings, Up-Type Yukawa Couplings and Higgs Doublet-Triplet Splitting }\label{TQMCMI}
Top quark Yukawa couplings in the string models derived from the $\mathbb{Z}_2\times \mathbb{Z}_2$ heterotic orbifold take the general form
\beq 
\lambda_t \mathbf{S}^{Q_L}\mathbf{S}^{u_R}\mathbf{V}^{H_u}.
\eeq 
It can be demonstrated that this coupling can come either from a coupling of the type $T_kT_kU_k$, $k=1,2,3$, or of the type $T_kT_lT_m$, $k\neq l\neq m=1,2,3$, where $T$ indicates a twisted sector and $U$ indicates the (untwisted) Neveu-Schwarz sector. 
The assignment of asymmetric boundary conditions determines which of the two 
couplings can appear at leading order in the string vacua \cite{tqmp}.

\sloppy The asymmetric boundary conditions for the internal worldsheet fermions $\{y^I,w^I \ |\ {\bar y}^I,{\bar w}^I\}$ induce a doublet--triplet splitting mechanism 
of the untwisted $\mathbf{5}$ and $\mathbf{\bar5}$ representations \cite{dtsm}. The mechanism is induced
by the basis vectors that break the $SO(10)$ symmetry to the Pati--Salam subgroup,
with respect to the three pairs of untwisted vectorial $\mathbf{5}$ and $\mathbf{\bar{5}}$ multiplets, where 
symmetric boundary conditions retain the colour triplets pairs, and project the 
electroweak doublets, 
whereas asymmetric boundary conditions project the triplets and retain the doublets. 
Thus, in the case of models with solely symmetric boundary conditions, only flipped $SU(5)$ models can produce cubic level couplings of the type
$T_kT_kU_k$, utilising the Higgs doublets from the NS sector.

Similar to the stringy doublet--triplet splitting mechanism that is determined by the assignment of asymmetric versus symmetric boundary conditions, the asymmetric/symmetric
assignment selects between up/down--quark Yukawa couplings at leading order
\cite{yukawa,tqmp, towards}. 
This Yukawa coupling selection mechanism operates in the basis vector that breaks the 
$SO(10)$ symmetry to the $SU(5)\times U(1)$ subgroup, where symmetric boundary 
conditions select a down--quark type Yukawa coupling, whereas asymmetric 
boundary conditions select an up--quark type Yukawa coupling. Hence, this
Yukawa coupling selection mechanism can be utilised
in flipped $SU(5)$ and standard--like string models. 


Given that we consider Flipped $SU(5)$ models, representations in the $\mathbf{5}$ and $\bar{\mathbf{5}}$ of $SU(5)$ arise from the NS sector generically. These representations yield the electroweak Higgs doublets and color Higgs triplets. Through asymmetric boundary condition assignments of the internal fermions under an extra Pati-Salam type breaking vector the doublets and triplets may be distinguished. However, in our case we will get both regardless of the GGSO configuration and boundary condition assignment from $\bm{A}$. Therefore top mass couplings of the form $T_kT_kU_k$ can arise in our models. 

As is familiar from the symmetric orbifold classification, couplings $T_kT_lT_m$ can also arise. In this case the Higgs doublet arises from the twisted sectors (\ref{Vs}). Therefore both types of couplings can give rise to a realistic Up-Type Yukawa Coupling and both will be analysed. Similar to the case of the couplings to the untwisted Higgs doublets, 
selection conditions of up--type versus down--type quark Yukawa couplings can be 
formulated \cite{tqmc}.


\subsection{Partition Function and Cosmological Constant for Asymmetric Orbifolds}\label{PF}

The analysis of the partition function for asymmetric orbifolds constructed in the free fermionic formulation as described in Sections \ref{FFModelBuilding} and \ref{setup} is largely similar to the symmetric case presented in \cite{so10tclass,PStclass}. However, there are some key differences and subtleties which are important to explicitly discuss. These arise for two main reasons, namely the asymmetric parings introduced by the basis vector $\bm{\gamma}$ and the appearance of half boundary conditions in the basis set (\ref{basis}). 

From the point of view of the partition function, the asymmetric pairings introduce imaginary GGSO phases, meaning that the fermionic partition function
\begin{equation}
  Z = \sum_{\bm{\alpha},\bm{\beta}} \CC{\bm{\alpha}}{\bm{\beta}} \prod_{f} Z
  \sqbinom{\bm{\alpha}(f)}{\bm{\beta}(f)},
  \label{ZFerm}
\end{equation}
will have imaginary terms which have to cancel. This cancellation is, however, ensured by modular invariance. In the case of symmetric orbifolds, since $Z\smb{a}{b}=\sqrt{\vartheta\smb{a}{b}}$, the fermionic part of the partition function can be expressed using the four standard Jacobi theta functions 
\begin{equation}\label{ThetaQ}
\vartheta\smb{a}{b} = \sum_{n\in\mathbb{Z}} q^{(n+a/2)^{2} / 2} e^{ 2 \pi i(n+a/2)b/2},
\end{equation}
with $a,b\in\{0,1\}$. In the presence of half boundary conditions there will be sixteen such theta functions with $a$ and $b$  now taking values in the set $a,b\in\{-1/2,0,1/2,1\}$.

To express the partition function of the models under consideration in the classification setup, it is beneficial to use the notation utilised in \cite{fr1,fr2,florakis}. This makes many properties immediately readable from the form of the partition function and allows us to economically express all models used in this paper in one compact form. Since the classification of asymmetric shifts depends on the exact form of the vector $\bm{\gamma}$ it is instructive to first write down the partition function of the subset $\{\mathds{1},\bm{e_i},\bm{S},\bm{b_1},\bm{b_2},\bm{b_3},\bm{z_1},\bm{x}\}$ without $\bm{\gamma}$. In this case all $\bm{e_i}$ are compatible and so we have 13 basis vectors giving
\begin{align}\label{BasePF}
Z=&\frac{1}{\eta^{10}\bar{\eta}^{22}}\;\frac{1}{2^4}\sum_{\substack{a,k,r,\rho\\b,l,s,\sigma}} \;\frac{1}{2^6} \sum_{\substack{H_i\\G_i}} \;\frac{1}{2^3} \sum_{\substack{h_1,h_2,H\\g_1,g_2,G}} (-1)^{\Phi\left[\begin{smallmatrix}
a&k&\rho&r&H_i&h_1&h_2&H\\b&l&s&\sigma&G_i&g_1&g_2&G
\end{smallmatrix} \right]}\nonumber\\[0.1cm]
&\times \vth\smb{a}{b} \;\, \vth\smb{a+h_1}{b+g_1} \vth\smb{a+h_2}{b+g_2} \vth\smb{a-h_1-h_2}{b-g_1-g_2} \\[0.2cm]
&\times \Gamma_{(6,6)}\left[\begin{smallmatrix}
r&H_i&h_1&h_2\\s&G_i&g_1&g_2
\end{smallmatrix} \right] \nonumber\\[0.2cm]
&\times \vthb\smb{k}{l}^5 \vthb\smb{k+h_1}{l+g_1} \vthb\smb{k+h_2}{l+g_2} \vthb\smb{k-h_1-h_2}{l-g_1-g_2} \vthb\smb{\rho}{\sigma}^4 \vthb\smb{\rho +H}{\sigma +G}^4, \nonumber
\end{align}
where all indices are summed over the set $\{0,1\}$. The phase $\Phi$, which is a polynomial in the summation indices, is chosen such that the entire partition function is modular invariant. The choice of this phase translates to a choice of GGSO matrix in the classification setup. Indices $k,l$ and $\rho,\sigma$ represent the sixteen complex right-moving fermions giving the fermionic representation of the $E_8\times E_8$ lattice of the underlying 10D heterotic theory. The non-freely acting $\mathbb{Z}_2\times \mathbb{Z}_2$ orbifold is represented by the parameters $h_i$ and $g_i$, where the $h_i$ give the various twists, while the $g_i$ implement the orbifold projections. The $H_i$ and $G_i$ correspond to the basis vectors $\bf{e_i}$ and hence are responsible for orbifold shifts along the six internal dimensions of the $T^6$. Finally, $H$ and $G$ break one of the $E_8$ factors in the hidden sector by a $\mathbb{Z}_2$ twist.

The internal lattice $\Gamma_{(6,6)}$ which corresponds to the $T^6$ is given by
\begin{align}\label{BaseLattice}
\Gamma_{(6,6)}\left[\begin{smallmatrix}r&H_i&h_1&h_2\\s&G_i&g_1&g_2\end{smallmatrix}\right] =& 
\;\; \Big| \; \vth_{y\bar{y}^1}\smb{r+h_1+H_1}{s+g_1+G_1} \vth_{y\bar{y}^2}\smb{r+h_1+H_2}{s+g_1+G_2} \vth_{y\bar{y}^3}\smb{r+h_2+H_3}{s+g_2+G_3}  \nonumber\\
&\times  \vth_{y\bar{y}^4}\smb{r+h_2+H_4}{s+g_2+G_4} \vth_{y\bar{y}^5}\smb{r+h_2+H_5}{s+g_2+G_5} \vth_{y\bar{y}^6}\smb{r+h_2+H_6}{s+g_2+G_6}  \nonumber\\[0.1cm]
&\times  \vth_{w\bar{w}^1}\smb{r-h_1-h_2+H_1}{s-g_1-g_2+G_1} \vth_{w\bar{w}^2}\smb{r-h_1-h_2+H_2}{s-g_1-g_2+G_2} \vth_{w\bar{w}^3}\smb{r-h_1-h_2+H_3}{s-g_1-g_2+G_3}  \\ 
&\times  \vth_{w\bar{w}^4}\smb{r-h_1-h_2+H_4}{s-g_1-g_2+G_4} \vth_{w\bar{w}^5}\smb{r+h_1+H_5}{s+g_1+G_5} \vth_{w\bar{w}^6}\smb{r+h_1+H_6}{s+g_1+G_6}\Big|, \nonumber
\end{align}
where $\left|\vth\smb{a}{b}\right|=\sqrt{\vth\smb{a}{b}\vthb\smb{a}{b}}$. The subscript on the $\vth$'s denote which worldsheet fermions the terms correspond to. We see that with this basis the internal lattice is left-right symmetric, meaning that all left moving $y$'s and $w$'s are paired with a right moving $\bar{y}$ or $\bar{w}$. This is why the internal lattice can be written as a magnitude.

The introduction of asymmetric parings via the vector $\bm{\gamma}$ introduces further complexity to the above partition function. Recall the notation introduced in Section \ref{setup}, where the most general consistent form of $\bm{\gamma}$ is written as in (\ref{GeneralGamma})
\begin{equation}
    \bm{\gamma} = \bm{A}+\{ \bar{\psi}^{1,...,5}=\bar{\eta}^{1,2,3}=\bar{\phi}^{1,2,6,7}=\frac{1}{2}\}+\bm{B},
\end{equation}
where 
\begin{align}
    \bm{B}=&\,\{B(\bar{\phi}^3),B(\bar{\phi}^4),B(\bar{\phi}^5),B(\bar{\phi}^8)\}, \nonumber\\
    \bm{A}=& 
    \begin{cases}
    \{A(y^1),\cdots,A(w^6)\ |  A(\bar{y}^1),\cdots,A(\bar{w}^6)\}  \qquad\qquad\quad \text{if $\bm{\gamma}$ bosonic;}  \\
    \{\psi^\mu,\chi^{12},A(y^1),\cdots,A(w^6)\ | \ A(\bar{y}^1),\cdots,A(\bar{w}^6)\} \quad \, \text{if $\bm{\gamma}$ fermionic.}
    \end{cases}
\end{align}
Also recall the vector $\bm{E}=(E_1,E_2,E_3,E_4,E_5,E_6)$ of (\ref{ED}), which qualifies which of the $\bm{e_i}$ are compatible with a specific choice of $\bm{\gamma}$ and hence appear in the basis set. That is, if $E_i=0$ then $\bm{e_i}\notin \mathcal{B}$ and vice-versa.

In terms of the above quantities we can now examine the effect of $\bm{\gamma}$ on the partition function (\ref{BasePF}) within the frame of the general classification setup. For simplicity we consider the case where $\bm{\gamma}$ is bosonic and hence has no action on $\psi^\mu$ and $\chi^{1-6}$.  The antiholomorphic hidden worldsheet fermions are affected by the choice of $\bm{B}$, while the specific choice of $\bm{A}$  will only change how the internal lattice is structured. Thus the partition function takes the form
\begin{align}\label{GammaPF}
Z=&\frac{1}{\eta^{10}\bar{\eta}^{22}} \; \frac{1}{2^4}\sum_{\substack{a,k,r,\rho\\b,l,s,\sigma}} \;\frac{1}{2^{\sum_i E_i}} \sum_{\substack{H_i\\G_i}} \;\frac{1}{2^3} \sum_{\substack{h_1,h_2,H\\g_1,g_2,G}} \;\frac{1}{4}  \sum_{\substack{H'\\G'}} \; (-1)^{\Phi\left[\begin{smallmatrix}
a&k&\rho&r&H_i&h_1&h_2&H&H'\\b&l&s&\sigma&G_i&g_1&g_2&G&G'
\end{smallmatrix} \right]}\nonumber\\[0.1cm]
&\times \vth\smb{a}{b} \vth\smb{a+h_1}{b+g_1} \vth\smb{a+h_2}{b+g_2} \vth\smb{a-h_1-h_2}{b-g_1-g_2}\nonumber\\[0.2cm]
&\times \Gamma_{(6,6)}^\gamma\left[\begin{smallmatrix}
r&H_i&h_1&h_2&H'\\s&G_i&g_1&g_2&G'
\end{smallmatrix} \right] \\[0.2cm]
&\times \vthb\smb{k+H'}{l+G'}^5 \vthb\smb{k+h_1+H'}{l+g_1+G'} \vthb\smb{k+h_2+H'}{l+g_2+G'} \vthb\smb{k-h_1-h_2+H'}{l-g_1-g_2+G'} \nonumber\\[0.1cm]
&\times \vthb\smb{\rho+H'}{\sigma+G'}^2 \vthb\smb{\rho +H+H'}{\sigma +G+G'}^2 \vthb\smb{\rho+2B(\bar{\phi}^3)H'}{\sigma+2B(\bar{\phi}^3)G'} \vthb\smb{\rho+2B(\bar{\phi}^4)H'}{\sigma+2B(\bar{\phi}^4)G'}\nonumber \\[0.1cm]
&\times \vthb\smb{\rho+H+2B(\bar{\phi}^5)H'}{\sigma+G+2B(\bar{\phi}^5)G'} \vthb\smb{\rho+H+2B(\bar{\phi}^8)H'}{\sigma+G+2B(\bar{\phi}^8)G'}, \nonumber
\end{align}
where the sum in the new indices $H'$ and $G'$ run over $\{-1/2,0,1/2,1\}$ as opposed to the other indices which still take values in $\{0,1\}$. This is because the half boundary conditions in $\bm{\gamma}$ introduce a new $\mathbb{Z}_4$ orbifold to the picture. The factor of two in front of some indices is a result of having both half and integer boundary conditions within the same basis vector, and hence this factor ensures that integer boundary conditions are correctly accounted for.

The form of the internal lattice $\Gamma_{(6,6)}^\gamma$ depends on the choice of asymmetric shifts in the internal degrees of freedom, {\it i.e.} $\bm{A}$. Consequently, this determines which of the symmetric $\mathbb{Z}_2$ shifts $\bm{e_i}$ are compatible with this choice, which fixes $\bm{E}$. The asymmetric shifts introduced by $\bm{\gamma}$ break the left-right symmetry of the lattice (\ref{BaseLattice}). To examine this further, we have to look at what happens to a set of internal fermions corresponding to one of the orbifold planes.  If we take the fist plane, {\it i.e.} the fermions $\{y^{3,4,5,6} \ | \ \bar{y}^{3,4,5,6}\}$, the corresponding part of the lattice is
\begin{align}
 \Gamma_1 =& \;\; \vth_{y^3}\smb{r+h_2+H_3}{s+g_2+G_3}^{1/2} 
 \vth_{y^4}\smb{r+h_2+H_4}{s+g_2+G_4}^{1/2} \vth_{y^5}\smb{r+h_2+H_5}{s+g_2+G_5}^{1/2} \vth_{y^6}\smb{r+h_2+H_6}{s+g_2+G_6}^{1/2} \nonumber\\ 
 &\times\vth_{\bar{y}^3}\smb{r+h_2+H_3}{s+g_2+G_3}^{1/2} \vthb_{\bar{y}^4}\smb{r+h_2+H_4}{s+g_2+G_4}^{1/2} \vthb_{\bar{y}^5}\smb{r+h_2+H_5}{s+g_2+G_5}^{1/2} \vthb_{\bar{y}^6}\smb{r+h_2+H_6}{s+g_2+G_6}^{1/2}.
\end{align}
Since the asymmetric shifts cannot mix the orbifold planes, we either have 0, 1 or 2 such shifts affecting these fermions. As an example, we consider what happens when $\bm{A}$ contains one such pairing, say $y^5y^6$. Firstly, this imposes that $\bm{E}=(1,1,1,1,0,0)$, {\it i.e.} $\bm{e_{5,6}}$ are no longer in the basis, so that $H_{5,6}$ and $G_{5,6}$ are not present. Secondly, it breaks the left-right symmetry of the $(y^5\bar{y}^5)$ and $(y^6\bar{y}^6)$ pairings which become  $(y^5\bar{y}^5)(y^6\bar{y}^6) \rightarrow (y^5y^6)(\bar{y}^5\bar{y}^6)$. Given the above factors, the internal lattice of the first orbifold plane becomes
\begin{align}
 \Gamma_1^{\gamma} =& \;\; \vth_{y^3}\smb{r+h_2+H_3}{s+g_2+G_3}^{1/2} 
 \vth_{y^4}\smb{r+h_2+H_4}{s+g_2+G_4}^{1/2} \vth_{y^{5,6}}\smb{r+h_2+2H'}{s+g_2+2G'}\nonumber\\ 
 &\times\vth_{\bar{y}^3}\smb{r+h_2+H_3}{s+g_2+G_3}^{1/2} \vthb_{\bar{y}^4}\smb{r+h_2+H_4}{s+g_2+G_4}^{1/2} \vthb_{\bar{y}^{5,6}}\smb{r+h_2}{s+g_2}.
\end{align}
If there are two such  asymmetric holomorphic pairings in the first plane then, regardless of the specific pairing, the lattice simply becomes
\begin{equation}
\Gamma_1^{\gamma} = \; \vth_{y^{3,4,5,6}}\smb{r+h_2+2H'}{s+g_2+2G'}^2 \vthb_{\bar{y}^{3,4,5,6}}\smb{r+h_2}{s+g_2}^2.
\end{equation}
The construction of the partition function for the remaining two planes is equivalent and can be straightforwardly done once a specific basis is taken.

Once a model is chosen and the partition function is fixed according to the above considerations, the cosmological constant can be calculated according to methods used in \cite{so10tclass,PStclass,DienesCoC}. This entails performing a $q$-expansion of the theta functions according to (\ref{ThetaQ}), which will result in the partition function taking the form
\begin{equation}\label{QExpPF}
    Z = \sum_{n,m} a_{mn} q^m \bar{q}^n, 
\end{equation}
where the $\eta$-functions have also been q-expanded. Written in this form, the $a_{mn}$ correspond to the Bose-Fermi degeneracy at a given mass level. That is, $a_{mn}=n_b-n_f$ at the mass level with conformal weights of $(m,n)$ for the holomorphic and anti-holomorhic sector respectively. The one-loop cosmological constant $\Lambda$ is then given by the integral of this partition function over the fundamental domain of the modular group
\begin{equation}
    \Lambda = \int_\mathcal{F}\frac{d^2\tau}{\tau_2^2}\, Z_B Z_F = \int_\mathcal{F}\frac{d^2\tau}{\tau_2^3}\, \sum_{n.m} a_{mn} q^m \bar{q}^n,
    \label{CoC}
\end{equation}
where $Z_B$ is the contribution from the bosonic degrees of freedom given by
\begin{equation}
  Z_B = \frac{1}{\tau_2} \frac{1}{\eta^2\bar{\eta}^2},
  \label{Z_B}
\end{equation}
and $Z_F$ is the contribution from the worldsheet fermions as given in (\ref{GammaPF}). Since the models under consideration are void of physical tachyons, the series expansion contains only finite terms and converges exponentially fast. It is important to note that the above expression (\ref{CoC}) gives the worldsheet vacuum energy $\Lambda_\text{WS}$ which is unitless. The spacetime cosmological constant is obtained by introducing the string-scale via \begin{equation}\label{WSvST}
\Lambda_\text{ST} =  -\frac{1}{2}\mathcal{M}^4\Lambda_{\text{WS}}.
\end{equation}
It is also interesting to note that all of the above models considered in the classification exhibit a form of misaligned supersymmetry discovered in \cite{MSUSYDienes1,MSUSYDienes2}. This is not unexpected as this phenomenon is a direct consequence of modular invariance \cite{MSUSYDienes1,MSUSYDienes2,MSUSYAngelantonj}, or a smaller subgroup of the modular group in some cases \cite{MSUSYFlavio1,MSUSYFlavio2}, and so heterotic asymmetric orbifolds should also respect this mechanism.

\section{Asymmetric Orbifold Class A}\label{ClassA}

The first Class of models we will choose is a pairing choice where all untwisted moduli are retained, {\it i.e.} $\bm{M}=(4,4,4)$. 
The pairing we choose is inspired by that used in the model of \cite{fny} and is given by $\bm{A}=\{y^3y^6,y^1w^6,w^1w^3\}$. The basis for this class of models is then
\begin{align}\label{basisA}
\bm{\mathds{1}}&=\{\psi^\mu,
\chi^{1,\dots,6},y^{1,\dots,6}, w^{1,\dots,6}\ | \ \overline{y}^{1,\dots,6},\overline{w}^{1,\dots,6},
\overline{\psi}^{1,\dots,5},\overline{\eta}^{1,2,3},\overline{\phi}^{1,\dots,8}\},\nonumber\\
\bm{S}&=\{{\psi^\mu},\chi^{1,\dots,6} \},\nonumber\\
\bm{e_2}&=\{y^{2},w^{2}\; | \; \overline{y}^{2},\overline{w}^{2}\}, 
\ \ \  \nonumber\\
\bm{e_4}&=\{y^{4},w^{4}\; | \; \overline{y}^{4},\overline{w}^{4}\}, 
\ \ \  \nonumber\\
\bm{e_5}&=\{y^{5},w^{5}\; | \; \overline{y}^{5},\overline{w}^{5}\}, 
\ \ \  \nonumber\\
\bm{b_1}&=\{\psi^\mu,\chi^{12},y^{34},y^{56}\; | \; \overline{y}^{34},
\overline{y}^{56},\overline{\eta}^1,\overline{\psi}^{1,\dots,5}\},\\
\bm{b_2}&=\{\psi^\mu,\chi^{34},y^{12},w^{56}\; | \; \overline{y}^{12},
\overline{w}^{56},\overline{\eta}^2,\overline{\psi}^{1,\dots,5}\},\nonumber\\
\bm{b_3}&=\{\psi^\mu,\chi^{56},w^{1234}\; | \; \overline{w}^{1234},\overline{\eta}^3,\overline{\psi}^{1,\dots,5}\},\nonumber\\
\bm{z_1}&=\{\overline{\phi}^{1,\dots,4}\},\nonumber\\
\bm{x}&=\{\overline{\psi}^{1,\dots,5},\overline{\eta}^{1,2,3}\},\nonumber\\
\bm{\gamma}&=\{y^3y^6,y^1w^6,w^1w^3 \ | \  \bar{\psi}^{1,2,3,4,5}=\bar{\eta}^{1,2,3}=\frac{1}{2},\bar{\phi}^{1,2,6,7}=\frac{1}{2} \}\nonumber
\end{align}
We can immediately note the following for this class
\begin{align}
    \begin{split}
        \bm{E}&=(0,1,0,1,1,0)\\
        \bm{\Delta}&=(1,1,1)\\
        \bm{D}&=(1,1,1)
    \end{split}
\end{align}
which will help us easily determine the key characteristics of the models in this class.

The vector bosons from the untwisted sector of these models generate the gauge symmetry group 
\begin{align} \label{GGA}
\text{Observable: } \ &SU(5)\times U(1) \times U(1)_{k=1,2,3} \times U(1)_{l=4,5,6}  \\
\text{Hidden: } \ &SU(2)\times U(1)_{H_1} \times SO(4)^2 \times SU(2)\times U(1)_{H_2}.
\end{align} 
where we note that $U(1)_{k=1,2,3}$ are generated by the antiholomorphic currents $\bar{\eta}^k\bar{\eta}^{k*}$ and the $U(1)_{l=4,5,6}$ are horizontal symmetries arising from the asymmetric pairings: $\bar{y}^3\bar{y}^6,\bar{y}^1\bar{w}^6$ and $\bar{w}^1\bar{w}^3$. Another important note is that for this Class of models we can apply eq. (\ref{JiJj}) and see that all the untwisted moduli are, indeed, retained. \\

From the discussion in Section \ref{SUSYSpace} we note that the space of $\mathcal{N}=1$ vacua is $2^{45}\sim 3.52\times 10^{13}$. It is important to note at this point that there are two imaginary phases $\CC{\mathds{1}}{\bm{\gamma}}=\pm i$ and $\CC{\bm{z_1}}{\bm{\gamma}}=\pm i$, consistent with modular invariance, and all other phases are real. Furthermore, we note that the latter of these, $\CC{\bm{z_1}}{\bm{\gamma}}$, and the following four phases do not play a role in the phenomenological constraints 
\beq 
\CC{\mathds{1}}{\bm{b_1}}, \ \CC{\mathds{1}}{\bm{b_2}}, \CC{\mathds{1}}{\bm{b_3}}, \ \CC{\mathds{1}}{\bm{z_1}}.
\eeq 
This leaves a space of $2^{40}\sim 1.1\times 10^{12}$ $\mathcal{N}=1$ GGSO phase configurations. 

\subsection{Class A Phenomenological Features}\label{APheno}
\subsubsection*{Observable Spinorial Representations}
From eq. (\ref{spin16sAll}) we can write the sectors producing fermions generations 
\begin{align}
    \begin{split} \bm{F}^1_{qr}&=\bm{b_1}+q\bm{e_4}+r\bm{e_5} \\
    \bm{F}^2_{qr}&=\bm{b_2}+q\bm{e_2}+r\bm{e_5} \\
    \bm{F}^3_{qs}&=\bm{b_3}+q\bm{e_2}+s\bm{e_4} 
    \end{split}
\end{align}
and $\bm{D}=(1,1,1)$ means that any of these sectors will produce one copy of all states in the $\mathbf{16}$ or $\overline{\mathbf{16}}$ when present in the massless spectrum. Therefore the number of generations (\ref{gens}) simplifies to 
\beq 
n_g=N_{16}-N_{\overline{16}}.
\eeq 

\begin{figure}[!htb]
\centering
\includegraphics[width=0.8\linewidth]{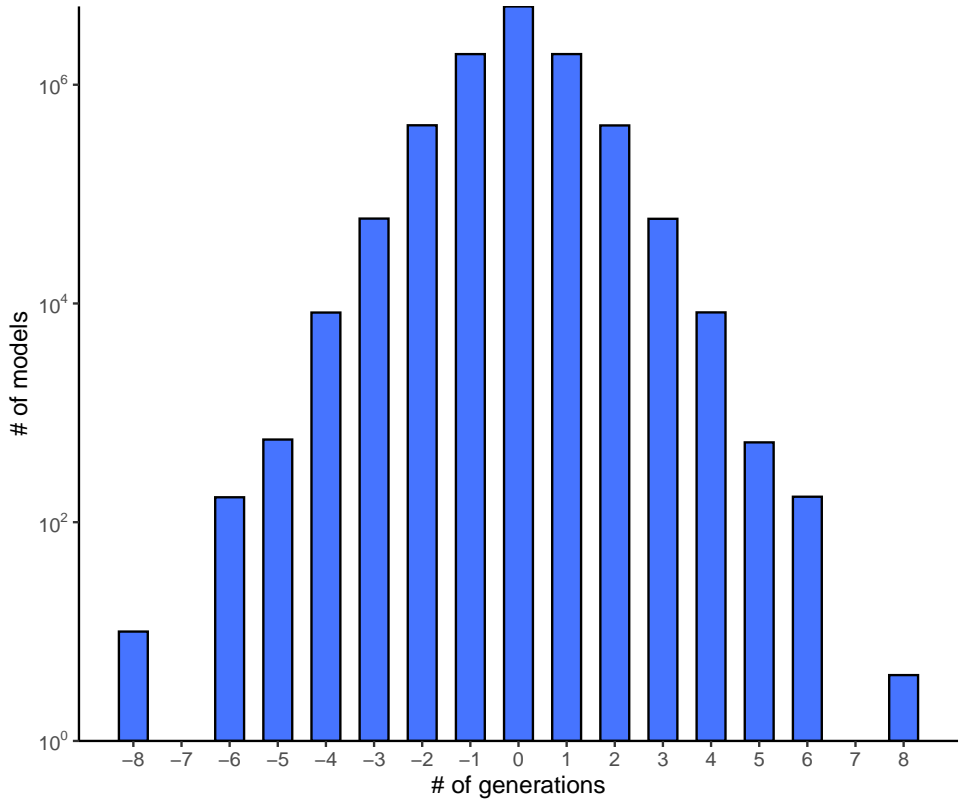}
\caption{\label{GensGraphA}\emph{Frequency plot for number of generations from a sample of $10^7$ Class A vacua.}}
\end{figure}

Encoding the condition for 3 generations (\ref{3gen}) for this class of models into Z3 returns \texttt{sat} to confirm  3 generation models are present for this class. In order to see the spread of generation number, $n_g$, we can generate a bar graph of generations for a random scan of Class A models. This graph is shown in Figure \ref{GensGraphA} for a sample of $10^7$ vacua with $N_{16}\geq N_{\overline{16}}$ so that models with $n_g \geq 0$ are plotted. From this sample we find 3 generations models with probability of approximately $6\times 10^{-3}$. 

\subsubsection*{Heavy Higgs}
From eq. (\ref{HeavyBs}) we can write the Heavy Higgs producing sectors for the Class A models as
\begin{align}
    \begin{split} \bm{B}^1_{qr}&=\bm{S}+\bm{b_1}+q\bm{e_4}+r\bm{e_5} \\
    \bm{B}^2_{qr}&=\bm{S}+\bm{b_2}+q\bm{e_2}+r\bm{e_5} \\
    \bm{B}^3_{qs}&=\bm{S}+\bm{b_3}+q\bm{e_2}+s\bm{e_4} 
    \end{split}
\end{align}
and note that each sector $\bm{B}^{k}_{pqrs}$, $k=1,2,3$, generates a $\mathbf{16}+\mathbf{\overline{16}}$, which correspond to the would-be superpartners of the fermionic states in the $\mathbf{16}/\mathbf{\overline{16}}$ and their CPT conjugates. Therefore, any sector $\bm{B}^{k}_{pq}$ that survives generates one Heavy Higgs $\left(\mathbf{10}, +\frac{1}{2}\right)+\left(\mathbf{\overline{10}}, -\frac{1}{2}\right)$, along with a further vector-like pair $\left(\mathbf{\bar{5}}, +\frac{3}{2}\right)+\left(\mathbf{1},\frac{5}{2}\right)+\left(\mathbf{5}, -\frac{3}{2}\right)+\left(\mathbf{1},-\frac{5}{2}\right)$. 
We can thus write the number of Heavy Higgs for a specific model as equal to the number of surviving sectors $\bm{B}^{k}_{pqrs}$
\beq 
n_{10H}=\sum_{\substack{k=1,2,3 \\ q,r,s=0,1}} \mathbb{P}_{\bm{B}^k_{qrs}}.
\eeq 

\subsubsection*{Top Quark Mass Couplings}
We note that we have possible TQMC from untwisted type couplings of the general form
\begin{align}
    \begin{split}\label{UTqmc}
        \bm{F}^1\bm{F}^1{\bar h}_1, \ \ \ \bm{F}^2\bm{F}^2{\bar h}_2, \ \ \  
        \bm{F}^3\bm{F}^3{\bar h}_3
    \end{split}
\end{align}
where ${\bar h}_k$, $k=1,2,3$, are the Higgs representations from the Neveu-Schwarz sector. In addition, there is also the possibility of twisted type couplings of the general form
\begin{align}\label{TTQMCA}
    \begin{split}
        &\bm{F}^1\bm{F}^2 \bm{V}^3_{\{\bar{\psi}^a\}}, \ \ \ 
        \bm{F}^1\bm{V}^2_{\{\bar{\psi}^a\}}\bm{F}^3,\ \ \ \bm{V}^1_{\{\bar{\psi}^a\}}\bm{F}^2\bm{F}^3
    \end{split}
\end{align}
In classifying vacua from Class A we will account for all 6 possibilities to check for any potentially viable TQMCs for a model. 

In particular, the presence of twisted Light Higgs is not a necessary condition for viable phenomenology in the FSU5 asymmetric models since with untwisted Higgs doublets generating a TQMC of untwisted type (\ref{UTqmc}) they are not necessary. However, the presence of such a coupling is not automatic \textit{a priori} and so for the analysis of whether a model contains a viable TQMC we will also have to check for TQMC from twisted-type coupling (\ref{TTQMCA}). 

Applying eq. (\ref{Vs}) we can write the sectors generating the Light Higgs representations as 
\begin{align}
    \begin{split} \bm{V}^1_{qr}&=\bm{S}+\bm{b_1}+\bm{x}+q\bm{e_4}+r\bm{e_5} \\
    \bm{V}^2_{qr}&=\bm{S}+\bm{b_2}+\bm{x}+q\bm{e_2}+r\bm{e_5} \\
    \bm{V}^3_{qs}&=\bm{S}+\bm{b_3}+\bm{x}+q\bm{e_2}+s\bm{e_4} 
    \end{split}
\end{align}
when accompanied by an antiholomorphic oscillator $\{\bar{\psi}^a\}$, $a\in\{1,2,3,4,5\}$. 

The projectors can be written as follows for these sectors
\begin{align}
    \begin{split}
        \mathbb{P}_{\{\bar{\psi}^a\}\bm{V}^1_{qr}}&=\frac{1}{2^4}\left(1 + \CC{\bm{e_2}}{\bm{V}^{(1)}_{qr}}\right)\left(1+ \CC{\bm{2\bm{\gamma} +\bm{x}}}{\bm{V}^{(1)}_{qr}}\right)
\prod_{a=1,2}\left(1+ \CC{\bm{z_a}}{\bm{V}^{(1)}_{qr}}\right)\\
\mathbb{P}_{\{\bar{\psi}^a\}\bm{V}^2_{qr}}&=\frac{1}{2^4}\left(1 + \CC{\bm{e_4}}{\bm{V}^{(2)}_{qr}}\right)\left(1+ \CC{\bm{2\bm{\gamma} +\bm{x}}}{\bm{V}^{(2)}_{qr}}\right)
\prod_{a=1,2}\left(1+ \CC{\bm{z_a}}{\bm{V}^{(2)}_{qr}}\right)\\
\mathbb{P}_{\{\bar{\psi}^a\}\bm{V}^3_{qs}}&=\frac{1}{2^4}\left(1 + \CC{\bm{e_5}}{\bm{V}^{(3)}_{qs}}\right)\left(1+ \CC{\bm{2\bm{\gamma} +\bm{x}}}{\bm{V}^{(3)}_{qs}}\right)
\prod_{a=1,2}\left(1+ \CC{\bm{z_a}}{\bm{V}^{(3)}_{qs}}\right)
    \end{split}
\end{align}
Using these we can write the number of Light Higgs states for a specific model as equal to the number of $\{\bar{\psi}^a\}\ket{\bm{V}^{k}_{qrs}}$ in the massless spectrum
\beq 
n_{5h}=\sum_{\substack{k=1,2,3 \\ q,r,s=0,1}} \mathbb{P}_{\bm{V}^k_{qrs}}.
\eeq

\subsubsection*{Tachyonic Sector Analysis}
 When classifying the $\mathcal{N}=0$ models we must ensure the projection of all on-shell tachyonic sectors. In addition to the model-independent tachyonic sectors (\ref{MItachs}), we have the following on-shell tachyonic sectors for Class A models that require an anitholomorphic oscillator 
\begin{equation}\label{Vectachs}
\footnotesize
T_1=\begin{Bmatrix}
\{ \bar{\lambda}\}_{\frac{1}{2}}: & \ket{\bm{e_2}} & \ket{\bm{e_4}} & \ket{\bm{e_5}} \\
\{ \bar{\lambda}\}_{\frac{1}{2}}: &\ket{\bm{e_2}+\bm{e_4}} & \ket{\bm{e_2}+\bm{e_5}} & \ket{\bm{e_4}+\bm{e_5}}\\
\{ \bar{\lambda}\}_{\frac{1}{2}}: &\ket{\bm{e_2}+\bm{e_4}+\bm{e_5}}&\ket{\bm{G}+\bm{e_2}+\bm{e_4}+\bm{e_5}}\\
\{ \bar{\lambda}\}_{\frac{1}{2}}: &\ket{(3)\bm{\gamma}}& \ket{\bm{x}+(3)\bm{\gamma}}\\
\{ \bar{\lambda}\}_{\frac{1}{4}}: &\ket{\bm{z_1}+(3)\bm{\gamma}}&\ket{\bm{z_2}+(3)\bm{\gamma}}&\ket{\bm{z_1}+\bm{x}+(3)\bm{\gamma}}&\ket{\bm{z_2}+\bm{x}+(3)\bm{\gamma}}
\end{Bmatrix}
\end{equation}
As well as the following on-shell tachyonic sectors which arise with no oscillator
\begin{equation}\label{SpintachsA}
\footnotesize
T_2=\begin{Bmatrix}
\ket{\bm{z_1}} & \ket{\bm{z_2}} & \\
\ket{\bm{e_i}+\bm{z_1}} & \ket{\bm{e_i}+\bm{z_2}} &  \\
\ket{\bm{e_i}+\bm{e_j}+\bm{z_1}} & \ket{\bm{e_i}+\bm{e_j}+\bm{z_2}} &  \\
\ket{\bm{e_i}+\bm{e_j}+\bm{e_k}+\bm{z_1}} & \ket{\bm{e_i}+\bm{e_j}+\bm{e_k}+\bm{z_2}} &  \\
\ket{\bm{G}+\bm{e_2}+\bm{e_4}+\bm{e_5}+\bm{z_1}} & \ket{\bm{G}+\bm{e_2}+\bm{e_4}+\bm{e_5}+\bm{z_2}} & \\
&& \\
\ket{\bm{x}+2\bm{\gamma}} &\ket{\bm{z_1}+\bm{x}+2\bm{\gamma}} &\\
\ket{\bm{e_i}+\bm{x}+2\bm{\gamma}}&\ket{\bm{e_i}+\bm{z_1}+\bm{x}+2\bm{\gamma}}\\ 
\ket{\bm{e_i}+\bm{e_j}+\bm{x}+2\bm{\gamma}}&\ket{\bm{e_i}+\bm{e_j}+\bm{z_1}+\bm{x}+2\bm{\gamma}}\\
\ket{\bm{e_2}+\bm{e_4}+\bm{e_5}+\bm{x}+2\bm{\gamma}}&\ket{\bm{e_2}+\bm{e_4}+\bm{e_5}+\bm{z_1}+\bm{x}+2\bm{\gamma}} &\\
\ket{\bm{G}+\bm{e_2}+\bm{e_4}+\bm{e_5}+\bm{x}+2\bm{\gamma}}&\ket{\bm{G}+\bm{e_2}+\bm{e_4}+\bm{e_5}+\bm{z_1}+\bm{x}+2\bm{\gamma}} &\\
&& \\
 \ket{\bm{z_2}+\bm{x}+2\bm{\gamma}} & \ket{\bm{z_1}+\bm{z_2}+\bm{x}+2\bm{\gamma}} \\
 \ket{\bm{e_i}+\bm{z_2}+\bm{x}+2\bm{\gamma}} & \ket{\bm{e_i}+\bm{z_1}+\bm{z_2}+\bm{x}+2\bm{\gamma}} \\
  \ket{\bm{e_i}+\bm{e_j}+\bm{z_2}+\bm{x}+2\bm{\gamma}} & \ket{\bm{e_i}+\bm{e_j}+\bm{z_1}+\bm{z_2}+\bm{x}+2\bm{\gamma}} \\
 \ket{\bm{e_2}+\bm{e_4}+\bm{e_5}+\bm{z_2}+\bm{x}+2\bm{\gamma}} & \ket{\bm{e_2}+\bm{e_4}+\bm{e_5}+\bm{z_1}+\bm{z_2}+\bm{x}+2\bm{\gamma}} \\
 \ket{\bm{G}+\bm{e_2}+\bm{e_4}+\bm{e_5}+\bm{z_2}+\bm{x}+2\bm{\gamma}} &\ket{\bm{G}+\bm{e_2}+\bm{e_4}+\bm{e_5}+\bm{z_1}+\bm{z_2}+\bm{x}+2\bm{\gamma}} \\
 &&\\
\ket{\bm{z_1}+\bm{z_2}+(3)\bm{\gamma}}&\ket{\bm{z_1}+\bm{z_2}+\bm{x}+(3)\bm{\gamma}}
\end{Bmatrix}
\end{equation}
where $i\neq j\in\{2,4,5\}$. 

All of these sectors, $\bm{t}\in T_1$ and $\bm{t}\in T_2$, must be projected from the spectrum through appropriate definitions of their generalised projectors $\mathbb{P}_{\bm{t}}=0$. Since there are so many sectors this is generally the most computationally expensive aspect of the classification methodology and is a key reason for introducing SMT methods into the program. 

For reasons of efficiency in projecting the tachyonic sectors we can split the projection into two steps. Firstly, since the SUSY generating vector $\bm{S}$ acts as a projector on all tachyonic sectors, we can implement this projection on all tachyonic sectors and see which sectors remain. Then we can construct and perform the other projections for the remaining sectors. 

\subsubsection*{Enhancements}

In classifying the Class A models we should ensure the absence of enhancements to the observable gauge factors coming from the class-Independent sectors given in eq. (\ref{ObsEnhs}) using the generalised projectors discussed in Section \ref{CIA}. We have further sectors giving possible observable enhancements through combinations with $\bm{\gamma}$. At the level $(\bm{\alpha}_L\cdot \bm{\alpha}_L,\bm{\alpha}_R\cdot \bm{\alpha}_R)=(0,6)$ we have the following sectors
\beq
\psi^\mu \{\bar{\lambda}\}_{\frac{1}{4}}
\begin{cases}
\ket{\bm{e_{136}}+(3)\bm{\gamma}}=:\bm{O_1}\\
\ket{\bm{e_{136}}+\bm{x}+(3)\bm{\gamma}}=:\bm{O_2}\\
\end{cases}
\eeq
 and at level $(0,8)$ there are the sectors 
\beq
\psi^\mu 
\begin{cases}
\ket{\bm{e_{136}}+\bm{z_1}+(3)\bm{\gamma}}=:\bm{O_3}\\
\ket{\bm{e_{136}}+\bm{z_1}+\bm{x}+(3)\bm{\gamma}}=:\bm{O_4}\\
\ket{\bm{e_{136}}+\bm{z_2}+(3)\bm{\gamma}}=:\bm{O_5}\\
\ket{\bm{e_{136}}+\bm{z_2}+\bm{x}+(3)\bm{\gamma}}=:\bm{O_6}
\end{cases}
\eeq
which should be projected to ensure the absence of observable enhancements. In order to construct the projectors we note that
\begin{align}\label{OUpsilons}
    \begin{split}
        \Upsilon(\bm{O_{1,2}})=&\{\bm{S},\bm{e_2},\bm{e_4},\bm{e_5},\bm{z_1}+\bm{z_2}+\bm{x}+2\bm{\gamma}\}\\
        \Upsilon(\bm{O_{3,4}})=&\{\bm{S},\bm{e_2},\bm{e_4},\bm{e_5},\bm{z_2}+\bm{x}+2\bm{\gamma}\}\\
        \Upsilon(\bm{O_{5,6}})=&\{\bm{S},\bm{e_2},\bm{e_4},\bm{e_5},\bm{z_1}+\bm{x}+2\bm{\gamma}\}
    \end{split}
\end{align}
and the projectors have the form 
\begin{align}
\mathbb{P}_{\bm{O}_{1,2}}&=\prod_{\bm{\xi}\in \Upsilon(\bm{O}_{1,2})}\frac{1}{2}\left( 1+\delta_{\bm{O}_{1,2}} \delta^{\psi^\mu}_{\bm{\xi}}\delta^{\bar{\lambda}}_{\bm{\xi}} \CC{\bm{O}_{1,2}}{\bm{\xi}}\right)\\
\mathbb{P}_{\bm{O}_{3,4,5,6}}&=\prod_{\bm{\xi}\in \Upsilon(\bm{O}_{3,4,5,6})}\frac{1}{2}\left( 1+\delta_{\bm{O}_{3,4,5,6}} \delta^{\psi^\mu}_{\bm{\xi}} \CC{\bm{O}_{3,4,5,6}}{\bm{\xi}}\right).
\end{align}
which gives three unique projectors from (\ref{OUpsilons}), on which we impose 
\beq \label{NoObsEnhA}
\forall \ \bar{\lambda}, \ \forall \ i\in[1,6]: \ \  \mathbb{P}_{\bm{O_i}}=0.
\eeq. 
\subsubsection*{Exotic Sectors}
Another important consideration for ensuring reasonable phenomenology is the absence of chiral exotics.


Along with the sectors (\ref{Exots}) there are 124 sectors at the level $(4,6)$ that can produce exotic massless states with a right moving oscillator such that $\nu_f=\frac{1}{2}$ or $\nu_{f^*}=-\frac{1}{2}$. These all arise in pairs with $+\bm{\gamma}$ and $+3\bm{\gamma}$ which contribute equal and opposite gauge charges and therefore do not contribute to any chiral anomaly. Similarly for the 212 exotic sectors at level $(4,8)$. 
Therefore we conveniently do not need to implement a condition on chiral exotics in the classification for this class of models.

\subsection{Class A Results}
Having defined the key phenomenological characteristics for models in Class A we now seek to classify a large space of both $\mathcal{N}=0$ and $\mathcal{N}=1$ vacua with reference to the following key classification criteria 

\begin{align}\label{ClassConstraints}
    \begin{split}
        &(1) \ \text{No On-Shell Tachyons as discussed in Section \ref{MIPheno} and \ref{APheno}} \\
        &(2) \ \text{No Observable Enhancements as given by eq. (\ref{NoObsEnhMI}) and (\ref{NoObsEnhA})} \\ 
        &(3) \ \text{Complete Generations: } \ n_g\neq 0 \text{ and }  n_{10}-n_{\overline{10}}=n_{\bar{5}}-n_5 \ \ \\
        &(4) \ \text{Three generations: }\ n_g=3: \ \ \\
        &(5) \ \text{Presence of Heavy Higgs: } \ n_{10H}\geq 1 \ \ \\ 
        &(6) \ \text{Presence of viable TQMC as discussed in Section \ref{TQMCMI} and \ref{TTQMCA}}\\
        &(7) \ \text{Super No-Scale Condition: } a_{00}=N_b^0-N_f^0=0 
    \end{split}
\end{align}
We note that determining whether a viable TQMC is present requires checking for either an untwisted or twisted type coupling. 

\begin{figure}[t]
\centering
\includegraphics[width=1\linewidth]{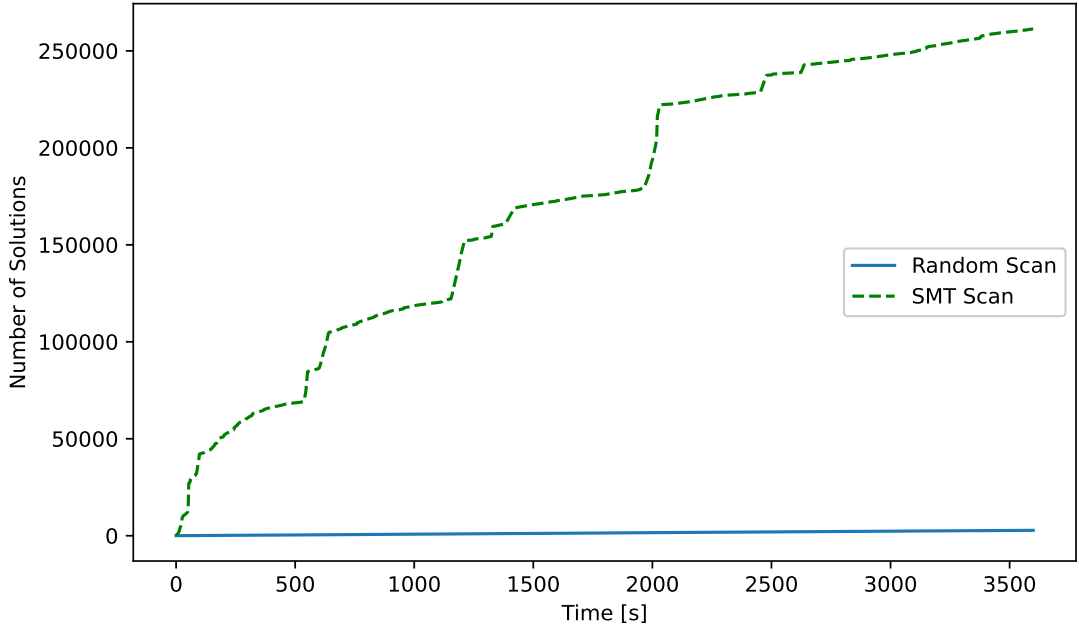}
\caption{\label{ClassATimings}\emph{Rate at which the Z3 SMT finds solutions satisfying constraints (1)-(7) compared with a random generation approach over a 1 hour period.}}
\end{figure}

The results of a classification of $10^9$ Class A models created through random generation is presented in Table \ref{StatstableA}. 
\begin{table}[H]
\small
\centering
\begin{tabular}{|c|l|r|c|c|c|r|}
\hline
 & \multicolumn{5}{|l|}{Total models in sample: $10^9$}   \\ \hline
  & SUSY or Non-SUSY: & $\mathcal{N}=1$ &Probability& $\mathcal{N}=0$& Probability  \\ \hline
&{ Total} & 15624051 &$1.56\times 10^{-2}$ &984375949& 0.984  \\  \hline
(1)&{+ Tachyon-Free} & \cellcolor{gray!25} &\cellcolor{gray!25}&30779240& $3.08\times 10^{-2}$   \\  \hline
(2)& {+ No Observable Enhancements} & 15135704 &$1.51\times 10^{-2}$&28581301&$2.86\times 10^{-2}$  \\ \hline
(3)&{+ Complete Generations} &15135704  &$1.51\times 10^{-2}$&28581301& $2.86\times10^{-2}$ \\  \hline
(4)&{+ Three Generations} & 89930 &$8.99\times 10^{-5}$&195716& $1.96\times 10^{-4}$  \\  \hline
(5)&{+ Heavy Higgs}& 89820 &$8.98\times10^{-5}$&129233&$1.29\times10^{-4}$  \\ \hline
(7)&{+ TQMC}& 89820 &$8.98\times10^{-5}$&129233& $1.29\times10^{-4}$ \\ \hline
(8)&{+ $a_{00}=N_b^0-N_f^0=0$}& \cellcolor{gray!25}&\cellcolor{gray!25}&388& $3.88\times10^{-7}$ \\ \hline
\end{tabular}
\caption{\label{StatstableA} \emph{Phenomenological statistics from sample of $10^8$ Class A models.  Note that the number of $a_{00}=0$ models is an estimate based on extrapolating from a sample of $2 \times 10^3$ of the 129233 $\mathcal{N}=0$ models satisfying (1)-(7).}}
\end{table}
As mentioned in Section \ref{SMTClass} we can employ our Z3 SMT Solver to efficiently find models satisfying the phenomenological criteria as well as to inform us of when criteria are in contradiction and no solutions can be found. As a test of efficiency we ran the SMT for 1 hour to see how many models it finds satisfying the criteria (1)-(7) in Table \ref{StatstableA} and compared it with the random generation method over the same time. The result of this comparison is displayed in Figure \ref{ClassATimings}. We find that the SMT is approximately 322 times faster than the random scan after 3 minutes but after 1 hour it levels out at approximately 93 times faster. This demonstrates that the Z3 SMT tool is especially effective as a fishing algorithm in finding pools of solutions very quickly, whereas its efficiency in complete enumeration of solutions reduces. If we are interested in more complete enumeration it may be instrumental to employ another SAT/SMT solver such as PicoSAT \cite{Pico} which is optimised for such complete enumeration.  

\begin{figure}[H]
\centering
\includegraphics[width=1\linewidth]{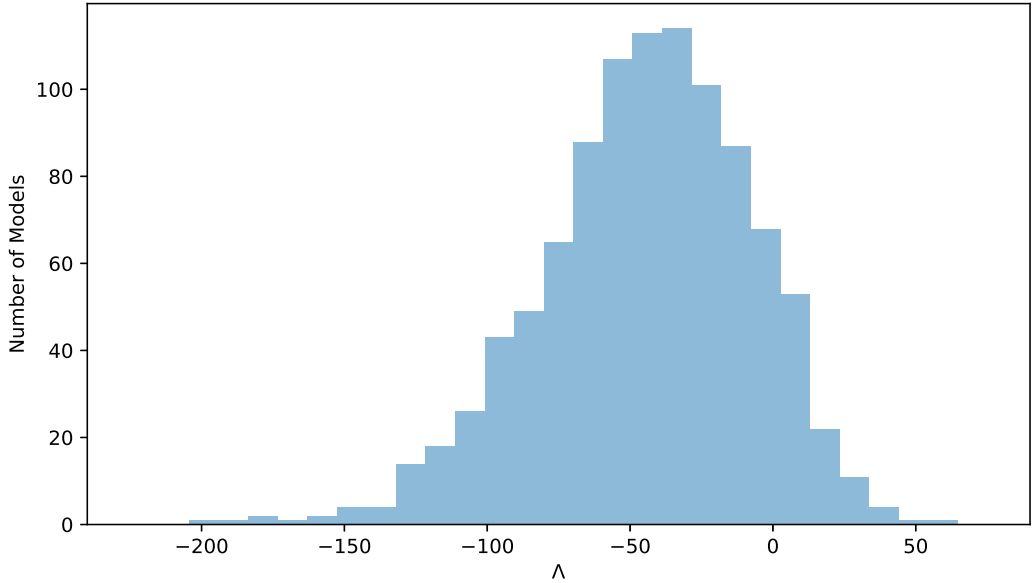}
\caption{\label{ClassACoC}\emph{The distribution of the cosmological constant $\Lambda_\text{ST}$ for a sample off $10^3$ Class A models satisfying conditions (1)-(7) of Table \ref{StatstableA}. To gain the physical value, a factor of $\mathcal{M}^4$ must be reinstated. These values are evaluated at the free fermionic point using methods discussed in Section \ref{PF}.}}
\end{figure}

\begin{figure}[H]
\centering
\includegraphics[width=1\linewidth]{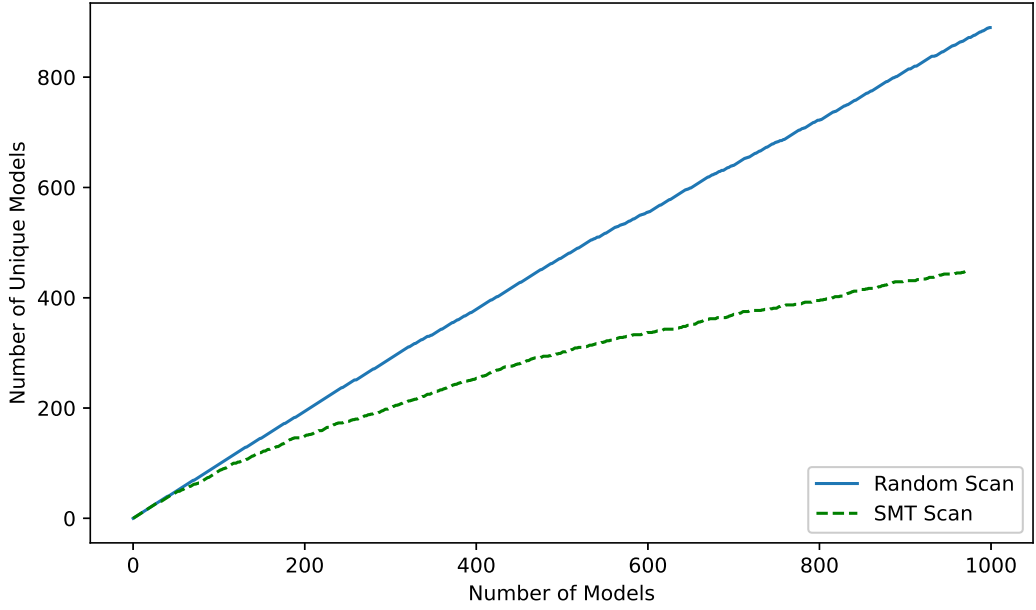}
\caption{\label{ClassADegeneracy}\emph{The degeneracy of models in a Random versus an SMT scan for Class A as seen from the partition function.}}
\end{figure}

We can also perform a statistical analysis at the level of the partition function. This includes the calculation of the $q$-expanded partition function and the evaluation of the one loop cosmological constant. In Figure \ref{ClassACoC}, we present the distribution of the cosmological constant for a sample of Class A models evaluated at the free fermionic point. This shows that there is a tendency towards negative values, even though positive values are not excluded. It is important to note that this is not guaranteed to be a stable point in moduli space as there may be flat directions, however, the analysis of the potential is outside the scope of this paper and is left for future work. It is also interesting to compare the effectiveness of the SMT and random scan algorithms in finding unique models from the point of view of the partition function. Form Figure \ref{ClassADegeneracy} we see that the SMT algorithm has a tendency to find more degenerate solutions as compared to a random scan. However, this does not conclude that random scans are more efficient. Indeed, comparing this to Figure \ref{ClassATimings}, we see that SMT algorithms still vastly outperform random scans by more than 2 orders of magnitude.

\subsection{Example Model Class A} \label{ClassAExample}
Having classified a random sample of Class A vacua, we can provide an example model satisfying criteria (1)-(7) of (\ref{ClassConstraints}). Consider a model defined by the basis set (\ref{basisA}) and choice of GGSO phases given by
{\begin{equation}
\small
\CC{\bm{v_i}}{\bm{v_j}}= 
\begin{blockarray}{ccccccccccccc}
&\mathbf{1}& \bm{S} & \bm{e_2}& \bm{e_4}&\bm{e_5}&\bm{b_1}& \bm{b_2}&\bm{b_3}&\bm{z_1}& \bm{x}&\bm{\gamma}& \\
\begin{block}{c(rrrrrrrrrrrr)}
\mathbf{1}&  1 &   1 & -1 & -1 & -1 &   1 & -1 &   1 &   1 &   1 &  i&\ \\
\bm{S}&   1 &   1 & -1 &   1 &   1 &   1 &   1 &   1 &   1 & -1 &   1 &\ \\
\bm{e_2}& -1 & -1 &   1 &   1 & -1 & -1 & -1 & -1 &   1 & -1 & -1 &\ \\
\bm{e_4}& -1 &   1 &   1 &   1 & -1 &   1 & -1 &   1 & -1 & -1 & -1 &\ \\
\bm{e_5}& -1 &   1 & -1 & -1 &   1 & -1 & -1 & -1 & -1 & -1 &   1 &\ \\
\bm{b_1}&   1 & -1 & -1 &   1 & -1 &   1 & -1 & -1 &   1 &   1 &   1 &\ \\
\bm{b_2}& -1 & -1 & -1 & -1 & -1 & -1 & -1 & -1 & -1 &   1 & -1 &\ \\
\bm{b_3}&   1 & -1 & -1 &   1 & -1 & -1 & -1 &   1 &   1 & -1 & -1 &\ \\
\bm{z_1}&   1 &   1 &   1 & -1 & -1 &   1 & -1 &   1 &   1 & -1 &  i&\ \\
\bm{x} &   1 & -1 & -1 & -1 & -1 & -1 & -1 &   1 & -1 & -1 & -1 &\ \\
\bm{\gamma}&    1 &   1 & -1 & -1 &   1 & -1 &   1 &   1 & -1 & -1 & -1 &\ \\
\end{block}
\end{blockarray}
\end{equation}}
This model has 3 fermion generations arising from $\bm{b_1}+\bm{e_4}$, $\bm{b_2}+\bm{e_2}$ and $\bm{b_3}+\bm{e_2}+\bm{e_4}$. As for all models in this class, there are untwisted Higgs states from all 3 orbifold planes. The top quark mass coupling arises on each plane from a coupling of the type $U_kF_kF_k$ discussed in Section \ref{TQMCMI}. The Heavy Higgs is provided by the sector $\bm{S}+\bm{b_2}+\bm{e_5}$ to ensure that the $SU(5)\times U(1)$ can be broken to the SM.

The partition function and for Class A models can be found using the methods discussed in Section \ref{PF}. Specifically, the internal lattice can be constructed by noting that the form of $\bm{A}$ introduces exactly one asymmetric pairing in each of the three orbifold planes. Thus the internal lattice takes the form
\begin{align}
 \Gamma^{\gamma}_{(6,6)}=& \; \Gamma_1^{\gamma} \times \Gamma_2^{\gamma} \times \Gamma_3^{\gamma} \nonumber\\
 =& \; \vth_{y^4}\smb{r+h_2+H_4}{s+g_2+G_4}^{1/2} \vth_{y^5}\smb{r+h_2+H_5}{s+g_2+G_5}^{1/2} \vth_{y^{3,6}}\smb{r+h_2+2H'}{s+g_2+2G'}\nonumber\\ 
 &\times\vth_{\bar{y}^4}\smb{r+h_2+H_4}{s+g_2+G_4}^{1/2} \vthb_{\bar{y}^5}\smb{r+h_2+H_5}{s+g_2+G_5}^{1/2} \vthb_{\bar{y}^{3,6}}\smb{r+h_2}{s+g_2} \nonumber\\[0.3cm]
 &\times \; \vth_{y^2}\smb{r+h_2+H_2}{s+g_2+G_2}^{1/2}  \vth_{w^5}\smb{r+h_2+H_5}{s+g_2+G_5}^{1/2} \vth_{y^1w^6}\smb{r+h_2+2H'}{s+g_2+2G'} \\ 
 &\times\vth_{\bar{y}^2}\smb{r+h_1+H_2}{s+g_1+G_2}^{1/2} \vthb_{\bar{w}^5}\smb{r+h_1+H_5}{s+g_1+G_5}^{1/2} \vthb_{\bar{y}^1\bar{w}^6}\smb{r+h_1}{s+g_1} \nonumber\\[0.3cm]
 &\times \; \vth_{w^2}\smb{r-h_1-h_2+H_2}{s-g_1-g_2+G_2}^{1/2}  \vth_{w^4}\smb{r-h_1-h_2+H_4}{s-g_1-g_2+G_4}^{1/2} \vth_{w^{1,3}}\smb{r-h_1-h_2+2H'}{s-g_1-g_2+2G'}\nonumber\\ 
 &\times\vth_{\bar{w}^2}\smb{r-h_1-h_2+H_2}{s-g_1-g_2+G_2}^{1/2} \vthb_{\bar{w}^4}\smb{r-h_1-h_2+H_4}{s-g_1-g_2+G_4}^{1/2} \vthb_{\bar{w}^{1,3}}\smb{r-h_1-h_2}{s-g_1-g_2}, \nonumber 
\end{align}
where $\Gamma_i^{\gamma}$ denotes the part corresponding to the $i^\text{th}$ orbifold plane. We can then use this expression together with (\ref{GammaPF}) and (\ref{QExpPF}) to gain the $q$-expanded partition function of this model which is 
\begin{align}
    Z =& \, 2\,q^{0}\bar{q}^{-1} -8\,q^{1/4}\bar{q}^{-3/4} -16\,q^{1/2}\bar{q}^{-1/2} +8\,q^{-1/2}\bar{q}^{1/2}  \nonumber\\
    &+176\,q^{1/8}\bar{q}^{1/8} +976\,q^{1/4}\bar{q}^{1/4} + 2048\,q^{3/8}\bar{q}^{3/8} + 2560\,q^{1/2}\bar{q}^{1/2},
\end{align}
including all terms up to at most $\mathcal{O}(q^{1/2})$ and $\mathcal{O}(\bar{q}^{1/2})$. The top line gives the off-shell tachyonic states required by modular invariance, while the bottom line gives all on-shell states. Note the presence of the off-shell model-independent term $2\,q^0\bar{q}^{-1}$ obtained from the so-called `proto-graviton' resulting from the state $\psi^\mu \ket{0}_L\otimes \ket{0}_R$. This provides a neat check to confirm correct normalisation of the partition function. We also see that this model is indeed of the super no-scale type, {\it i.e.}s has $a_{00}=n_b^0-n_f^0=0$. Integrating this expansion over the fundamental domain of the modular group via (\ref{CoC}) yields the spacetime cosmological constant
\begin{equation}
\Lambda_\text{ST} = 13.34 \times \mathcal{M}^4,
\end{equation}
which was calculated to 4$^\text{th}$ order $q$ and $\bar{q}$. It is important to note that this value is not calculated at a minimum in the moduli space, but rather at a maximally symmetric self dual point where the orbifold theory admits a free fermionic description.  

Whether the cosmological constant can indeed be suppressed requires more in-depth analysis and in these Class A models all untwisted moduli being retained complicates this analysis, which motivates the study of a different class of models where some moduli are projected that we turn to in the next section. Through a translation to a $\mathbb{Z}_2^n$ orbifold in the bosonic picture the dependence on some of these geometric moduli can be reinstated and a systematic investigation of the one-loop potential can be attempted as done in \cite{fr1,fr2} for symmetric orbifolds, however its implementation for asymmetric models is left for future work. 

\section{Asymmetric Orbifold Class B}\label{ClassB}
The second Class of models we study is an example where all untwisted moduli on the second and third tori are projected and only $h_{11},h_{12},h_{21}$ and $h_{22}$ are retained. 
From Table \ref{PairingsClassB} and \ref{PairingsClassF} we can see there are 12 possible pairings in both the bosonic and fermionic cases that give rise to just $h_{11},h_{12},h_{21}$ and $h_{22}$, whilst allowing for odd number generations. These all have $\bm{E}=(1,1,0,0,0,0)$. The possible pairings can be grouped into 3 types according to their $\bm{\Delta}=(\Delta_1,\Delta_2,\Delta_3)$ and degeneracies $\bm{D}=(D_1,D_2,D_3)$, which for the bosonic case are
\begin{equation}\label{Pairs3Ds}
    \bm{A}=\begin{cases}
    \{\bar{w}^{3456}\},  \{y^{34},w^{34},\bar{y}^{34},\bar{w}^{56}\},  \ \ \ \ \ \bm{\Delta}=(0,1,1), \ \ \bm{D}=(8,1,1)\\ 
      \{y^{3456},w^{3456},\bar{y}^{3456}\},  \{y^{56},w^{56},\bar{y}^{56},\bar{w}^{34}\}\\
    \{\bar{y}^{56},\bar{w}^{34}\}, \{y^{56},w^{56},\bar{w}^{3456}\}, \ \ \ \ \ \bm{\Delta}=(1,0,1), \ \ \bm{D}=(4,2,1)\\ 
      \{y^{34},w^{34},\bar{y}^{3456}\},  \{y^{3456},w^{3456},\bar{y}^{34},\bar{y}^{56}\}\\
    \{\bar{y}^{34},\bar{w}^{56}\}, \{y^{34},w^{34},\bar{y}^{34}\bar{w}^{56}\}, \ \ \  \ \ \bm{\Delta}=(1,1,0), \ \ \bm{D}=(4,1,2)\\
    \{y^{3456},w^{3456},\bar{y}^{3456}\}, \{y^{56},w^{56},\bar{y}^{56},\bar{w}^{34}\}
    \end{cases}
\end{equation}
As mentioned in Section \ref{CIA}, the condition for odd number generations \ref{OddGen} 
is a necessary but not sufficient condition for the possibility of having 3 generation models within a class. We can check which of the 3 pairing possibilities in (\ref{Pairs3Ds}) can give rise to 3 generations by checking whether eq. (\ref{3gen}) is satisfiable with our SMT solver for each $\bm{A}$. Doing this tells us that none of the pairings can give rise to 3 generation models. 
Despite this we will choose the pairing $\bm{A}=\{\bar{w}^{34},\bar{w}^{56}\}$ with  $\bm{D}=(4,2,1)$ to classify systematically and in Section \ref{No3Gen} we will demonstrate the origin of the absence of three generations. 

The basis for this class of models will then be
\begin{align}\label{basisB}
\bm{\mathds{1}}&=\{\psi^\mu,
\chi^{1,\dots,6},y^{1,\dots,6}, w^{1,\dots,6}\ | \ \overline{y}^{1,\dots,6},\overline{w}^{1,\dots,6},
\overline{\psi}^{1,\dots,5},\overline{\eta}^{1,2,3},\overline{\phi}^{1,\dots,8}\},\nonumber\\
\bm{S}&=\{{\psi^\mu},\chi^{1,\dots,6} \},\nonumber\\
\bm{e_1}&=\{y^{1},w^{1}\; | \; \overline{y}^{1},\overline{w}^{1}\}, 
\nonumber\\
\bm{e_2}&=\{y^{2},w^{2}\; | \; \overline{y}^{2},\overline{w}^{2}\}, 
\nonumber\\
\bm{b_1}&=\{\psi^\mu,\chi^{12},y^{34},y^{56}\; | \; \overline{y}^{34},
\overline{y}^{56},\overline{\eta}^1,\overline{\psi}^{1,\dots,5}\},\nonumber\\
\bm{b_2}&=\{\psi^\mu,\chi^{34},y^{12},w^{56}\; | \; \overline{y}^{12},
\overline{w}^{56},\overline{\eta}^2,\overline{\psi}^{1,\dots,5}\},\\
\bm{b_3}&=\{\psi^\mu,\chi^{56},w^{1234}\; | \; \overline{w}^{1234},\overline{\eta}^3,\overline{\psi}^{1,\dots,5}\},\nonumber\\
\bm{z_1}&=\{\overline{\phi}^{1,\dots,4}\},\nonumber\\
\bm{x}&=\{\overline{\psi}^{1,\dots,5},\overline{\eta}^{1,2,3}\},\nonumber\\
\bm{\gamma}&=\{ \bar{y}^{56},\bar{w}^{34},\bar{\psi}^{1,...,5}=\bar{\eta}^{1,2,3}=\bar{\phi}^{1,2,6,7}=\frac{1}{2},\bar{\phi}^8\}\nonumber
\end{align}
where we have the same $\bm{z_2}$ combination as eq. (\ref{z2}) and the untwisted gauge group is  
\begin{align} \label{GGB}
\text{Observable: } \ &SU(5)\times U(1) \times U(1)_{i=1,2,3} \times U(1)_{j=4,5} \\
\text{Hidden: } \ &SU(2)\times U(1)_{H_1} \times SO(4) \times U(1)_{H_2}\times  SU(2)\times U(1)_{H_3} \times U(1)_{H_4}.
\end{align} 
There are two horizontal symmetries associated to the antiholomorphic currents from the pairings $\bar{y}^{5,6}$ and $\bar{w}^{3}\bar{w}^{4}$.
Since there are 10 basis vectors we naively have $2^{45}$ independent GGSO configurations but the following 10 phases do not affect the projection criteria for the phenomenological criteria we investigate 
\beq
\CC{\mathds{1}}{\bm{b_1}},\CC{\mathds{1}}{\bm{b_2}},\CC{\mathds{1}}{\bm{b_3}},\CC{\mathds{1}}{\bm{z_1}},\CC{\mathds{1}}{\bm{\gamma}},\CC{\bm{S}}{\bm{\gamma}},\CC{\bm{b_1}}{\bm{\gamma}},\CC{\bm{b_3}}{\bm{\gamma}},\CC{\bm{z_1}}{\bm{\gamma}},\CC{\bm{x}}{\bm{\gamma}}.
\eeq 
This leaves just 35 free GGSO phases generating a space of $2^{35}\sim 3.4 \times 10^{10}$ independent configurations to classify. 
The supersymmetric subspace  of which is subject to conditions (\ref{SUSYconstraint}) and (\ref{SUSYChiral}).

\subsection{Class B Phenomenological Features}\label{BPheno}
\subsubsection*{Observable Spinorials Representations and Absence of Three Generation}\label{No3Gen}
The following sectors give rise to the fermion generations
\begin{align}
   \bm{F}^1_{t}&=\bm{b_1}+t\bm{e_{3456}}\\ 
   \bm{F}^2_{pq}&=\bm{b_2}+p\bm{e_1}+q\bm{e_2}\\
   \bm{F}^3_{pq}&=\bm{b_3}+p\bm{e_1}+q\bm{e_2}
\end{align}
and the degeneracies $\bm{D}$ tell us that $\bm{F}^1_{0}$ generate 4 copies of the $\mathbf{16}$ or $\mathbf{\overline{16}}$, $\bm{F}^1_{1}$ generate 2 copies of the $\mathbf{16}$ \textit{and} 2 copies of the $\mathbf{\overline{16}}$, whilst $\bm{F}^2_{pq}$ generate 2 copies of either
 $\left(\mathbf{10}, +\frac{1}{2}\right)$, $\left(\mathbf{\bar{5}}, -\frac{3}{2} \right) + \left(\mathbf{1},\frac{5}{2}\right)$,
$\left(\mathbf{\overline{10}}, -\frac{1}{2}\right)$ or $\left(\mathbf{5}, +\frac{3}{2}\right) +\left(\mathbf{1},-\frac{5}{2}\right)$. Lastly, $\bm{F}^3_{pqrs}$ generates 1 copy of the $\mathbf{16}$ or $\mathbf{\overline{16}}$. 

As mentioned above, three generation models do not arise in this class, and to see why it will be useful to write the projection equations for these spinorial sectors. We can first construct the projectors for these sectors by utilising eq. (\ref{FUpsilons})
\begin{align}
        \mathbb{P}_{\bm{F}^1_{t}}&=\frac{1}{2^5}\prod_{i=1,2}\left(1 - \CC{\bm{F}^{(1)}_{t}}{\bm{e_i}}\right)\left(1- \CC{\bm{F}^{(1)}_{t}}{\bm{2\bm{\gamma} +\bm{x}}}\right)
\prod_{a=1,2}\left(1- \CC{\bm{F}^{(1)}_{t}}{\bm{z_a}}\right)\label{PF1}\\
\mathbb{P}_{\bm{F}^2_{pq}}&=\frac{1}{2^3}\left(1- \CC{\bm{F}^2_{pq}}{\bm{2\bm{\gamma} +\bm{x}}}\right)
\prod_{a=1,2}\left(1- \CC{\bm{F}^2_{pq}}{\bm{z_a}}\right)\label{PF2}\\
\mathbb{P}_{\bm{F}^3_{pq}}&=\frac{1}{2^3}\left(1- \CC{\bm{F}^3_{pq}}{\bm{2\bm{\gamma} +\bm{x}}}\right)
\prod_{a=1,2}\left(1- \CC{\bm{F}^3_{pq}}{\bm{z_a}}\right)\label{PF3}
\end{align}

Next we can apply eq. (\ref{ChProjs}) to get the chirality phases
\begin{align}
\begin{split}
\bm{X}^1_{t=0} &= -\CC{\bm{F}^1_{0}}{\bm{b_2}}^*, \nonumber\\
\bm{X}^2_{pq} &= -\CC{\bm{F}^2_{pq}}{\bm{b_1}}^*\\
\bm{X}^3_{pq} &= -\CC{\bm{F}^3_{pq}}{\bm{b_1}}^*\nonumber
\end{split}
\end{align}
where we have chosen $\text{ch}(\psi^\mu)=+1$ for the spacetime fermion chirality and note the $\bm{F}^1_1$ does not have a chirality operator as it gives rise to 2 copies of the $\mathbf{16}$ and the $\overline{\mathbf{16}}$. By applying eq. (\ref{FSU5Ns}) we can write the quantum numbers of the $SU(5)\times U(1)$ representations as 
\begin{align} 
\begin{split}
    n_{10}=&\sum_{\substack{ t=0,1}} 2
\mathbb{P}_{\bm{F}^1_{t}}\frac{1}{2}\left(1 +t+(1-t) \bm{X}^1_{t}\right)+\sum_{\substack{ p,q=0,1}} 2
\mathbb{P}_{\bm{F}^2_{pq}}\frac{1}{4}\left(1 + \bm{X}^2_{pq}\right)\left(1+\CC{\bm{F}^2_{pq}}{\bm{\gamma}}\right)\\
&+\sum_{\substack{ p,q=0,1}} 
\mathbb{P}_{\bm{F}^3_{pq}}\frac{1}{2}\left(1 + \bm{X}^3_{pq}\right) \\
    n_{\bar{5}}=&\sum_{\substack{ t=0,1}} 2
\mathbb{P}_{\bm{F}^1_{t}}\frac{1}{2}\left(1 +t+(1-t) \bm{X}^1_{t}\right)+\sum_{\substack{ p,q=0,1}} 2
\mathbb{P}_{\bm{F}^2_{pq}}\frac{1}{4}\left(1 + \bm{X}^2_{pq}\right)\left(1-\CC{\bm{F}^2_{pq}}{\bm{\gamma}}\right)\\
&+\sum_{\substack{ p,q=0,1}} 
\mathbb{P}_{\bm{F}^3_{pq}}\frac{1}{2}\left(1 + \bm{X}^3_{pq}\right)\\
    n_{\overline{10}}=&\sum_{\substack{ t=0,1}} 2
\mathbb{P}_{\bm{F}^1_{t}}\frac{1}{2}\left(1+t -(1-t) \bm{X}^1_{t}\right)+\sum_{\substack{ p,q=0,1}} 2
\mathbb{P}_{\bm{F}^2_{pq}}\frac{1}{4}\left(1 - \bm{X}^2_{pq}\right)\left(1+\CC{\bm{F}^2_{pq}}{\bm{\gamma}}\right)\\
&+\sum_{\substack{ p,q=0,1}} 
\mathbb{P}_{\bm{F}^3_{pq}}\frac{1}{2}\left(1 - \bm{X}^3_{pq}\right)\\
    n_{5}=&\sum_{\substack{ t=0,1}} 2
\mathbb{P}_{\bm{F}^1_{t}}\frac{1}{2}\left(1 +t-(1-t) \bm{X}^1_{t}\right)+\sum_{\substack{ p,q=0,1}} 2
\mathbb{P}_{\bm{F}^2_{pq}}\frac{1}{4}\left(1 - \bm{X}^2_{pq}\right)\left(1-\CC{\bm{F}^2_{pq}}{\bm{\gamma}}\right)\\
&+\sum_{\substack{ p,q=0,1}} 
\mathbb{P}_{\bm{F}^3_{pq}}\frac{1}{2}\left(1 - \bm{X}^3_{pq}\right),
\end{split}
\end{align}
where we note the singlets have the same projection as $\mathbf{5}$ and $\mathbf{\bar{5}}$. 
Imposing the condition for complete generations $n_{10}-n_{\overline{10}}=n_{\overline{5}}-n_5$ results in the condition
\beq \label{completegens}
\sum_{p,q}\mathbb{P}_{\bm{F}^2_{pq}}\CC{\bm{F}^2_{pq}}{\bm{\gamma}}X^2_{pq}=0
\eeq 
and $n_{10}-n_{\overline{10}}=3$ for three generations tells us
\beq \label{3genFs}
3= \sum_{t} 2 \mathbb{P}_{F^1_t}X^1_t+\sum_{p,q}2\mathbb{P}_{F^2_{pq}}\frac{1}{2}\left(1+\CC{\bm{F}^2_{pq}}{\bm{\gamma}}\right)X^2_{pq}+\sum_{p,q}\mathbb{P}_{\bm{F}^3_{pq}}X^3_{pq}
\eeq 
which is only possible if
\beq 
\sum_{p,q}\mathbb{P}_{\bm{F}^3_{pq}}X^3_{pq}\in\{1,3\}
\eeq 
but $\sum_{p,q}\mathbb{P}_{\bm{F}^3_{pq}}X^3_{pq}=3$ we can show is impossible by inspecting (\ref{PF3}) which only depends on nine phases
\beq 
\CC{\bm{b_3}}{\bm{z_1}}, \ \CC{\bm{b_3}}{\bm{z_2}}, \ \CC{\bm{b_3}}{\bm{x}}, \ \CC{\bm{e_1}}{\bm{z_1}}, \ \CC{\bm{e_1}}{\bm{z_2}}, \ \CC{\bm{e_1}}{\bm{x}}, \ \CC{\bm{e_2}}{\bm{z_1}}, \ \CC{\bm{e_2}}{\bm{z_2}}, \ \CC{\bm{e_2}}{\bm{x}}
\eeq 
and if 3 of the 4 sectors $\bm{F}^3_{pq}$ have $\mathbb{P}_{\bm{F}^3_{pq}}=1$ then all 9 phases are fixed and ensures the fourth also has $\mathbb{P}_{\bm{F}^3_{pq}}=1$. 

Therefore the only way to satisfy (\ref{3genFs}) is if $\sum_{p,q}\mathbb{P}_{\bm{F}^3_{pq}}X^3_{pq}=1$. This further implies $\sum_{p,q}\mathbb{P}_{\bm{F}^2_{pq}}\CC{\bm{F}^2_{pq}}{\bm{\gamma}}X^2_{pq}\in \{2,4\}$ from (\ref{completegens}). If we assume $\sum_{p,q}\mathbb{P}_{\bm{F}^2_{pq}}\CC{\bm{F}^2_{pq}}{\bm{\gamma}}X^2_{pq}=2$ then the constraints this imposes on the phases in $\mathbb{P}_{\bm{F}^2_{pq}}$ necessitates
\beq 
\sum_{p,q}\mathbb{P}_{\bm{F}^3_{pq}}X^3_{pq}\in\{0,2\}
\eeq 
making 3 generations impossible. Similarly if $\sum_{p,q}\mathbb{P}_{\bm{F}^2_{pq}}\CC{\bm{F}^2_{pq}}{\bm{\gamma}}X^2_{pq}=4$ this imposes 
\beq 
\sum_{p,q}\mathbb{P}_{\bm{F}^3_{pq}}X^3_{pq}\in\{0,4\}
\eeq 
which again makes 3 generations impossible. 

\begin{figure}[!htb]
\centering
\includegraphics[width=0.7\linewidth]{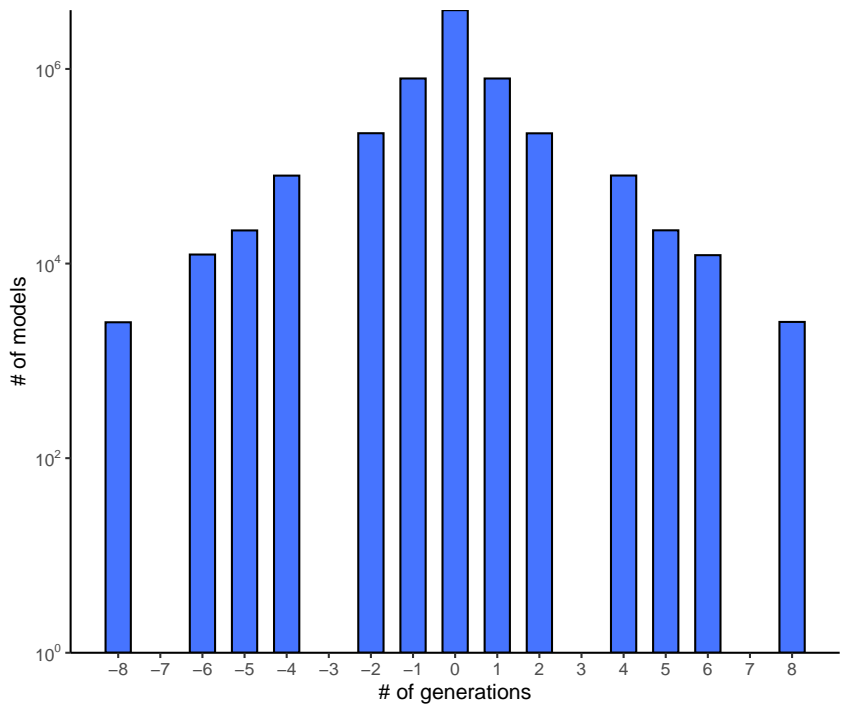}
\caption{\label{GensGraphB}\emph{Frequency plot for number of generations from a sample of $10^7$ Class B vacua.}}
\end{figure}

Not only does the Z3 SMT solver confirm the unsatisfiability of 3 generation configrations, it also generates a proof written in computer language \footnote{Available at https://github.com/BenjaminPercival/AsymmetricOrbifolds.git}. There are also additional tools available in Z3 to explore unsatisfiability such as identifying a minimal `unsatisfiable core' \cite{UnsatCore}, isolating the contradiction by giving a (locally) minimal subset of constraints, where dropping either of them results in a satisfiable constraint system.
In Figure \ref{GensGraphB} the distribution of $n_g$ is plotted for a random sample of $10^7$ Class B models showing empirically the absence of $n_g=3$ models. 

The origin of this contradiction can be traced to the projection of moduli in the 2nd and 3rd tori which means there is no $\bm{e_i}$ vectors to project $\bm{F}^2_{pq}$ and $\bm{F}^3_{pq}$. This results in constraining and correlating their presence in the massless spectrum. 

\subsubsection*{Heavy Higgs}
As in Class A we will demand the presence of at least one heavy Higgs to break the FSU5 gauge symmetry in our classification. The number of Heavy Higgs $\# \left[ (\mathbf{10},\frac{1}{2})+(\overline{\mathbf{10}},-\frac{1}{2})\right]$ can again be calculated through the projections of sectors $\bm{S}+\bm{F}^k_{pq}$.

\subsubsection*{Top Quark Mass Couplings}

As in Class A, in classifying vacua from Class B we will account for all the three untwisted type TQMC and all 3 twisted type when checking whether a potentially viable TQMC arises from a model. 

To check the presence of a viable TQMC we, again, account for the twisted Light Higgs sectors, which can be checked through analogous projection conditions for Class B as in Class A. It is simply the number of $\left[ (\mathbf{5},-1)+(\mathbf{\bar{5}},+1)\right]$ from the sectors $\bm{V}^k_{pq}$ in the massless spectrum.
\subsubsection*{Tachyonic Sector Analysis}
Class A models have significantly fewer tachyonic sectors than Class B. Specifically there are 27 sectors producing on-shell tachyons for Class B, compared with the 78 of Class A. 

The following 3 sectors will produce on-shell tachyons with a right-moving oscillator should they be present in the spectrum of a model
\begin{equation}\label{VectachsB}
\footnotesize
T_1=\begin{Bmatrix}
\{ \bar{\lambda}\}_{\frac{1}{2}}: & \ket{\bm{e_1}} & \ket{\bm{e_2}}  \\
\{ \bar{\lambda}\}_{\frac{1}{2}}: &\ket{\bm{e_1}+\bm{e_2}} 
\end{Bmatrix}
\end{equation}

Further to this, the following on-shell tachyonic sectors arise with no oscillator
\begin{equation}\label{SpintachsB}
\footnotesize
T_2=\begin{Bmatrix}
\ket{\bm{z_1}} & \ket{\bm{z_2}} & \ket{\bm{x}+2\bm{\gamma}} \\
\ket{\bm{e_i}+\bm{z_1}} & \ket{\bm{e_i}+\bm{z_2}} & \ket{\bm{e_i}+\bm{x}+2\bm{\gamma}} \\
\ket{\bm{e_1}+\bm{e_2}+\bm{z_1}} & \ket{\bm{e_1}+\bm{e_2}+\bm{z_2}} & \ket{\bm{e_1}+\bm{e_2}+\bm{x}+2\bm{\gamma}} \\
&& \\
\ket{\bm{z_1}+\bm{x}+2\bm{\gamma}} & \ket{\bm{z_2}+\bm{x}+2\bm{\gamma}} & \ket{\bm{z_1}+\bm{z_2}+\bm{x}+2\bm{\gamma}} \\
\ket{\bm{e_i}+\bm{z_1}+\bm{x}+2\bm{\gamma}} & \ket{\bm{e_i}+\bm{z_2}+\bm{x}+2\bm{\gamma}} & \ket{\bm{e_i}+\bm{z_1}+\bm{z_2}+\bm{x}+2\bm{\gamma}} \\
\ket{\bm{e_1}+\bm{e_2}+\bm{z_1}+\bm{x}+2\bm{\gamma}} & \ket{\bm{e_1}+\bm{e_2}+\bm{z_2}+\bm{x}+2\bm{\gamma}} & \ket{\bm{e_1}+\bm{e_2}+\bm{z_1}+\bm{z_2}+\bm{x}+2\bm{\gamma}} 
\end{Bmatrix}
\end{equation}
where $i\in\{1,2\}$. 

The condition for the absence of such tachyonic sectors can be compactly written
\beq \label{NoTachsB}
\forall \  t\in T_1\cup T_2: \ \ \mathbb{P}_t=0.
\eeq 

\subsubsection*{Enhancements}
As in Class A we will ensure the absence of enhancements to the observable gauge factors given from sectors listed in eq. (\ref{ObsEnhs}) as well as the model-dependent sectors 

\beq
\psi^\mu \{\bar{\lambda}\}_{\frac{1}{4}}
\begin{cases}
\ket{\bm{z_1}+(3)\bm{\gamma}}=:\mathbf{O}_1\\
\ket{\bm{z_1}+\bm{x}+(3)\bm{\gamma}}=:\mathbf{O}_2\\
\ket{\bm{z_1}+\bm{z_2}+(3)\bm{\gamma}}=:\mathbf{O}_3\\
\ket{\bm{z_1}+\bm{z_2}+\bm{x}+(3)\bm{\gamma}}=:\mathbf{O}_4
\end{cases}
\eeq
 and as in Class A we ensure the generalised projectors of these sectors are zero, which can be written 
 \beq \label{NoObsEnhB}
 \forall \ i\in[1,4]: \ \ \mathbb{P}_{\mathbf{O}_i}=0.
 \eeq 
\subsubsection*{Exotics}
Along with the $(\bm{\alpha}_L\cdot \bm{\alpha}_L,\bm{\alpha}_R\cdot \bm{\alpha}_R)=(4,4)$ exotic sectors (\ref{Exots}),
there are 112 sectors at the level $(4,6)$ that can produce exotic massless states with a right moving oscillator with $\nu_f=\frac{1}{2}$ or $\nu_{f^*}=-\frac{1}{2}$. As in Model A these all arise in pairs with $+\bm{\gamma}$ and $+3\bm{\gamma}$ with equal and opposite gauge charges and therefore do not contribute to any chiral anomaly. Similarly for 176 sectors at level $(4,8)$. 
Therefore we conveniently do not need to implement a condition on chiral exotics in the classification.

\subsection{Class B Results}
We wish to implement the constraints listed in (\ref{ClassConstraints}) for the case of Class B. However, the absence of 3 generation models in this class means all models break at constraint (4). For completeness, we still present the reduced results in Table \ref{StatstableB}. In order to do a complete scan, we choose to impose condition (\ref{SUSYChiral2}) such that for $\mathcal{N}=0$ models SUSY is broken by phases beyond the NAHE-set. This condition reduces the parameter space to $2^{31}\sim 2.15\times 10^{9}$. We then enumerate all possible configurations of these 31 phases that give both $\mathcal{N}=1$ and $\mathcal{N}=0$ models. 
\begin{table}[!htb]
\small
\centering
\begin{tabular}{|c|l|r|c|c|c|r|}
\hline
 & \multicolumn{5}{|l|}{Total models in sample: $2^{31}=2147483648$}   \\ \hline
  & SUSY or Non-SUSY: & $\mathcal{N}=1$ &Probability& $\mathcal{N}=0$& Probability  \\ \hline
&{Total} & 134217728 &$6.25\times 10^{-2}$&2013265920& $9.38\times 10^{-1}$  \\  \hline
(1)&{+ Tachyon-Free} & \cellcolor{gray!25} &\cellcolor{gray!25}&518921216&$2.42\times 10^{-1}$   \\  \hline
(2)& {+ No Obs. Enhancements} & 121896960 &$5.68\times 10^{-2}$&478915840&$2.23\times 10^{-1}$  \\ \hline
(3)&{+ Complete Generations} &74317824  &$3.46\times 10^{-2}$&271702016&$1.27\times 10^{-1}$   \\  \hline
(8)&{+ $a_{00}=N_b^0-N_f^0=0$}& \cellcolor{gray!25} &\cellcolor{gray!25}& 326042&$1.51\times 10^{-4}$  \\ \hline
\end{tabular}
\caption{\label{StatstableB} \emph{Phenomenological statistics from a complete scan of $2^{31}$ Class B models. Note that the number of $a_{00}=0$ models is an estimate based on extrapolating from a sample of $2.5 \times 10^3$ of the 1245265024 $\mathcal{N}=0$ models satisfying (1)-(3).}}
\end{table}

\begin{figure}[H]
\centering
\includegraphics[width=1\linewidth]{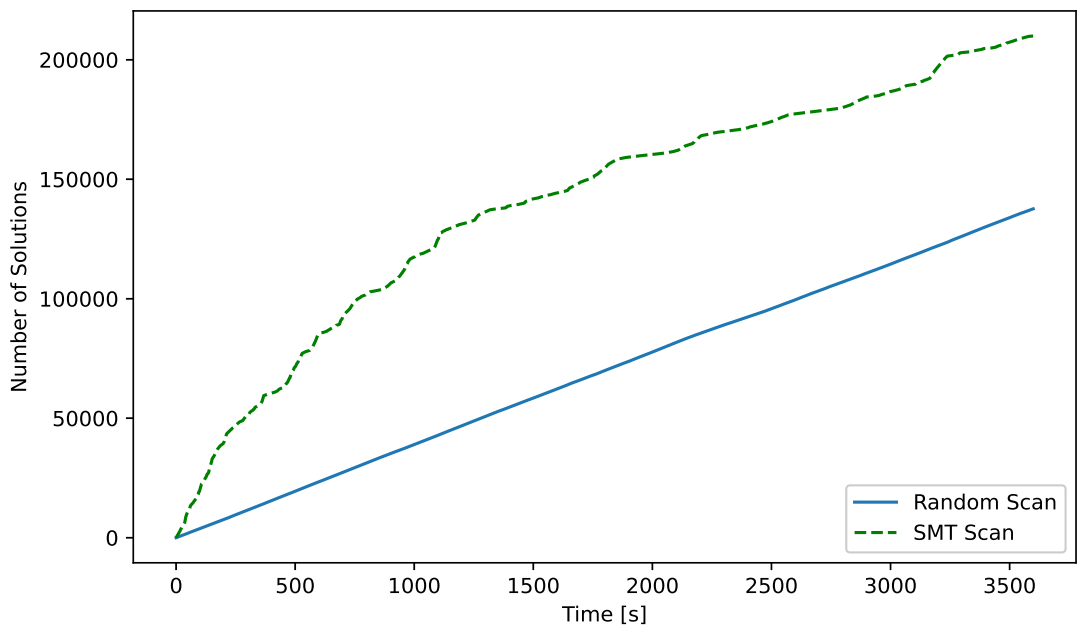}
\caption{\label{ClassBTimings}\emph{Rate at which the Z3 SMT finds 4 generation models satisfying constraints (1)-(3) and (5)-(7) compared with a random generation approach over a 1 hour period.}}
\end{figure}

In order to compare the efficiency of the SMT solver to that of a random scan we can search for four generation models rather than three that satisfy criteria (1)-(3) and (5)-(7) from (\ref{ClassConstraints}). The results of this comparison are shown in Figure \ref{ClassBTimings}. We see that the efficiency gained from the SMT is lower for Class B than the Class A case with efficiency approximately 5.5 times higher compared to the random approach after 3 minutes, reducing to approximately 1.5 times after 1 hour. This reduced efficiency for Class B seems to result from the fewer constraints imposed from the absence of tachyons evidence by the probability $2.42\times 10^{-1}$ for Table \ref{StatstableB} compared to $3.08\times 10^{-2}$ for Table \ref{StatstableA}, as well as the smaller space of models and higher degeneracy meaning the SMT algorithm's search saturates more quickly than in Class A. 

\begin{figure}[H]
\centering
\includegraphics[width=1\linewidth]{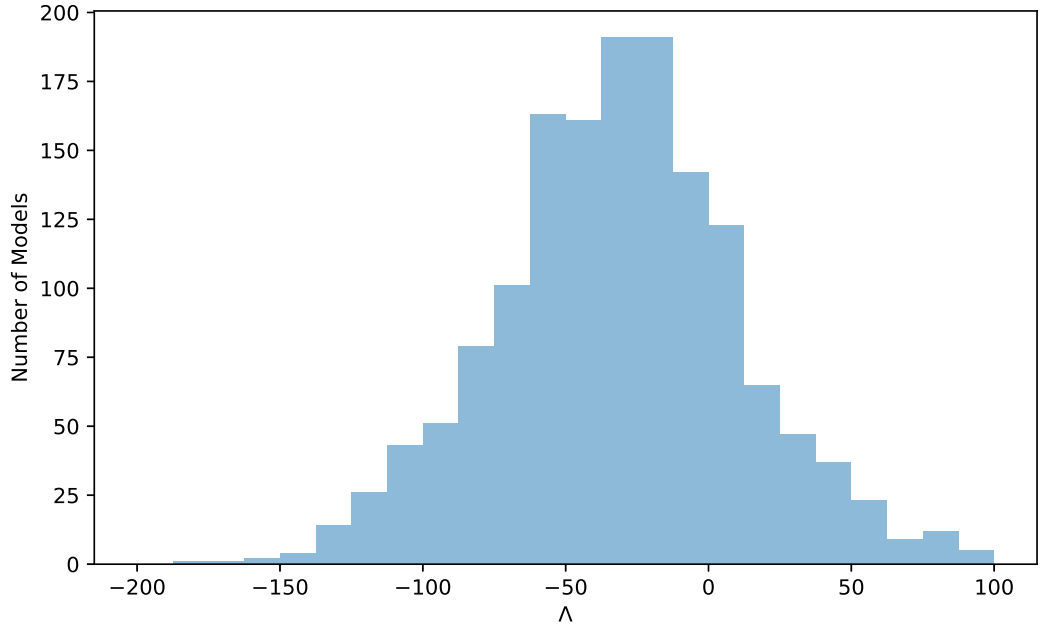}
\caption{\label{ClassBCoC}\emph{The distribution of the cosmological constant $\Lambda_\text{ST}$ for a sample off $1.5\times 10^3$ Class B models satisfying conditions (1)-(3) of Table \ref{StatstableB}. To gain the physical value, a factor of $\mathcal{M}^4$ must be reinstated. These values are evaluated at the free fermionic point using methods discussed in Section \ref{PF}.}}
\end{figure}

\begin{figure}[H]
\centering
\includegraphics[width=1\linewidth]{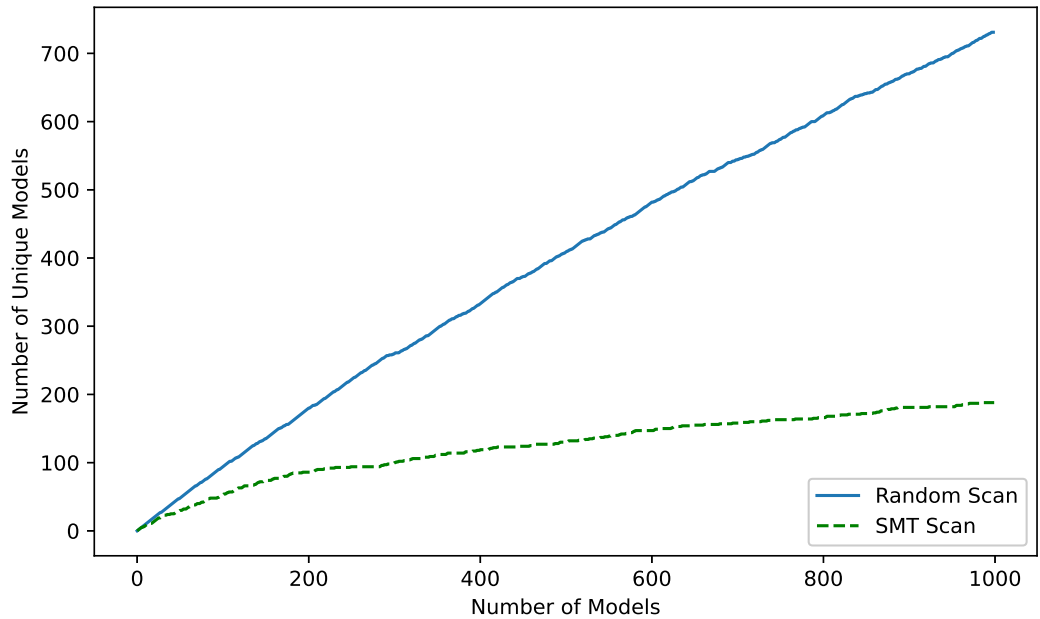}
\caption{\label{ClassBDegeneracy}\emph{The degeneracy of models in a Random versus an SMT scan for Class B.}}
\end{figure}

As in the case of Class A models, it is also interesting to perform a statistical analysis at the level of the partition function. Figure \ref{ClassBCoC} shows the distribution of the cosmological constant for batch of $1.5 \times 10^3$ Class B models satisfying conditions (1)-(3) of Table \ref{StatstableB}. We again note the slight tendency to negative values even though positive values are not excluded. In Figure \ref{ClassBDegeneracy} we see that the SMT algorithm finds relatively more degenerate models as compared to the Class A case. This is mostly due to the reduced number of constraints on the GGSO phases and the increased frequency of solutions as discussed above.

\subsection{Class B Example Model with 4 Generations}
Having discussed the absence of three generation models in this class, we give an example four generation model and discuss its key characteristics. We emphasize that, although this class of models is not phenomenological, they are of particular interest due to the fact that the untwisted moduli of the 2nd and 3rd tori are fixed. The chosen model is defined by the basis (\ref{basisB}) and the GGSO phases 
{\begin{equation}
\small
\CC{\bm{v_i}}{\bm{v_j}}= 
\begin{blockarray}{cccccccccccc}
&\mathbf{1}& \bm{S} & \bm{e_1}& \bm{e_2}&\bm{b_1}& \bm{b_2}&\bm{b_3}&\bm{z_1}& \bm{x}&\bm{\gamma}& \\
\begin{block}{c(rrrrrrrrrrr)}
\mathbf{1}&  1 & -1 &   1 &   1 & -1 & -1 & -1 & -1 &   1 &  -i& \ \\
\bm{S}& -1 & -1 &   1 &   1 & -1 & -1 & -1 &   1 &   1 & -1 & \ \\
\bm{e_1}&   1 &   1 & -1 &   1 & -1 &   1 & -1 & -1 &   1 &   1 & \ \\
\bm{e_2}&   1 &   1 &   1 & -1 & -1 &   1 &   1 &   1 &   1 &   1 & \ \\
\bm{b_1}& -1 &   1 & -1 & -1 & -1 & -1 & -1 & -1 & -1 & -1 & \ \\
\bm{b_2}& -1 &   1 &   1 &   1 & -1 & -1 &   1 &   1 & -1 &  -i& \ \\
\bm{b_3}& -1 &   1 & -1 &   1 & -1 &   1 & -1 & -1 &   1 & -1 & \ \\
\bm{z_1}& -1 &   1 & -1 &   1 & -1 &   1 & -1 & -1 &   1 &  -i& \ \\
\bm{x} &   1 &   1 &   1 &   1 &   1 &   1 & -1 &   1 & -1 & -1 & \ \\
\bm{\gamma}&   1 & -1 &   1 &   1 & -1 & -1 & -1 &   1 & -1 &  -i& \ \\
\end{block}
\end{blockarray}
\end{equation}}
The states from sector $\bm{b_1}$ generate four copies of fermion generations in the $\mathbf{16}$. 
We obtain a Heavy Higgs from the sector $\bm{S}+\bm{b_3}+\bm{e_1}+\bm{e_2}$. 
There is an untwisted-type TQMC from the first orbifold plane and we note the existence of an additional hidden sector gauge boson from $\psi^\mu \{\bar{y}^1\}\ket{z_1}$ which enhances the hidden gauge group
\begin{align}
\begin{split}
& SU(2)\times U(1)_{H_1}\times SO(4) \times U(1)_{H_2}\times  SU(2)\times U(1)_{H_3} \times U(1)_{H_4}\  \\
&\rightarrow \ U(1)_{H_1}\times  SO(5)\times SU(2) \times U(1)_{H_2}\times  SU(2)\times U(1)_{H_3} \times U(1)_{H_4}. 
\end{split}
\end{align}
 
The partition function for this model can be calculated similarly to the Class A model presented in Section \ref{ClassAExample}. The main difference in this case is that the asymmetric shifts introduced by $\bm{A}$ only explicitly include the anti-holomorphic part of the internal lattice in the first and third orbifold plane. That is, the lattice becomes
\begin{align}
 \Gamma^{\gamma}_{(6,6)}=& \;\; \Gamma_1^{\gamma} \times \Gamma_2^{\gamma} \times \Gamma_3^{\gamma} \nonumber\\[0.1cm]
 =& \quad\;\; \vth_{y^{3,4}}\smb{r+h_2}{s+g_2}  \vth_{y^{5,6}}\smb{r+h_2}{s+g_2}\nonumber\\ 
 &\times \; \vth_{\bar{y}^{3,4}}\smb{r+h_2}{s+g_2} \vthb_{\bar{y}^{5,6}}\smb{r+h_2+2H'}{s+g_2+2G'} \nonumber\\[0.3cm]
 &\times \; \vth_{y^1}\smb{r+h_1+H_1}{s+g_1+G_1}^{1/2} \vth_{y^2}\smb{r+h_1+H_2}{s+g_1+G_2}^{1/2} \vth_{w^{5,6}}\smb{r+h_1}{s+g_1}\\[0.1cm]
 &\times \; \vthb_{\bar{y}^1}\smb{r+h_1+H_1}{s+g_1+G_1}^{1/2} \vthb_{\bar{y}^2}\smb{r+h_1+H_2}{s+g_1+G_2}^{1/2} \vthb_{\bar{w}^{5,6}}\smb{r+h_1}{s+g_1}\nonumber\\[0.3cm]
 &\times \; \vth_{w^1}\smb{r-h_1-h_2+H_1}{s-g_1-g_2+G_1}^{1/2}  \vth_{w^2}\smb{r-h_1-h_2+H_2}{s-g_1-g_2+G_2}^{1/2} \vth_{w^{3,4}}\smb{r-h_1-h_2}{s-g_1-g_2}\nonumber\\ 
 &\times\vth_{\bar{w}^1}\smb{r-h_1-h_2+H_1}{s-g_1-g_2+G_1}^{1/2} \vthb_{\bar{w}^2}\smb{r-h_1-h_2+H_2}{s-g_1-g_2+G_2}^{1/2} \vthb_{\bar{w}^{3,4}}\smb{r-h_1-h_2+2H'}{s-g_1-g_2+2G'}, \nonumber 
\end{align}
where $\Gamma_i^{\gamma}$ again denotes the terms corresponding to the $i^\text{th}$ orbifold plane. We see that indeed $\Gamma^\gamma_2$ remains left-right symmetric and that the lack of $\bm{e_{3,4,5,6}}$ simplifies the lattice. Based on this lattice, we can gain the $q$-expansion of the model, which is now given by
\begin{align}
    Z =& \, 2\,q^{0}\bar{q}^{-1}  +56\,q^{1/2}\bar{q}^{-1/2} +208\,q^{-1/2}\bar{q}^{1/2}  \nonumber\\
    &+8\,q^{0}\bar{q}^{0} -192\,q^{1/8}\bar{q}^{1/8} +1280\,q^{1/4}\bar{q}^{1/4} -5632\,q^{1/2}\bar{q}^{1/2},
\end{align}
including all terms up to at most $\mathcal{O}(q^{1/2})$ and $\mathcal{O}(\bar{q}^{1/2})$. We note again the presence of the proto-graviton term with the correct factor and the presence of a constant term $q^{0}\bar{q}^{0}$. There was no model found with $N_b=N_f$ in a sample of $2.5 \times 10^3$ 4 generation models.  Integrating this expansion over the fundamental domain of the modular group via (\ref{CoC}) gives the spacetime cosmological constant
\begin{equation}
\Lambda_\text{ST} = 31.86 \times \mathcal{M}^4,
\end{equation}
which was calculated to $\mathcal{O}(q^{4}\bar{q}^4)$. As in the Class A case, this value is evaluated at the free fermionic self dual point in moduli space. While some moduli are projected by the asymmetric shifts, some of the geometric moduli remain unfixed and require further analysis.

\section{Conclusion}

In this paper we initiated the extension of the fermionic $\mathbb{Z}_2\times \mathbb{Z}_2$ orbifold 
classification method to string vacua with asymmetric boundary conditions. There are 
notable phenomenological advantages for string models with asymmetric boundary conditions, 
among them the stringy Higgs doublet--triplet splitting mechanism \cite{dtsm}
and the top--bottom quark mass hierarchy \cite{tqmp}. Perhaps most notable is the 
fact that asymmetric boundary conditions fix many of the untwisted moduli by projecting
out the moduli fields from the massless spectrum \cite{moduli}. In this respect
we note that there exist cases in which all the untwisted moduli are projected out
\cite{moduli}, as well as cases in which it has been argued the string vacuum is entirely 
fixed, {\it i.e.} cases in which the twisted, as well as the supersymmetric moduli 
are fixed \cite{cleaverwithmanno}. We note that from the point of view of the free fermionic classification methodology, these cases are futile because it entails that
they are not compatible with any of the $\bm{ e_i}$ vectors discussed in Section \ref{Pairings}. 
Our purpose here was therefore to analyse configurations in which some, but not all, of 
the moduli are fixed. This approach is particularly suited in the search for string vacua 
with positive cosmological constant, {\'a} la references \cite{fr1,fr2}. In these 
cases the potential of some of the remaining unfixed moduli is analysed away
from the self--dual point with the aim of finding a vacuum state with a positive
vacuum energy at a stable minimum. Thus, whereas in the case of \cite{fr1,fr2} the
other moduli are unfixed, in the case of vacua with asymmetric boundary conditions 
the possibility exists of finding such vacua in which the other moduli are fixed. 

We would like to remark that the issue of dilaton stabilisation cannot be addressed in the
perturbative heterotic--string limit that we have been analysing in this paper. Indeed, 
it is well known that in this limit the dilaton potential exhibits a run away behavior 
\cite{dineseiberg}. Stabilisation of the dilaton potential at finite value therefore
requires utilisation of nonperturbative effects. It is also known that in the 11 dimensional
supergravity limit of $M$--theory, the finite value of the dilaton may be interpreted as 
the radius of an extra dimension. For our purpose here, we note that the dilaton may be 
stabilised at finite value by using the so--called racetrack mechanism \cite{racetrack}, 
in which gaugino condensation of two competing hidden sector gauge groups are balanced
against each other. We note from eqs. (\ref{GGA}) and (\ref{GGB}) that the hidden sectors 
in the models explored here do indeed contain non--Abelian group factors with similar 
order beta functions that are, in principle, suitable for implementation of the 
racetrack mechanism. Detailed analysis of dilaton stabilisation is beyond our scope in 
this paper, but we note the more detailed implementation of the racetrack mechanism 
in the context of string inspired phenomenological models \cite{rossetal}.

We comment further here on the issue of physical tachyonic states in the string spectrum. 
The models scanned by the free fermionic classification method are analysed at the 
free fermionic point in the moduli space. At that point the models presented are 
free of physical tachyons, which are projected out by the GGSO projections. However, 
in principle, it is not guaranteed that moving away from that in the moduli scape will
regenerate physical tachyonic states. Two comments are in order. The first is that naively 
we expect the free fermionic point to have the maximal number of physical tachyons. The reason
being that this is the most symmetric point in the moduli space and the maximal number
of physical massless states are generated at this point in the moduli space. Moving away 
from the free fermionic point entails that some radii are increased from their minimal 
value at the self--dual free fermionic point and hence their contribution to the masses 
of the physical states is increased. This argument is in fact in accordance with the 
results presented in ref. \cite{fr1,fr2} which presented free fermionic heterotic--string 
models that are tachyonic at the free fermionic point but are tachyon free when the moduli
are moved from that point. 

In the classification of vacua with asymmetric boundary conditions, there exist 
a variation in the pairings of the holomorphic worldsheet fermions. We presented a 
complete classification of all the possible pairings, consistent with modular 
invariance and worldsheet supersymmetry, and picked two of these choices for 
detailed classification. We showed the existence of three generation quasi-realistic 
models in the first case, whereas the second case did not produce any three generation 
models. In both cases, the incorporation of asymmetric boundary conditions was done
in a single basis vector, whereas the remaining basic set, aside from the set 
of the $\bm{e_i}$ basis vectors that are compatible with the given pairings, were 
identical in the two cases. We note that in principle this can be relaxed, {\it e.g.}
by not including the vector $\bm{z_1}$ in the basis, and that three generation model
might be obtainable with this variation, we leave such variations for future work. We note, however, that the program initiated herein opens the door to the systematic investigation
of quasi--realistic vacua that are intrinsically non--geometric. We furthermore demonstrated effective applications of SMT algorithms to the space of free fermionic models under investigation. Not only do they provide significant efficiency increases, as demonstrated in Figure \ref{ClassATimings} and \ref{ClassBTimings}, but they also allowed for immediate evaluation of unsatisfiable constraints, such as proving the absence of three generation models in Class B. 

Other than the systematic study of the one-loop potential for asymmetric models mentioned as a key motivation for this work, future work classifying Standard-like models (SLMs) with asymmetric boundary conditions is a natural extension of this work. In that context the role of asymmetric pairings in the (untwisted) Doublet-Triplet splitting mechanism \cite{dtsm} will be evident, in a way it is not for the FSU5 models studied here. The space of asymmetric SLMs will be larger and phenomenologically viable models more sparsely distributed, thus the application of SMT algorithms could prove instrumental in effective searches of this landscape. The analysis of Section \ref{Pairings} can be extended so that the SMT can explicitly interpret phenomenological constraints as a function of all asymmetric pairings and provide generic results, including no-go theorems, over a varied space of models. It will furthermore be interesting to explore different possibilities for how to implement the asymmetric boundary conditions other than solely through the $SO(10)$ breaking vector as studied in this work. 

\section*{Acknowledgments}

We would like to thank Sven Schewe for fruitful discussions regarding the application of SMT algorithms.
AEF is supported in part by a Weston visiting professorship at the Weizmann Institute 
of Science and would like to thank Doron Gepner and the Department of Particle Physics 
and Astrophysics for hospitality. 
The work of BP is supported in part by STFC grant ST/N504130/1 
and the work of VGM is supported in part by EPSRC grant EP/R513271/1.

\newpage

\bibliographystyle{unsrt}

\begin{thebibliography}{50}


\bibitem{z2z2}
\AEF, \PLB{326}{1994}{62};\\
E. Kiritsis and C. Kounnas, \NPB{503}{1997}{117};\\
\AEF, S. Forste and C. Timirgaziu, \JHEP{0608}{2006}{057};\\
R. Donagi and K. Wendland, {\it J.Geom.Phys.}\/ {\bf  59} (2009) 942;\\
P. Athanasopoulos, \AEF, S. Groot Nibbelink and V.M. Mehta, \JHEP{1604}{2016}{038}.
                              
\bibitem{panosdic} P. Athanasopoulos, \AEF, S.G. Nibbelink, and V. Mehta, \JHEP{1007}{2016}{38}.

\bibitem{fff}
I. Antoniadis, C. Bachas, and C. Kounnas, \NPB{289}{1987}{87};\\
H. Kawai, D.C. Lewellen, and S.H.-H. Tye, \NPB{288}{1987}{1};\\
I. Antoniadis and C. Bachas, \NPB{298}{1988}{586}.

\bibitem{fsu5} I. Antoniadis, J. Ellis, J. Hagelin and D.V. Nanopoulos,
\PLB{231}{1989}{65}.


\bibitem{fny} 
A.E. Faraggi, D.V. Nanopoulos and K. Yuan, \NPB{335}{1990}{347};\\
\AEF, \PRD{46}{1992}{3204};\\
G.B. Cleaver, A.E. Faraggi and D.V. Nanopoulos, \PLB{455}{1999}{135}.

\bibitem{slm} 
A.E. Faraggi, \PLB{278}{1992}{131}; \NPB{387}{1992}{239};\\
\AEF, E. Manno and C.M. Timirgaziu, \EJP{50}{2007}{701}.                        

\bibitem{alr} 
I. Antoniadis. G.K. Leontaris and J. Rizos,\PLB{245}{1990}{161};\\
G.K. Leontaris and J. Rizos, \NPB{554}{1999}{3}.

\bibitem{lrs}
G.B. Cleaver, A.E. Faraggi and C. Savage, \PRD{63}{2001}{066001};\\
G.B. Cleaver, D.J Clements and A.E. Faraggi, \PRD{65}{2002}{106003}.

\bibitem{fknr} 
\AEF, C. Kounnas, S.E.M Nooij and J. Rizos, \NPB{695}{2004}{41}.

\bibitem{fkr1}
\AEF, C. Kounnas and J. Rizos, \PLB{648}{2007}{84}.             

\bibitem{fkr2}
\AEF, C. Kounnas and J. Rizos, \NPB{774}{2007}{208}; \NPB{799}{2008}{19}. 

\bibitem{svd} T.Catelin--Julian \etal, \NPB{812}{2009}{103};\\
C. Angelantonj, \AEF~and M. Tsulaia, \JHEP{1007}{2010}{314};\\
\AEF, I. Florakis, T. Mohaupt and M. Tsulaia, \NPB{848}{2011}{332}. 

\bibitem{acfkr}
B. Assel \etal,   \PLB{683}{2010}{306}; \NPB{844}{2011}{365}; \\
C. Christodoulides, A.E. Faraggi and J. Rizos, \PLB{702}{2011}{81}.

  
\bibitem{frs}
\AEF, J. Rizos and H. Sonmez, \NPB{886}{2014}{202}; \\
H. Sonmez, \PRD{93}{2016}{125002}.

\bibitem{slmclass} 
\AEF, J. Rizos and H. Sonmez, \NPB{927}{2018}{1}.

\bibitem{lrsclass} 
\AEF, G. Harries and J. Rizos, \NPB{936}{2018}{472}. 

\bibitem{su62} 
L. Bernard \etal, \NPB{868}{2013}{1}.

\bibitem{frzprime}
\AEF \ and J. Rizos, \NPB{895}{2015}{233}.

\bibitem{so10tclass} 
\AEF, V.G. Matyas and B. Percival, \NPB{961}{2020}{115231}.

\bibitem{PStclass}  
\AEF, V.G. Matyas and B. Percival, \PRD{104}{2021}{046002}.

\bibitem{type0}  
\AEF, V.G. Matyas and B. Percival, \IJMP{36}{2021}{2150174}.

\bibitem{type0bar}  
\AEF, V.G. Matyas and B. Percival, \PLB{814}{2021}{136080}.

\bibitem{NSUSYBranes1}
I. Antoniadis, E. Dudas and A. Sagnotti, \PLB{464}{1999}{38}.

\bibitem{NSUSYBranes2}
C. Angelantonj, \NPB{566}{2000}{126}.

\bibitem{NSUSYBranes3}
C. Angelantonj, I. Antoniadis, G. D'Appollonio, E. Dudas and A. Sagnotti, \NPB{572}{2000}{36}. 

\bibitem{NSUSYBranes4}
S. Parameswaran and F. Tonioni, \JHEP{12}{2020}{174}.

\bibitem{SS1}
C. Kounnas and M. Porrati, \NPB{310}{1988}{355}.

\bibitem{SS2}
J. Scherk and J. H. Schwarz, \PLB{82}{1979}{60}. 

\bibitem{SS3}
J. Scherk and J. H. Schwarz, \NPB{153}{1979}{61}.

\bibitem{CDC2}
S. Ferrara, C. Kounnas and M. Porrati \NPB{304}{1988}{500}.

\bibitem{CDC4}
S. Ferrara, C. Kounnas, M. Porrati and F. Zwirner \NPB{318}{1989}{75}.

\bibitem{CDC1}
C. Kounnas and B. Rostand, \NPB{341}{1990}{641}.

\bibitem{CDC3}
S. Ferrara, C. Kounnas and M. Porrati, \PLB{206}{1988}{25}.

\bibitem{Interpol1}
H. Itoyama and T.R. Taylor, \PLB{186}{1987}{129}.

\bibitem{Interpol2}
\AEF and M. Tsulaia, \PLB{683}{2010}{314}.

\bibitem{Interpol3}
S. Abel, K.R. Dienes and E. Mavroudi, \PRD{91}{2015}{126014}.

\bibitem{Interpol4}
B. Aaronson, S. Abel and E. Mavroudi, \PRD{95}{2017}{106001}.

\bibitem{CoCSuppression1}
H. Itoyama and S. Nakajima, \PLB{816}{2021}{136195}.

\bibitem{CoCSuppression2}
H. Itoyama and S. Nakajima, \NPB{958}{2020}{115111}.

\bibitem{CoCSuppression3}
H. Itoyama and S. Nakajima, \emph{Prog. of Theor. and Exp. Phys.} \textbf{2019} 12 (2019) 123B01.

\bibitem{CoCSuppression4}
S. Abel and R.J. Stewart, \PRD{96}{2017}{106013}.

\bibitem{SO16}
L. Alvarez--Gaume, P.H. Ginsparg, G.W. Moore and C. Vafa, \PLB{171}{1986}{155}.

\bibitem{tach10d}
L.J. Dixon, J.A. Harvey, \NPB{274}{1986}{93};\\
H. Kawai, D.C. Lewellen and S.H.H. Tye, \PRD{34}{1986}{3794}. 

\bibitem{spwsp} 
\AEF, \EJP{79}{2019}{703}.

\bibitem{stable} 
\AEF, V.G. Matyas and B. Percival \EJP{80}{2020}{337}.

%




\bibitem{aafs} 
J.M. Ashfaque, P. Athanasopoulos, \AEF~and H. Sonmez, \EJP{76}{2016}{208}.

\bibitem{ferlrs}
\AEF, G. Harries, B. Percival and J. Rizos \NPB{953}{2020}{1169}.

\bibitem{genalgo}
S. Abel and J. Rizos, \JHEP{08}{2014}{010}.



\bibitem{MLReview}{\it For review and references see e.g.}: F. Ruehle, \PRT{839}{2020}{1}.  

\bibitem{fpsw} 
\AEF, B. Percival, S. Schewe and D. Wojtczak (2021), \PLB{816}{2021}{136187}.

\bibitem{Reps} 
D. C. Lewellen, \NPB{337}{1990}{61};\\
K. R. Dienes, J. March-Russell, \NPB{479}{1996}{113}.

\bibitem{nongeomreview}
E. Plauschinn, \PRT{798}{2019}{1}. 

\bibitem{moduli}
\AEF, \NPB{728}{2005}{83}

\bibitem{yukawa}
\AEF, \PRD{47}{1993}{5021}. 

\bibitem{dtsm}
\AEF, \NPB{428}{1994}{111}; \PLB{520}{2001}{337}.

\bibitem{tqmp}
\AEF, \PLB{274}{1992}{47}; \PLB{377}{1996}{43}. 

\bibitem{KP}
C. Kounnas and H. Partouche, \NPB{919}{2017}{41};\\
C. Kounnas and H. Partouche, \NPB{913}{2016}{593}.

\bibitem{ADM}
S. Abel, K. R. Dienes and E. Mavroudi, 
Phys. Rev. D 91 (2015) 126014.

\bibitem{fr1} 
I.~Florakis and J.~Rizos, \NPB{913}{2016}{5}.

\bibitem{fr2} 
I.~Florakis, J.~Rizos and K. Violaris-Gountonis, arXiv:2110.06752. 




\bibitem{florakis}
I. Florakis, “Th\'eorie de Cordes et Applications Ph\'enom\'enologiques et Cosmologiques,” PhD Thesis, Universit\'e Pierre et Marie Curie, 2011.

\bibitem{Z3}
L. de Moura, N. Bjørner, Tools and Algorithms for the Construction and Analysis of Systems, TACAS 2008, Lecture Notes in Computer Science, vol 4963. Springer, Berlin, Heidelberg; DOI: https://doi.org/10.1007/978-3-540-78800-3\_24.
%


\bibitem{NAHE}
\AEF~and D.V. Nanopoulos, \PRD{48}{1993}{3288}.

\bibitem{towards}
\AEF, \IJMP{14}{1999}{1663}.

\bibitem{fsu5masses} 
I. Antoniadis, J. Rizos and K. Tamvakis, \PLB{278}{1992}{257}.

\bibitem{ckm} \AEF~and E. Halyo, \NPB{416}{1994}{63}.

\bibitem{tqmc}
J. Rizos, \EJP{74}{2014}{2905}.

\bibitem{DienesCoC}  K.R. Dienes, \PRL{65}{1990}{1979}.

\bibitem{MSUSYDienes1} K.R. Dienes, \NPB{429}{1994}{533}.

\bibitem{MSUSYDienes2}  K.R. Dienes, M. Moshe and R.C. Myers, \PRL{74}{1995}{4767}.

\bibitem{MSUSYAngelantonj} C. Angelantonj, M. Cardella, S. Elitzur and E. Rabinovici,  \JHEP{2}{2011}{24}.

\bibitem{MSUSYFlavio1} N. Cribiori, S. Parameswaran, F. Tonioni and T. Wrase, \JHEP{04}{2021}{099}.

\bibitem{MSUSYFlavio2} N. Cribiori, S. Parameswaran, F. Tonioni and T. Wrase,  \JHEP{01}{2022}{127}.

\bibitem{Pico}
{\it Program and documentation found at}: https://github.com/zimmski/picosat.



\bibitem{UnsatCore}
A. Cimatti, A. Griggio and R. Sebastiani, 
{\it Journal Artificial Intelligence Res.}  {\bf 40}, 1 (2001) 701–728.

\bibitem{cleaverwithmanno} G.B Cleaver, \AEF, E. Manno and C. Timirgaziu, 
\PRD{78}{2008}{046009}. 

\bibitem{dineseiberg} M. Dine and N. Seiberg, \PRL{55}{1985}{366}; 
                                              \PLB{162}{1985}{299}. 
                                            
\bibitem{racetrack} N.V. Krasnikov, \PLB{193}{1987}{37}; \\ 
 L.J. Dixon, {\it Supersymmetry breaking in string theory}, in The Rice Meeting Proceedings, B. Bonner and H. Miettinen, World Scientific (Singapore) 1990. 
 
 \bibitem{rossetal} 
 J.A. Casas, Z. Lalak, C. Munoz and G.G. Ross, \NPB{347}{1990}{243};\\ 
 B. de Carlos, J.A. Casas and C. Munoz, \NPB{399}{1993}{623}.                                              

\end{thebibliography}

\end{document}